\documentclass[12pt]{article}
\input epsf
\usepackage{color,graphics,latexsym,amsfonts}
\newcommand{\sect}[1]{\section{#1}\setcounter{equation}{0}}

\begin{document}
\def\aie{{\em i.e.$\,$}}
\def\aeg{{\em e.g.$\,$}}
\def\aEg{{\em E.g.$\,$}}
\newcommand{\ul}[1]{\underline{#1}}
\newcommand{\ol}[1]{\overline{#1}}
\def\boxit#1#2{\fbox{\begin{minipage}[c]{#1}#2\end{minipage}}}
\newcommand\bmp[1]{\begin{minipage}[c]{#1\textwidth}}
\newcommand\emp{\end{minipage}}
\def\half{\mbox{\scriptsize{${{1}\over{2}}$}}}
\def\halff{\mbox{\scriptsize{${{5}\over{2}}$}}}
\def\quarter{\mbox{\scriptsize{${{1}\over{4}}$}}}
\def\eighth{\mbox{\scriptsize{${{1}\over{8}}$}}}
\def\gs{g_{\rm{s}}}
\def\ls{\ell_{\rm{s}}}
\def\lP{\ell_{\rm{P}}}
\def\ld{\ell_{{d}}}
\def\TH{T_{\rm{H}}}
\def\SBH{S_{\rm{BH}}}
\def\lC{\lambda_{{C}}}
\def\rH{r_{{H}}}
\def\RH{R_{{H}}}
\def\gYM{g_{\rm{YM}}}
\def\Gopen{G_{\rm{open}}}
\def\dnopen{d_n^{\rm{open}}} 
\def\lra{\leftrightarrow}
\def\bs{\vspace{5pt}}
\def\be{\begin{equation}}
\def\ee{\end{equation}}
\def\ba{\begin{array}}
\def\ea{\end{array}}
\def\bc{\begin{center}}
\def\ec{\end{center}}
\def\bfig{\begin{figure}}
\def\efig{\end{figure}}
\def\bZ{{\mathbb{Z}}}
\def\bR{{\mathbb{R}}}
\def\bP{{\mathbb{P}}}
\def\cN{{\cal{N}}}
\def\eql{=}
\def\pls{+}
\def\mns{-}
\def\pr{Phys.~Rev.}
\def\prl{Phys.~Rev.~Lett.}
\def\np{Nucl.~Phys.}
\def\pl{Phys.~Lett.}
\def\cqg{Class.~Quant.~Grav.}
\def\jhep{JHEP}
\def\atmp{Adv.~Theor.~Math.~Phys.}
\def\prpt{Phys.~Rept.}
\def\cmp{Comm.~Math.~Phys.}
\def\ijmp{Int.~J.~Mod.~Phys.}
\def\mpl{Mod.~Phys.~Lett.}
\def\mnras{Mon.~Not.~R.~Astron.~Soc.}
\def\anp{Ann.~Phys.}
\def\npps{Nucl.~Phys.~Proc.~Suppl.}
\hyphenation{Schwarz-schild}
\hyphenation{Be-ken-stein}
\hyphenation{Haw-king}
\begin{titlepage}

\begin{flushright}
hep-th/0008241\\
UTPT-00-10\\
NSF-ITP-99-145
\end{flushright}

\vfil

\bc

{\bf\Huge{TASI Lectures \\on\\ Black Holes in String Theory\\}}

\vskip0.5truein

Amanda W. Peet\\
{\it Department of Physics, \\University of Toronto, \\
60 St.$\!\!$ George St., \\ Toronto, ON, M5S 1A7, \\ Canada.}

\vfill

\abstract{This is a write-up of introductory lectures on black holes
in string theory given at TASI-99. Topics discussed include: black
holes, thermodynamics and the Bekenstein-Hawking entropy, the
information problem; supergravity actions, conserved quantum numbers,
supersymmetry and BPS states, units, duality and dimensional
reduction, solution-generating; extremal M-branes and D-branes,
smearing, probe actions, nonextremal branes, the Gregory-Laflamme
instability; breakdown of supergravity and the Correspondence
Principle, limits in parameter space, singularities; making black
holes with branes, intersection-ology, the harmonic function rule,
explicit $d{=}5,4$ examples; string computations of extremal black
hole entropy in $d{=}5,4$, rotation, fractionation; non-extremality
and entropy, the link to BTZ black holes, Hawking radiation and
absorption cross-sections in the string/brane and supergravity
pictures.}

\vfill

\ec

\end{titlepage}
\newpage
\tableofcontents

\vskip3.0truein

\bc
{\boxit{0.9\textwidth}{
\bc
{\it  ``Whaia te iti kahura{\b{k}}i;\\
Ki te tuohu koe, me he mau{\b{k}}a teitei''}\\
``Aspire to the highest pinnacles;\\
If you should bow, let it be to a lofty mountain'' \\
\ec 
In Maori culture\footnote{Maori are the indigenous peoples in Aotearoa
(New Zealand)}, and in others throughout the world, the mountain is
revered and respected for its mana, awesome presence and sheer
majesty.  This proverbial saying, then, encapsulates all that my
grandfather has meant to me; he has been my lofty mountain. His
wisdom, knowledge and guidance encouraged me throughout my life in the
pursuit of excellence.  I therefore dedicate this review to him.  
}}
\ec

\vfill
\noindent Dedication crafted by Maurice Gray, Kaumatua, Te
Runa{\b{k}}a Ki Otautahi O {\b{K}}ai Tahu, and gifted to the author.

\newpage\setcounter{footnote}{0}
\sect{GR black holes, and thermodynamics}\label{sectone}

Black holes have long been objects of interest in theoretical physics,
and more recently also in experimental astrophysics.  Interestingly,
study of them has led to new results in string theory.  Here we will
study black holes and their $p$-brane cousins in the context of string
theory, which is generally regarded as the best candidate for a
unified quantum theory of all interactions including gravity.  Other
approaches to quantum gravity, such as ``quantum geometry'', have been
recently discussed in works such as \cite{ashtekarrovellidyntriang}.
Other relatively recent reviews of black hole entropy in string theory
have appeared in \cite{garyreviews,juanthesis,awpcqgetal}.

Black holes may arise in string theory with many different conserved
quantum numbers attached.  We will begin our discussion by studying
two basic black holes of General Relativity; they are special cases of
the string theory black holes.

Note that the units we will use throughout are such that only
$\hbar=c=k_B=1$; we will not suppress powers of the string coupling
$\gs$, the string length $\ls$, or the Newton constant $G$.

\subsection{Schwarzschild black holes}

The Schwarzschild metric is a solution of the $d=4$ action
\be
S= {{1}\over{16\pi{}G_4}}\int d^4x \sqrt{-g}R[g] \,.
\ee
The field equations following from this action are
the source-free ($T_{\mu\nu}=0$) Einstein equations
\be
R_{\mu\nu} - \half g_{\mu\nu}R = 0 \,.
\ee
In standard Schwarzschild coordinates, the metric takes the form
\be 
ds^2 = -\left(1 - {{\rH}\over{r}}\right) dt^2 + {\left(1 -
{{\rH}\over{r}}\right)}^{-1} dr^2 + r^2 d\Omega_2^2 \,.
\ee

Astrophysical black holes formed via gravitational collapse have a
lower mass limit of a few solar masses.  However, we will be
interested in all sizes of black holes, for theoretical reasons; we
will not discuss any mechanisms by which `primordial' black holes
might have formed.  When we move to discussion of charged black holes,
we will also ignore the fact that any astrophysical charged black hole
discharges on a very short timescale via Schwinger pair production.
The reader unhappy with this should simply imagine that the charges we
put on our black holes are not carried by light elementary quanta in
nature such as electrons.

Not all massive objects are black holes.  In order for a small object
to qualify as a black hole, we need at a minimum that its
Schwarzschild radius be larger than its Compton wavelength,
$\rH\!\gg\!\lC{\eql}m^{-1}$.  This implies that
$m\!\gg\!G_4^{-1/2}{\eql}m_{\rm{Planck}}$.  So the electron, which is
about $10^{-23}$ times lighter than the Planck mass, does not qualify.

The event horizon of a stationary black hole geometry occurs where
\be
g^{rr} = 0 \,.
\ee
For the Schwarzschild solution, the above condition is the same as the
condition $g_{tt}=0$ but in general, \aeg for the Kerr black hole, the
two conditions do not coincide.  Note also that for an evolving
geometry the event horizon does not even have a local definition; it
is a global concept.  In the present static case, solving for the event
horizon locus we find a sphere, and the radius is in Schwarzschild
coordinates
\be
\rH = 2 G_4 M \,.
\ee
Although metric components blow up at $r=\rH$, the horizon is only a
coordinate singularity, as we can see by computing curvature
invariants.  Note that the source-free Einstein equations imply that
the Ricci scalar ${\cal{R}}=0$ and so the Ricci tensor $R_{\mu\nu}=0$.
For the Riemann tensor we get
\be
R^{\mu\nu\lambda\sigma}R_{\mu\nu\lambda\sigma} = {{12\rH^2}\over{r^6}}
\rightarrow \left\{ 
\ba{lll} 
12\rH^{-4} & {\mbox{at}} & r=\rH \cr
 \infty & {\mbox{at}} & r=0 
\ea  \right. \,.
\ee
Therefore, the curvature at the horizon of a big black hole is weak,
and it blows up at $r=0$, the physical singularity.

The Carter-Penrose diagram in Fig.\ref{schwpenr} shows the causal
structure of the eternal Schwarzschild black hole spacetime.  Note
that, following tradition, only the $(t,r)$ plane is drawn, so that
there is an implicit $S^2$ at each point.  In gravitational collapse
only part of this diagram is present, and it is matched onto a region
of Minkowski space.  In collapse situations there is of course no time
reversal invariance, and so the Carter-Penrose diagram is not
symmetric.

\bfig \hskip0.2\textwidth\epsfysize=1.0truein
\epsfbox{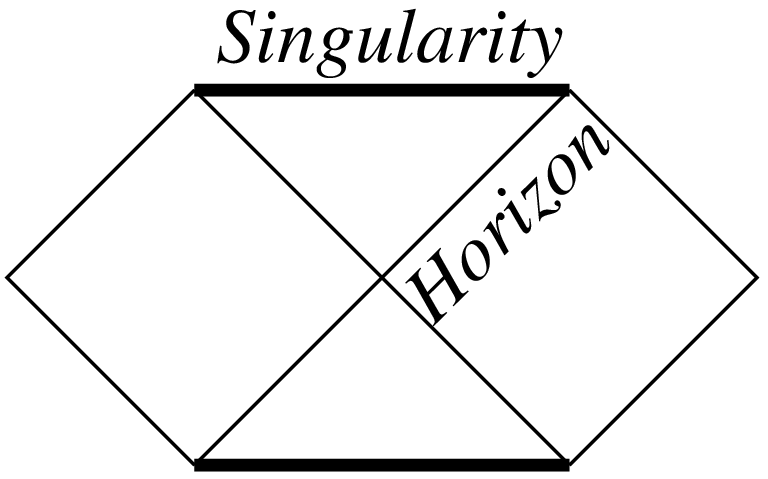}
\caption{\small{The Penrose diagram for an eternal 
Schwarzschild black hole.}}
\label{schwpenr}
\efig

The Schwarzschild geometry is asymptotically flat, as can be seen by
inspection of the metric at large-$r$.  Let us now inspect the
geometry near the horizon.  Define $\eta$ to be the proper distance,
\aie $g_{\eta\eta}=1$.  Then
\be
\eta = \sqrt{r(r-\rH)} +  \rH {\mbox{arccosh}}(\sqrt{r/\rH}) \,.
\ee
Near $r=\rH$, $\eta\sim 2\sqrt{\rH(r-\rH)}$.  Now rescale time,
\be
\omega = {{t}\over{2\rH}}\,;
\ee
the metric becomes 
\be
ds^2 \sim -\eta^2d\omega^2 + d\eta^2 + \rH^2d\Omega_2^2 \,.
\ee
{}From this form of the metric it is easy to see that if we Wick
rotate $\omega$, we will avoid a conical singularity if we identify
the Euclidean time $i\omega$ with period $2\pi$.  Now, in field theory
applications, we have the formal identification of the Euclidean
Feynman path integral with a statistical mechanical partition
function, and the periodicity in Euclidean time is identified as the
inverse temperature.  Tracing back to our original coordinate system,
we identify the black hole temperature to be
\be\label{schwhaw}
\TH = {{1}\over{8\pi G_4M}} \,.
\ee
This is the Hawking temperature of the black hole.

The use of Euclidean methods in quantum gravity has been discussed in,
for example, \cite{euclidean1}.  There can be subtleties in doing a
Wick rotation, however, which may mean that it is not a well-defined
operation in quantum gravity in general.  One thing which can go wrong
is that there may not exist a Euclidean geometry corresponding to the
original geometry with Lorentzian signature.  In addition, smooth
Euclidean spaces can turn into singular Lorentzian ones upon Wick
rotation.  

In any case, the result for the Hawking temperature as derived here
can easily be replicated by other calculations, see \aeg the recent
review of \cite{waldreview}.  These results also tell us that the
black hole radiates with a thermal spectrum, and that the Hawking
temperature is the physical temperature felt by an observer at
infinity.

Notice from (\ref{schwhaw}) that $\TH$ increases as $M$ decreases, so
that the specific heat is negative.  This gives rise to runaway
evaporation of the black hole at low mass.  We can compute the
approximate lifetime of the black hole from its luminosity, using the
fact that it radiates (roughly) like a blackbody,
\be
-{{dM}\over{dt}} \sim {\rm{(Area)\ }} \TH^4 \sim 
\left(G_4M\right)^{2-4}
\quad\Rightarrow\quad \Delta t \sim G_4^2 M^3 \,.
\ee
For astrophysical-sized black holes, this is much longer than the age
of the Universe.  For small black holes, however, there is a more
pressing need to identify the endpoint of Hawking radiation.  We will
have more to say about this topic later when we discuss the
Correspondence Principle.  To find some numbers on what constitutes a
`small' vs. `large' black hole in the context of evaporation, let us
restore the factors of $\hbar,c$.  We obtain an extra factor in the
denominator of $c^4\hbar$ in the expression for $\Delta t$.  The
result is that the mass of the black hole whose lifetime is the age of
the universe, roughly 15 billion years, is $\sim\!10^{12}$kg.  Such a
black hole has a Schwarzschild radius of about a femtometre.

\subsection{Reissner-Nordstr{\"{o}}m black holes}

For the case of Einstein gravity coupled to a $U(1)$ gauge field, both
the metric and gauge field can be turned on
\be\ba{rl}
\bs
ds^2 =&  -\Delta_+(\rho) \Delta_-(\rho) dt^2 
+ \Delta_+(\rho)^{-1} \Delta_-(\rho)^{-1} d\rho^2 + \rho^2d\Omega_2^2
\,\cr
F_{t\rho} =& {\displaystyle{ {{Q}\over{\rho^2}}  }} \,\cr
\Delta_\pm(\rho) &= {\displaystyle{ 
 		\left( 1 - {{r_\pm}\over{\rho}}\right) }} \,\qquad
r_\pm = G_4 \left( M \pm \sqrt{M^2-Q^2}\right) \,.
\ea\ee
There are two horizons, located at $r=r_+$ and $r=r_-$.

Cosmic censorship requires that the singularity at $r=0$ be hidden
behind a horizon, i.e.
\be
M \geq |Q| \,.
\ee
The Hawking temperature is
\be
\TH = {{\sqrt{M^2-Q^2}}\over{2\pi G_4
\left(M+\sqrt{M^2-Q^2}\right)^2}} \,.
\ee
Notice that the extremal black hole, with $r_+{\eql}r_-$,
i.e. $M{\eql}|Q|$, has zero temperature.  It is a stable object, as it
does not radiate.  A phenomenon closely related to this and our
previous result for Schwarzschild black holes is that the specific
heat at constant charge $c_{Q}$ is not monotonic.  Specifically,
\be\ba{lll}
c_Q > 0 & {\mbox{for}} & M-|Q|\ll |Q|  \,,\cr
c_Q < 0 & {\mbox{for}} & M\gg|Q| 
\qquad\qquad{\mbox{like Schwarzschild}} \,.
\ea\ee
Consider the extremal geometry, and let the double
horizon be at $r_0$.  Change the radial coordinate to
\be
r \equiv \rho - r_0 \,;
\ee
then 
\be
\Delta_\pm = 1 - {{r_0}\over{\rho}} = 
\left( 1 + {{r_0}\over{r}} \right)^{-1}
\equiv H(r)^{-1}
\quad{\mbox{and}}\quad
\rho^2 = r^2\left(1+{{r_0}\over{r}}\right)^2  \,,
\ee
so that
\be\label{rniso}
ds^2_{\rm{ext}} = -H(r)^{-2} dt^2 + H(r)^2
\left(dr^2+r^2d\Omega_2\right)^2
\,.
\ee
We see that in these coordinates there is manifest $SO(3)$ symmetry;
they are known as isotropic coordinates.

The extremal black hole geometry has an additional special property.
Near the horizon $r=0$,
\be\ba{rl}
\bs
ds^2 = &\! {\displaystyle{
  -\left({{r}\over{r+r_0}} \right)^2 dt^2 +
\left(1+{{r_0}\over{r}}\right)^2\left(dr^2+r^2d\Omega_2\right)^2 
}} \cr
       & \rightarrow {\displaystyle{
-{{r^2}\over{r_0}^2}dt^2 + {{r_0^2}\over{r^2}}dr^2 
+ r_0^2d\Omega^2 }} \,.
\ea\ee
Defining yet another new coordinate
\be 
z \equiv {{r_0^2}\over{r}} \,,
\ee
so that $dz/z=dr/r$, we find a direct product of an anti-deSitter
spacetime with a sphere:
\be\ba{rl}
ds^2 & \rightarrow 
{\underbrace{ {{r_0^2}\over{z^2}}\left( -dt^2 + dz^2 \right)}} +
{\underbrace{ r_0^2d\Omega_2^2 }} \,. \\
{} & \qquad\qquad AdS_2\qquad \times\quad S^2 \,.
\nonumber
\ea\ee
Since the Reissner-Nordstr{\"{o}}m spacetime is also asymptotically
flat, we see that it interpolates between two maximally symmetric
spacetimes \cite{gibbons1}.  In the units we use here, $M=|Q|$, which
is a special relationship between the bosonic fields in the
Lagrangian.  It turns out that this means that the RN black hole
possesses a supersymmetry, something about which we will have more to
say in subsection (\ref{susysubsec}).

\subsection{Semiclassical gravity and black hole thermodynamics}

Given some assumptions about the field content of the Lagrangian,
classical no-hair theorems for black holes can be derived; see
e.g. \cite{nohairrefs} for a modern treatment.  For example, if there
is a $U(1)$ gauge field minimally coupled to Einstein gravity in
$d{\eql}4$, then the no-hair theorem states that an observer outside
the black hole can measure only the mass $M$, charge $Q$, and angular
momentum $J$ of a black hole.  These are the conserved quantum numbers
associated to the long-range fields in the Lagrangian.  The very
limited amount of long-range hair means that, classically, we have a
very limited knowledge of the black hole from the outside.  Also, a
black hole could have been formed via a wide variety of processes.
This suggests that a black hole will possess a degeneracy of states,
and hence an entropy, as a function of its conserved quantum numbers.

In the late 1960's and early 1970's, laws of classical black hole
mechanics were discovered \cite{bhmechlaws}, which bear a striking
resemblance to the laws of thermodynamics.  The zeroth black hole law
is that the surface gravity ${\hat{\kappa}}$ is constant over the
horizon of a stationary black hole.  The first law is
\be\label{smarr}
dM = {\hat{\kappa}} {{dA}\over{8\pi}} + \omega_{\rm H} dJ +
\Phi_{\rm e}dQ \,, 
\ee
where $\omega_{\rm H}$ is the angular velocity at the horizon and
$\Phi_{\rm e}$ the electrostatic potential.  The second law says that
the horizon area $A$ must be nondecreasing in any (classical) process.
Lastly, the third law says that it is impossible to achieve
${\hat{\kappa}}{\eql}0$ via a physical process such as emission of
photons.

{}From (\ref{smarr}) and other arguments, Bekenstein proposed
\cite{bekenstein} that the entropy of the black hole should be
proportional to the area of the event horizon.  Hawking's
semiclassical calculation of the black hole temperature
\be
T_H = {\frac{\hbar{\hat{\kappa}}}{2\pi}} \,.
\ee
made the entropy-area identification precise by fixing the
coefficient.  (In the semiclassical approximation, the spacetime is
treated classically, while matter fields interacting with it are
treated quantum-mechanically.)  In the reference frame of an
asymptotically faraway observer, Hawking radiation is emitted at the
horizon as a perfect blackbody. The thermal emission spectrum is then
filtered by potential barriers encountered by the outgoing radiation,
which arise from the varying gravitational potential, and give rise to
``greybody factors''.

The Bekenstein-Hawking or Black Hole entropy is in any spacetime
dimension $d$
\be
\SBH = {{A_d}\over{4\hbar G_d}} \,,
\ee
where $A_d$ is the area of the event horizon, and $G_d$ is the
$d$-dimensional Newton constant, which in units $\hbar=c=1$ has
dimensions of (length)$^{d-2}$.  This is a {\em universal} result for
any black hole, applicable to any theory with Einstein gravity as its
classical action.  Note that the black hole entropy is a humongous
number, \aeg for a four-dimensional Earth-mass black hole which has a
Schwarzschild radius of order 1cm, the entropy is $\SBH\!\sim\!
10^{66}$.

Up to constants, the black hole entropy is just the area of the
horizon in Planck units.  As it scales like the area rather than the
volume, it violates our naive intuition about extensivity of
thermodynamic entropy which we gain from working with quantum field
theories.  The area scaling has in fact been argued to be evidence for
``holography''.  There are several versions of holography, but the
basic idea is that since the entropy scales like the area rather than
the volume, the fundamental degrees of freedom describing the system
are characterised by a quantum field theory with one fewer space
dimensions and with Planck-scale UV cutoff.  This idea was elevated to
a principle by 't Hooft and Susskind.  The ``AdS/CFT correspondence''
does in fact provide an explicit and precise example of this idea.
For more details, including references, see Susskind's lectures on the
Holographic Principle at this School \cite{lennytasi}.

As we will see later on in explicit examples, there are systems where
the entropy of a zero-temperature black hole is nonzero.  Note that
this does not imply a violation of the third law of thermodynamics if
the analogy between black hole mechanics and thermodynamics is indeed
exact.  There is no requirement in the fundamental laws of
thermodynamics that the entropy should be zero at zero temperature;
that version of the third law is a statement about equations of state
for ordinary types of matter.

A subtlety which we have suppressed until now in discussing black hole
thermodynamics is that an asymptotically flat black hole cannot really
be in equilibrium with a heat bath.  This is problematic if we wish to
work in the canonical thermal ensemble.  The trouble is the Jeans
instability: even a low-density gas distributed throughout a flat
spacetime will not be static but it will undergo gravitational
collapse.  Technical ways around this problem have been devised, such
as putting the black hole in a box and keeping the walls of the box at
finite temperature via the proverbial reservoir.  This physical setup
puts in an infrared cutoff which gets rid of the Jeans problem.  It
also alters the relation between the black hole energy and the
temperature at the boundary (the walls of the box rather than
infinity).  This in turn results in a positive specific heat for the
black hole.  For a large box, which is appropriate if we wish to
affect properties of the spacetime as little as possible, the black
hole is always the entropically preferred state, but for a small
enough box hot flat space results.  For more details see
\cite{brownyork}.

As the black hole Hawking radiates, it loses mass, and its horizon
area decreases, thereby providing an explicit quantum mechanical
violation of the classical area-increase theorem.  Since the area of
the horizon is proportional to the black hole entropy, it might appear
that this area decrease signals a violation of the second law.  On the
other hand, the entropy in the Hawking radiation increases, providing
a possible way out.  Defining a generalised entropy, which includes
the entropy of the black hole plus the other stuff such as Hawking
radiation,
\be
S_{\rm{tot}} = \SBH + S_{\rm{other}} \geq 0 \,,
\ee
was argued by Bekenstein to fix up the second law.  

Using gedankenexperiments involving gravitational collapse and
infalling matter, Bekenstein also argued that the entropy of a system
of a particular volume is bounded above by the entropy of the black
hole whose horizon bounds that volume.  The Bekenstein bound is
however not a completely general bound, as pointed out by Bekenstein
himself.  The system to which it applies must be one of ``limited
self-gravity'', and it must be a whole system not just a subsystem.
Examples of systems not satisfying the bound include a closed FRW
universe, or a super-horizon region in a flat FRW universe.  In these
situations, cosmological expansion drives the overall dynamics and
self-gravity is not limited; the entropy in a big enough volume in
such spacetimes will exceed the Bekenstein bound.  Also, certain
regions inside a black hole horizon violate the bound.

Bousso \cite{boussos99} has formulated a more general, covariant,
entropy bound.  A new ingredient in this construction is to use {\em
null} hypersurfaces bounded by the area $A$.  The surfaces used are
``light-sheets'', which are surfaces generated by light rays leaving
$A$ which have nonpositive expansion everywhere on the sheet.  The
Bousso bound then says that the entropy on the sheets must satisfy
\be
\SBH \le {{A}\over{4}} \,.
\ee
A proof of this bound was given in \cite{eannadonbob}; one or other of
two conditions on the entropy flux across the light-sheets was
required.  These conditions are physically reasonable conditions for
normal matter in semi-classical regimes below the Planck scale.  The
conditions can be violated, and so the bound does not follow from
fundamental physical principles.  It does however hold up in all
semiclassical situations where light-sheets make sense, as long as the
semiclassical approximation is used in a self-consistent fashion
\cite{boussos99}.  The generalised second law then works with the
entropy defined as above.  A recent discussion of semiclassical black
hole thermodynamics \cite{waldreview} points out that there are no
known gedankenexperiments which violate this generalised bound.  See
also a different discussion of holography in \cite{smolin0003}.

While the Bousso bound is a statement that makes sense only in a
semiclassical regime, it may well be more fundamental, in that the
consistent quantum theory of gravity obeys it.

\subsection{The black hole information problem}

Identifying the Bekenstein-Hawking entropy as the physical entropy of
the black hole gives rise to an immediate puzzle, namely the nature of
the microscopic quantum mechanical degrees of freedom giving rise to
that thermodynamic entropy.  Another puzzle, the famous information
problem \cite{infoloss}, arose from Hawking's semiclassical
calculation which showed that the outgoing radiation has a purely
thermal character, and depends only on the conserved quantum numbers
coupling to long-range fields.  This entails a loss of information,
since an infalling book and a vacuum cleaner of the same mass would
give rise to the same Hawking radiation according to an observer
outside the horizon.  In addition, since the classical no-hair
theorems allow observers at infinity to see only long-range hair,
which is very limited, black holes are in the habit of gobbling up
quantum numbers associated to all global symmetries.

In the context of string theory, which is a unified quantum theory of
all interactions including gravity, information should not be lost.
As a consequence, the information problem must be an artifact of the
semiclassical approximation used to derive it.  Information must
somehow be returned in subtle correlations of the outgoing radiation.
This point of view was espoused early on by workers including
\cite{thooftstu,lenny} and collaborators following the original
suggestion of Page \cite{pageone}.  Information return requires a
quantum gravity theory with subtle nonlocality, a property which
string theory appears to possess.  The AdS/CFT correspondence is one
context in which we have an explicit realization in principle of the
information retention scenario, as discussed in Susskind's lectures at
this School.  The information problem is therefore shifted to the
problem of showing precisely how semiclassical arguments break down.
This turns out to be a very difficult problem, and solving it is one
of the foremost challenges in this area of string theory.

At the ITP Conference on Quantum Aspects of Black Holes in 1993,
however, there were several scenarios on the market for solving the
information problem.  Nowadays, it is fair to say that the mainstream
opinion in the string theory community has zeroed in on the
information return scenario.  Let us briefly mention some aspects of
various scenarios to illustrate some useful physics points.

The idea that information is just lost in quantum gravity
\cite{infoloss} has many consequences apt to make high energy
theorists queasy, so we will not dwell on it.  We just mention that
one of them is that it usually violates energy conservation, although
a refinement is possible \cite{andycluster} where clustering is
violated instead.  Another scenario is that all of the information
about what fell into the black hole during its entire lifetime is
stored in a remnant of Planckian size.  The main problem with this
scenario is that energy is needed in order to encode information, and
a Planck scale remnant has very little energy.  In addition, remnants
as a class would need an enormous density of states in order to be
able to keep track of all information that fell into any one of the
black holes giving rise to the remnant.  This enormous density of
states of Planck scale objects is incompatible with any object known
in string theory.  Such a huge density of states could also lead to a
phenomenological disaster if tiny virtual remnants circulate in
quantum loops.  Remnants also cause trouble in the thermal atmosphere
of a big black hole \cite{lennytrouble}. A possibility for remnants is
that they are baby universes.  One approach to baby universes was to
consider in a Euclidean approach our large universe and the effects on
its physics due to a condensate of tiny (Planck-sized) wormholes.
Arguments were made \cite{colemanredherrings} that the tiny wormholes
lead to no observable loss of quantum coherence in our universe.
Nonetheless, the baby universe story does involve in-principle
information loss in our universe, and the physics depends on selection
of the wavefunction of the universe.  It is also difficult
\cite{lennywilly} to be sure that there are only tiny wormholes
present.  Lastly, as we mentioned previously, Wick rotation from
Euclidean to Lorentzian signature is not in general a well-understood
operation in quantum gravity.

The original Hawking radiation calculation is semiclassical, \aie the
black hole is treated classically while the matter fields are treated
quantum-mechanically.  The computation of the thermal radiation
spectrum, and subsequent computations, use at some point unwarranted
assumptions about physics above the Planck scale.  This is a fatal
flaw in the argument for information loss.  It has however been argued
in \cite{waldreview} that the precise nature of this super-Planckian
physics does not impact the Hawking spectrum very much, and that as
such it is a robust semiclassical result.  The information return
devil is in the details, however.  Susskind's analogy between the
black hole and a lump of coal fired on by a laser beam makes this
quite explicit.
 
An interesting piece of semiclassical physics \cite{niceslice} is that
all except a small part of the black hole spacetime near the
singularity can be foliated by Cauchy surfaces called ``nice slices'',
which have the property that both infalling matter and outgoing
Hawking radiation have low energy in the local frame of the slice.  An
adiabatic argument \cite{niceslice} then led to the conclusion that
return of information in the framework of local field theory is
difficult to reconcile with the existence of nice slices.  One
possibility is that the singularity plays an important role in
returning information, although it can hardly do so in a local manner.

Showing that the information return scenario is inconsistent turns out
to be very difficult.  One of the salient features of a black hole is
that it has a long information retention time, as argued in
\cite{lennylarus,lennytasi} using a result of Page \cite{pagetwo} on
the entropy of subsystems.  The construction involves a total system
made up of two subsystems, black hole and Hawking radiation; it is
assumed that the black hole horizon provides a true dividing line
between the two.  The entropy of entanglement encodes how entangled
the quantum states of the two subsystems are.  In the literature there
has been some confusion about the physical significance of the
entanglement entropy.  Let us just mention that although in the above
example it is bounded above by the statistical entropy of the black
hole, it is generally {\em not} identical to the black hole entropy.

The information return scenario does require that we give up on
semiclassical gravity as a way of understanding quantum gravity.  We
also consider it unlikely that the properties needed to resolve the
information problem are visible in perturbation theory at low (or
indeed all) orders \cite{bglBDHM}.  It is therefore very interesting
to search for precisely which properties of string theory will help us
solve the information problem.  Although perturbative string theory
obeys cluster decomposition, it does not obey the same axioms as local
quantum field theory.  In addition, the only truly gauge-invariant
observable in string theory is the S-matrix.  Some preliminary
investigations of locality and causality properties of string theory
were made in \cite{stringcausality}.

The conclusions we wish the reader to draw from the discussion of this
subsection are twofold.  Firstly, the violations of locality needed in
order to return information must be subtle, in order not to mess up
known low-energy physics.  Secondly, we will have to understand
physics at and beyond the Planck scale to understand precisely why
black holes do not gobble up information.  It is likely that there is
a subtle interplay between the IR and the UV of the theory in quantum
gravity, entailing breakdown of the usual Wilsonian QFT picture of the
impact of UV physics on IR physics.  It is also possible that the
fundamental rules of quantum mechanics need to be altered, although
there is no clear idea yet of how this might occur.  Recent studies of
non-commutative gauge theories such as \cite{natietal} show that those
theories whose commutative versions are not finite possess IR/UV
mixing.  We await further exciting progress in this subject and look
forward to applications.

We now move to the subject of the $p$-brane cousins of black holes in
the string theory context.

\sect{Quantum numbers, and solution-generating}\label{secttwo}

In order to discover which black holes and $p$-branes occur in string
theories, we need to start by identifying the actions analogous to the
Einstein action and thereby the quantum numbers that the black holes
and $p$-branes can carry.  In this section we stick to classical
physics; we will discuss quantum corrections in section
(\ref{sectfour}).

\subsection{String actions and $p$-branes}

In Clifford Johnson's lectures at this School, you saw how the
low-energy Lagrangians for string theory are derived.  Here, for
simplicity, we will discuss only in the Type IIA and IIB
supergravities.  These supergravity theories possess $\cN{\eql}2$
supersymmetry in $d=10$, \aie they have 32 real supercharges.  Type
IIB is chiral as its two Majorana-Weyl 16-component spinors have the
same chirality, while IIA is nonchiral as its spinors have opposite
chirality.

There are two sectors of massless modes of Type-II strings: NS-NS and
R-R.  In the NS-NS sector we have the string metric\footnote{Not to be
confused with the Einstein tensor, which we never use here.}
$G_{\mu\nu}$, the two-form potential $B_{\mathit 2}$, and the scalar
dilaton $\Phi$.  In the R-R sector we have the $n$-form potentials
$C_{\mathit n}$, $n$ even for IIB and odd for IIA.  For Type IIA, the
independent R-R potentials are $C_{\mathit 1},C_{\mathit 3}$.  The
low-energy effective action of IIA string theory is $d=10$ IIA
supergravity:
\be\ba{rcl}\label{iiaaction}
\bs
S_A &=& {\displaystyle{
{\frac{1}{(2\pi)^7 l_s^8}} \int d^{10}x \sqrt{-G}
\left\{ {{e^{-2\Phi}}\over{\gs^2}} 
        \left[ R_G + 4\left(\partial\Phi\right)^2 -
{\frac{3}{4}} \left(\partial B_{\mathit 2} \right)^2 \right]
  + {\rm (fermions)}
\right. }} \nonumber\\ & & {\displaystyle{
\quad - \left. {\frac{1}{4}} \left(2 \partial
C_{\mathit 1}\right)^2 - {\frac{3}{4}} 
\left(\partial C_{\mathit 3} - 2 \partial
B_{\mathit 2} C_{\mathit 1} \right)^2 \right\} + 
{\frac{1}{64}} \epsilon \partial
C_{\mathit 3} \partial C_{\mathit 3} B_{\mathit 2} }}\,.
\ea\ee
We have shifted the dilaton field so that it is zero at infinity.
Aside from the signature convention, we have used conventions of
\cite{BHO}\footnote{Typos in the T-duality formul\ae\ are fixed in
\cite{myerspuff}.}, where antisymmetrisation is done with weight one,
\aeg
\be
(\partial A)_{\mu\nu}\equiv \half \left(\partial_\mu A_\nu -
\partial_\nu A_\mu\right) \,.
\ee
In the action we could have used the Hodge dual `magnetic'
$(8{\mns}n)$-form potentials instead of the `electric' ones, \aeg a
6-form NS-NS potential instead of the 2-form.  However, we cannot
allow both the electric and magnetic potentials in the same
Lagrangian, as it would result in propagating ghosts.  The funny cross
terms, such as $\partial C_{\mathit 3}\wedge\partial C_{\mathit
3}\wedge B_{\mathit 2}$, are required by supersymmetry.  In many cases
there is a consistent truncation to an action without the cross terms,
but compatibility with the field equations has to be checked in every
case.  

For Type IIB string theory, the R-R 5-form field strength
$F_5^{+}\equiv\partial C_{\mathit 4}$ is self-dual, and so there is no
covariant action from which the field equations can be derived.
Define ${\tilde{H}}_{\mathit 3}\equiv\partial C_{\mathit 2}$,
$\ell\equiv C_{\mathit 0}$; $H_{\mathit 3}\equiv\partial B_{\mathit
2}$.  Then the equations of motion for the metric is
\be\ba{rl}
R_{\mu\nu} = & {\displaystyle{
2\nabla_\mu\partial_\nu\Phi - {{9}\over{4}}
H_{(\mu}^{\lambda\rho}H^{\ }_{\nu)\lambda\rho} -
e^{2\Phi}{{1}\over{2}}\left(\partial_\mu\ell\partial_\nu\ell-
{{1}\over{2}}G_{\mu\nu}(\partial\ell)^2\right)   }} \cr
& + {\displaystyle{ {{9}\over{4}}e^{2\Phi}\left[
({\tilde{H}}-\ell H)_{(\mu}^{\lambda\rho}H^{\ }_{\nu)\lambda\rho} 
- {{1}\over{6}}G_{\mu\nu}({\tilde{H}}-\ell H)^2 \right]
+ {{25}\over{6}}e^{2\Phi}\left(
F_{\mu\lambda\rho\sigma\kappa}F_{\nu}^{\lambda\rho\sigma\kappa}\right) }}
\,;
\ea\ee
while for the scalars they are
\be\ba{rl}
\bs
\nabla^2\Phi = & {\displaystyle{
(\partial\Phi)^2 + {{1}\over{4}}R_G + {{3}\over{16}}H^2  }}
\,,\cr
\nabla^2\ell = & {\displaystyle{ - {{3}\over{2}}H^{\mu\nu\lambda}
\left({\tilde{H}}-\ell H\right)_{\mu\nu\lambda} }}\,,\cr
\ea\ee
and for the gauge fields 
\be\ba{rl}
\bs
\nabla^\mu &\! {\displaystyle{
\left[(\ell^2+e^{-2\Phi})H-\ell{\tilde{H}}\right]_{\mu\nu\rho} }}
= {\displaystyle{ +{{10}\over{3}} F_{\nu\rho\sigma\lambda\kappa} 
{\tilde{H}}^{\sigma\lambda\kappa} }} \,,\cr
\bs
\nabla^\mu &\! {\displaystyle{
\left[{\tilde{H}}-\ell H\right]_{\mu\nu\rho} }}
= {\displaystyle{
-{{10}\over{3}}
F_{\nu\rho\sigma\lambda\kappa}H^{\sigma\lambda\kappa} }}\,,\cr
& F_{\mathit 5}^+ = {}^*F_{\mathit 5}^+ \,.
\ea\ee

Now recall that in $d=4$ electromagnetism, an electrically charged
particle couples to $A_{\mathit 1}$ (or its field strength $F_{\mathit
2}$), while the dual field strength ${}^*F_{\mathit 2}$ gives rise to
a magnetic coupling to point particles.  By analogy, a $p$-brane in
$d{\eql}10$ couples to $C_{\mathit n=p+1}$ ``electrically'', or
$C_{\mathit 7-p}$ magnetically.  As a result, we find 1-branes ``F1''
and 5-branes ``NS5'' coupling to the NS-NS potential $B_{\mathit 2}$,
and $p$-branes ``D$p$'' coupling to the R-R potentials $C_{\mathit
p+1}$ (or their Hodge duals).  Reviews of $p$-branes in string theory
can be found in \cite{duffreview,youmreview}.

Not all aspects of the physics of the R-R gauge fields can be gleaned
from the action / equations of motion given for IIA and IIB above.
The reader is referred to the recent work of \aeg \cite{wittenmoorerr}
for discussion of subtle effects involving charge quantisation, global
anomalies, self-duality, and the connection to K-theory.  We will
stick to putting branes on $\bR^d$ or $T^d$ where these effects will
not bother us.

\subsection{Conserved quantities: mass, angular momentum, charge}
\label{mandjandq}

There is a large variety of objects in string theory carrying
conserved quantum numbers.  These conserved quantities include the
energy which, if there is a rest frame available, becomes the mass
$M$.  In $D$ dimensions, we also have the skew matrix $J^{[\mu\nu]}$
with $[\half(D{\mns}1)]$ eigenvalues which are the independent angular
momenta, $J_i$.  The last type of conserved quantity couples to the
long range R-R gauge field; it is gauge charge $Q$.  All of these are
defined by integrating up quantities which are conserved courtesy of
the equations of motion.

The low-energy approximation to string theory yielded the supergravity
actions we saw in the previous subsection.  When a $p$-brane is
present and thereby sources the supergravity fields, there is an
additional term in the action encoding the collective modes of the
brane.  The low-energy action for the bulk supergravity with brane is
then
\be
S = S_{\rm SUGRA} + S_{\rm brane} \,;
\ee
Such a combined action is well-defined for classical string theory.
For fundamental quantum string theory, a different representation of
degrees of freedom would be necessary.  See \aeg \cite{arkadysing} for
a discussion of some of these issues.  The second term is an integral
only over the $p{\pls}1$ dimensions of the $p$-brane worldvolume, while
the first term is an integral over the $d{\eql}10$ bulk.  If we then
vary this action with respect to the bulk supergravity fields we
obtain delta-function sources on the right hand sides of the
supergravity equations of motion, at the location of the brane.

Let us consider the mass and angular momenta first.  In $d{\eql}10$,
$p$-branes of codimension smaller than 3 give rise to spacetimes which
are not asymptotically flat; there are not enough space dimensions to
allow the fields to have Coulomb tails.  We do not have space to
review these cases here; we refer the reader to \aeg sections 5.4 and
5.5 of \cite{kellyreview} where Scherk-Schwarz reduction is also
discussed, and to \cite{cveticdomainwalls}.  For the $p$-branes this
means we will consider only $p<7$.

The mass for an isolated gravitating system can be defined by
referring its spacetime to one which is nonrelativistic and weakly
gravitating \cite{robangmom}.  Let us go to Einstein frame, \aie
where
\be
S = \int d^Dx \left( {{\sqrt{-g} R[g]}\over{16\pi G_D}} +
{\cal{L}}_{\rm matter} \right) \,,
\ee
where the Einstein metric $g$ is given in terms of the string metric
$G$ as
\be\label{einsteinstring}
g_{\mu\nu} = e^{-4\Phi/(D-2)} G_{\mu\nu} \,.
\ee
Notice that in the action (\ref{iiaaction}), the dilaton field had the
``wrong-sign'' action in string frame; however, it becomes
``right-sign'' in this Einstein frame.  Also, recall that we have
defined the dilaton field $\Phi$ to be zero at infinity; we keep track
of the asymptotic value of the string coupling by keeping explicit
powers of $\gs$ where required.

The field equation for the Einstein metric is in $D$ dimensions
\be
R_{\mu\nu}-\half g_{\mu\nu}R = 8\pi G_D T_{\mu\nu}^{\rm (matter)} \,,
\ee
where $R_{\mu\nu}$ is the Ricci tensor and $T_{\mu\nu}^{\rm (matter)}$
is the energy-momentum tensor.  Far away, the metric becomes flat.
Let us linearise about the Minkowski metric $\eta_{\mu\nu}$
\be
g_{\mu\nu}= \eta_{\mu\nu} + h_{\mu\nu} \,,
\ee
\aie consider only first order terms in the deviation $h$.  (To
this order in algebraic quantities, we raise and lower indices with
the Minkowski metric.)  
We also impose the condition that the system be non-relativistic, so
that time derivatives can be neglected and
$T_{00}\!\gg\!T_{0i}\!\gg\!T_{ij}$.
Under coordinate transformations $\delta x^\mu = \xi^\mu$, the metric
deviation $h$ transforms as
${\delta}h_{\mu\nu}=-2\partial_{(\mu}\xi_{\nu)}$.  Let us (partially)
fix this symmetry by demanding
\be 
\partial_\nu \left( h^{\mu\nu} - \half \eta^{\mu\nu} h_\lambda^\lambda
\right) = 0 \,.
\ee
This is called the harmonic gauge condition.  
Then the field equation for the deviation $h$ becomes
\be\label{deviationeqn}
\left(\partial^i \partial_i \right) h_{\mu\nu}
= 16\pi G_D \left[ T_{\mu\nu}^{\rm (matter)} - {{1}\over{(D-2)}}
\eta_{\mu\nu} T^{{\rm (matter)}\ \lambda}_\lambda \right]
\equiv -16\pi G_D {\tilde{T}}_{\mu\nu} \,.
\ee
The indices $i=1\ldots (D-1)$ are contracted on the left hand side of
this equation with the flat metric.  The field equation in harmonic
gauge (\ref{deviationeqn}) is a Laplace equation and it has the
solution
\be
h_{\mu\nu}(x) = {{16\pi G_D}\over{(D-3)\Omega_{D-2}}} \ 
\int d^{D-1}{\vec{y}} \ 
{{{\tilde{T}}_{\mu\nu}\left(\left|{\vec{x}}-{\vec{y}}\right|\right)}
\over
{\left|{\vec{x}}-{\vec{y}}\right|^{D-3}}}  \,,
\ee
where the prefactor comes from the Green's function and
$\Omega_n=$area$(S^n)$.  Now let us expand this in moments,
\be
h_{\mu\nu}(x) = {{16\pi G_D}\over{(D-3)\Omega_{D-2}}} \left\{
  {{1}\over{r^{D-3}}} \int d^{D-1}y {\tilde{T}}_{\mu\nu}(y) 
+ {{x^j}\over{r^{D-1}}} \int d^{D-1}y y^j {\tilde{T}}_{\mu\nu}(y) 
+ \cdots \right\} \,.
\ee
On the other hand, the definitions of the ADM linear and angular
momenta are
\be
P^\mu =      \int d^{D-1}y T^{\mu 0} \quad\, \qquad 
J^{\mu\nu} = \int d^{D-1}y 
               \left( y^\mu T^{\nu 0} - y^\nu T^{\mu 0} \right)\,.
\ee
(Notice that the stress tensors appearing in these formul\ae\, are
not the tilde'd versions of $T$.)  Evaluating in rest frame yields
some simplifications, and gives the following relations from which we
can read off the mass and angular momenta of our spacetime:
\be\ba{rl}
\bs
g_{tt} \longrightarrow &\! -1 + {\displaystyle{ 
{{16\pi G_D}\over{(D-2)\Omega_{D-2}}} {{M}\over{r^{D-3}}} +\cdots}}\,;
\cr
\bs
g_{ij} \longrightarrow &\! 1 + {\displaystyle{ 
{{16\pi G_D}\over{(D-2)(D-3)\Omega_{D-2}}}{{M}\over{r^{D-3}}}+\cdots}}
\,; \cr
g_{ti} \longrightarrow &\! {\displaystyle{ 
{{16\pi G_D}\over{\Omega_{D-2}}} {{x^j J^{ji}}\over{r^{D-1}}}+\cdots}}
\,.
\ea\ee
For spacetimes which are not asymptotically flat (\aeg D$p$-branes
with $p\geq 7$), we must use different procedures, which we do not
have space to review here.

We now move to the analysis of conserved charges carried by branes.
For this, we need to know not only the bulk action but also the
relevant piece of the brane action.  For D$p$-branes, the part we need
is
\be S_{\rm brane} = -{{1}\over{(2\pi)^p\ls^{p+1}}} \int 
C_{\mathit p+1} + \ldots \,.
\ee
Here and in the following, to save carrying around clunky notation, we
are using $C_{\mathit p+1}$ to refer to either the usual R-R potential
or its Hodge dual, as appropriate according to the brane.  For a
single type of brane it is consistent to ignore the funny cross-terms
in the supergravity action, and so the relevant piece of the bulk
action is
\be
S_{\rm SUGRA} = {{-1}\over{(2\pi)^7\ls^8}}
\int d^{10}x \sqrt{-G} {{\left| (p+2) [\partial C]_{\mathit p+2}
\right|^2}\over{2(p+2)!}} + \ldots \,.  \ee
The field equation for the potential $C$ is then 
\be
d {\,}^*\left(dC_{\mathit p+1}\right) = 
(2\pi)^7\ls^8 {\,}^* \left(J_{\mathit p+1}\right) \,,
\ee
where the conserved $p+1$-form current $J$ is 
\be
J_{\mathit p+1}(x) = -(2\pi)^p\ls^{p+1} \int dX^0 \ldots
dX^p\delta^{10}(X-x)\,.
\ee
The physics is easiest to see in static gauge
\be
X^{\mu_i} (\sigma) = \sigma^{\mu_i} \,,\quad i=0\ldots p \,.
\ee
The Noether charge is the integral of the current, and using the field
equations we see that it is
\be
Q_p = \int_{S^{8-p}} {\,}^* (dC)_{\mathit p+2} \,.
\ee
(If these were NS-type branes, there would be a prefactor of
$e^{-2\Phi}/\gs^2$ in the integrand.)

In addition to the field equation for $C$ there is the Bianchi
identity, 
\be
d\left( [dC]_{p+2}\right) =0 \,,
\ee
from which we deduce the existence of a topological charge,
\be
P_{7-p} = \int_{S^{p+2}} (dC)_{p+2} \,.
\ee
As discussed in \cite{joebigbook}, these obey the Dirac quantisation
condition 
\be
Q_p P_{7-p} = 2\pi n \,,\quad n \in \bZ \,.
\ee
Here we have concentrated on D$p$-branes because they have proven to
be of great importance in recent years in studies of the physics of
black holes in the context of string theory.

\subsection{The supersymmetry algebra}\label{susysubsec}

The supersymmetry algebra is of central importance to a supergravity
theory.  Indeed, many of the properties of the supergravity theory can
be worked out from it, see \aeg \cite{susyalgebra1}.  An introduction
to the mechanics of supersymmetry can be found in \cite{toinereview}.
The (anti-)commutators involving two supersymmetry generators $Q$ are
\be\label{susyalgebra}
\{ Q_\alpha, Q_\beta \} \sim \left( {\cal{C}} \Gamma^\mu
\right)_{\alpha\beta} P_\mu + a\left( {\cal{C}}
\Gamma^{\mu_1\ldots\mu_p} \right)_{\alpha\beta} Z_{[\mu_1\ldots\mu_p]}
\,,
\ee
where ${\cal{C}}$ is the charge conjugation matrix, $\Gamma$'s are
antisymmetrised products of gamma matrices, $Z$ are $p$-brane charges,
and $P^\mu$ is the momentum vector.  If there is a rest frame, then
\be
\{ Q_\alpha, Q_\beta \} \sim 
\left( {\cal{C}} \Gamma^0 \right)_{\alpha\beta}
M + a\left( {\cal{C}} \Gamma^{1\ldots p} \right)_{\alpha\beta}
Z_{[1\ldots p]} \,.
\ee
Let us sandwich a physical states $|$phys$\rangle$ around this algebra
relation.  The state $Q|$phys$\rangle$ has nonnegative norm, and a bit
of algebra gives 
\be
M \geq a |Z| \,,
\ee
which is know as the Bogomolnyi bound.  This bound can also be derived
by analysing the supergravity Lagrangian, via the Nester procedure;
see \cite{nesterbog} for examples of the derivation for $\cN{\eql}1,2$
supergravity in $d{\eql}4$.  In this derivation it is important that
boundary conditions for bulk fields at infinity are specified.

The constant $a$ in the Bogomolnyi bound depends on the theory and its
couplings.  States saturating the bound must be annihilated by at
least one SUSY generator $Q$, so they are supersymmetric or ``BPS
states''.  It turns out that the relation $M=a|Z|$ is not renormalised
by quantum corrections, although generically both the mass and the
charge may be renormalised.  The statistical degeneracy of states is
also unrenormalised.  (For sub-maximal supersymmetry, jumping
phenomena, whereby new multiplets appear at a certain value of the
coupling constant, are not ruled out in general.  However, they are
not known to occur in any example involving black holes that we will
discuss.)
  
In the supergravity theory the supersymmetry transformations of the
fields have a spinorial parameter $\epsilon$.  For preserved
supersymmetries, the SUSY relation gives the projection condition,
again schematic,
\be
\left( 1 + \left[{\rm{sgn}}(Z)\right]\,\Gamma^{01\cdots{}p} \right)
\epsilon=0 \,.
\ee

For the special case of $d=11$ supergravity, the matrix on the left
hand side of the anticommutator relation (\ref{susyalgebra}), which is
real and symmetric and therefore has 528 components, can be regarded
as belonging to the adjoint representation of the group $Sp(32;\bR)$.
The decomposition of this representation with respect to the $d=11$
Lorentz group $SO(1,10)$ goes as ${\underline{528}} \rightarrow
{\underline{11}} \oplus {\underline{55}} \oplus {\underline{462}}$.
The purely spatial components of the two central charges $Z$, which
have two and five indices respectively, correspond to charge carried
by the M2- and M5-branes.  In a similar fashion, inspection of the
momentum vector yields the existence of the massless gravitational
wave, often denoted MW in the literature.  The remaining ten
components of the two-index central charge, which may involve of
course only one temporal index, correspond to the Hor{\v{a}}va-Witten
domain walls in the construction of $E_8\times E_8$ heterotic string
from M theory, while the remaining 210 components of the five-index
central charge, involving again just one temporal component by
antisymmetry, correspond to the $d=11$ Kaluza-Klein monopole, denoted
MK, which possesses NUT charge.  The details, including the
identification of preserved supersymmetries, are presented very nicely
in \cite{susyalgebra1}.  In $d=10$ supergravity the analogs of MK and
MW are denoted W and KK.

The above facts can be used with some work to identify the theory- and
object-dependent constant in the schematic SUSY bound $M\geq a|Z|$,
\be\label{theas}
a_{F1} \sim 1 \quad , \qquad a_{Dp} \sim {\frac{1}{\gs}} 
\quad , \qquad
a_{NS5} \sim {\frac{1}{\gs^2}} \,.
\ee
Since the charges $Z$ are integer-quantized in the quantum theory (but
not in supergravity), we see from these relations and the mass-charge
formula that for weak string coupling the F1-branes are the lightest
degrees of freedom.  Therefore, in perturbative string theory, they
are the fundamental degrees of freedom, while the D$p$ and NS5 are two
qualitatively different kinds of soliton.  However, in other regions
of parameter space F1's will not be ``fundamental'', as they will no
longer be the lightest degrees of freedom.  This gives rise to the
notion of `$p$-brane democracy' \cite{pbdemocracy}.

By analogy with the Reissner-Nordstr{\"{o}}m black holes we met in the
section \ref{sectone}, we can have extremal black $p$-brane
spacetimes, which have zero Hawking temperature.  Generally, for these
extremal spacetimes there is some unbroken supersymmetry in the bulk,
but this is not required to happen unless there is only one type of
brane present.

\subsection{Unit conventions, dimensional 
reduction and dualities}\label{simplereduction}

For units, we will be using the conventions of the textbook
\cite{joebigbook}.  The fundamental string tension is
\be
\tau_{{\rm F}1} = {{1}\over{2\pi\alpha^\prime}} \equiv 
{{1}\over{2\pi\ls^2}} \,.
\ee
while the D$p$-brane tension (mass per unit $p$-volume) is
\be
\tau_{{\rm D}p} = {{1}\over{\gs(2\pi)^p\ls^{p+1}}}\,,
\ee
and the NS5-brane tension is
\be
\tau_{{\rm NS}5} = {{1}\over{\gs^2(2\pi)^5\ls^6}}\,.
\ee
In ten dimensions the Newton constant $G$ is related to the
gravitational coupling $\kappa$ and $\gs,\ls$ by
\be
16\pi G_{10} \equiv 2\kappa_{10}^2 = (2\pi)^7 \gs^2 \ls^8 \,.
\ee
To get units convenient for T-duality, we define any volume $V$ to
have implicit $2\pi$'s in it.  If the fields of the theory are
independent of $(10-d)$ coordinates, then the integration measure
factorizes as $\int d^{10}x = \left[(2\pi)^{10-d}V_{10-d} \right] \int
d^{d} x$.  We can use this directly to find any lower-dimensional
Newton constant from the ten-dimensional one, as follows:
\be
G_d = {{G_{10}}\over{(2\pi)^{10-d}V_{10-d}}} \,,
\ee
The Planck length in $d$ dimensions, $\ld$, is defined by
\be
16\pi G_d \equiv (2\pi)^{d-3} \ld^{d-2} \,.
\ee
{}From these facts we can see that there is a neat interdimensional
consistency in the expression for the Bekenstein-Hawking entropy.  Let
us take a black $p$-brane and wrap it on $T^p$ to make $d=10-p$ black
hole.  Translational symmetry along the $p$-brane means that the
horizon has a product structure, and so the entropy  is
\be\ba{rl}
\SBH = & {\displaystyle{ {{A_{d+p}}\over{4 G_{d+p}}}  =
{{A_{d}(2\pi)^pV_p }\over{4 G_{d+p}}}  }}\cr
=& {\displaystyle{ {{A_{d}}\over{4 G_d}} }}  \,,
\ea\ee
which is the same as the black hole entropy.  

As a reminder, we mention that the event horizon area in the
Bekenstein-Hawking formula must always be computed in the Einstein
frame, which is the frame where the kinetic term for the metric is
canonically normalized,
\be
S_{\rm{grav}}= {{1}\over{16\pi G_d}}\int\sqrt{-g}R[g] \,.
\ee
The relation between the Einstein and string metrics was shown in
eqn (\ref{einsteinstring}), $g_{\mu\nu}=e^{-4\Phi/(D-2)}G_{\mu\nu}$.  

Figuring out the constants is only one small part of the mechanics of
dimensional reduction.  We now move to a simple example of
Kaluza-Klein reduction of fields in string frame, by reducing on a
circle of radius $R$.  More complicated toroidal reduction equations
may be found in standard references such as \cite{maharanaschwarzsen}.

Label the $d$ dimensional system with no hats and the $(d-1)$ system
with hats.  Split the indices as $\{x^\mu\}=\{x^{\hat{\mu}},z\}$.  The
vielbeins decompose as
\be
\left(E_\mu^a \right) = 
\left(\ba{cc}{\hat{E}}_{\hat{\mu}}^{\hat{a}} & 
e^{\hat{\chi}} {\hat{A}}_{\hat{\mu}} \cr 0 & e^{\hat{\chi}}\ea\right)
\qquad\Rightarrow\qquad
\left(G_{\mu\nu}\right) = 
\left(\ba{cc}{\hat{G}}_{\hat{\mu}\hat{\nu}} +
e^{2{\hat{\chi}}}{\hat{A}}_{\hat{\mu}}{\hat{A}}_{\hat{\nu}} &
e^{2{\hat{\chi}}}{\hat{A}}_{\hat{\mu}} \cr
{\hat{A}}_{\hat{\nu}}e^{2{\hat{\chi}}} & e^{2{\hat{\chi}}} \ea\right)
\,,
\ee
and
\be
\Phi = {\hat{\Phi}} + {{1}\over{2}}{\hat{\chi}} \,;
\ee
which yield
\be\ba{l}
\bs
{\displaystyle{ 
{{1}\over{16\pi G_d}} \int d^dx \sqrt{-G}e^{-2\Phi}R_G
}} = \cr {\displaystyle{ {{1}\over{16\pi G_{d-1}}}\int
d^{d-1}x \sqrt{-{\hat{G}}}e^{-2{\hat{\Phi}}}\left[ R_{\hat{G}} +
4(\partial{\hat{\Phi}})^2 - (\partial{\hat{\chi}})^2 -
{{1}\over{4}}e^{2{\hat{\chi}}}\left(2\partial{\hat{A}}\right)^2 
\right] }}
\,.
\ea\ee
More generally, reduction on several directions on tori or Calabi-Yau
manifolds leads to large U-duality groups.  \aeg $E_{(7,7)}$ for Type
II on $T^6$, $E_{(6,6)}$ for Type II on $T^5$.  A survey of
supergravities in diverse dimensions can be found in \cite{sidd}.

The Kaluza-Klein procedure can also be done in Einstein frame.
Taking the metric
\be
ds^2 = e^{2\alpha{\hat{\chi}}} d{\hat{s}}^2 +
e^{2\beta{\hat{\chi}}}\left(dz+{\hat{A}}_{\hat{\mu}}
dx^{\hat{\mu}}\right)^2
\,,
\ee
with $\beta=(2-D)\alpha$ and $\alpha^2=1/[2(D-1)(D-2)]$
\cite{kellyreview} gives
\be
\sqrt{-g}R_g = \sqrt{-\hat{g}} \left(R_{\hat{g}}
- \half (\partial{\hat{\chi}})^2 -
\quarter e^{-{2(D-1)\alpha{\hat{\chi}}}}F^2\right)\,,
\ee
where $F$ is the field strength of ${\hat{A}}$.  

We now turn to a very quick reminder on some common and useful
dualities.

\smallskip
\noindent{\ul{Type IIA $\lra$ M-theory}}
\smallskip

\noindent The 11th coordinate $x^\natural$ is compactified on a circle
of radius
\be
R_\natural = \gs \ls \,.
\ee
The supergravity fields decompose as
\be\ba{rl}
\bs
ds_{11}^2 = & e^{-2\Phi/3}dS_{10}^2 +
e^{4\Phi/3}\left(dx^\natural+C_{{\mathit 1}\mu}dx^\mu \right)^2 \,\cr
(\partial A_{\mathit 3}) = & e^{4\Phi/3}
\left(\partial C_{\mathit 3} -2H_{\mathit 3}C_{\mathit 1}\right)
+{{1}\over{2}}e^{\Phi/3} \left(\partial B_{\mathit 2}\right)
dx_\natural\,.
\ea\ee
We can turn M-theory objects into Type IIA objects by pointing them
in the 11th direction ($\swarrow$) or not ($\downarrow$).
\be\ba{rrrr} {\rm W} & {\rm M2} & {\rm M5} & {\rm KK}
\cr \swarrow\quad\downarrow
&\swarrow\quad\downarrow & \swarrow\quad\downarrow
&\swarrow\quad\downarrow \cr 
{\rm D0}\ \  {\rm W} & {\rm F1}\ \ {\rm D2} 
		     & {\rm D4}\ \ {\rm NS5} 
                     & {\rm D6}\ \ {\rm KK}  \ea \, . \ee

\smallskip
\noindent{\ul{S-duality of IIB}}
\smallskip

\noindent The low-energy limit of IIB string theory, IIB supergravity,
possesses a SL(2,$\bR$) symmetry (it is broken to SL(2,$\bZ$) in the
full string theory).  Define
\be
\lambda \equiv C_{\mathit 0} + i e^{-\Phi} \quad {\rm and\ }\quad 
H \equiv \pmatrix{ \partial B_{\mathit 2} \cr \partial C_{\mathit 2} }
\,.
\ee
Under an SL(2,$\bR$) transformation represented by the matrix
\be
U = \pmatrix{ a & b \cr c & d } \in {\rm SL(2},\bR) \,,
\ee
the fields transform as 
\be 
H \rightarrow U\,H \,\quad \lambda \rightarrow
{{a\lambda+b}\over{c\lambda+d}} \,.
\ee
The $d=10$ Einstein metric and the self-dual five-form field strength
are invariant.   

A commonly considered $\bZ_2$ subgroup obtains when $C_{\mathit 0}=0$.
The $\bZ_2$ flips the sign of $\Phi$, and exchanges $B_{\mathit 2}$
and $C_{\mathit 2}$.  The result is
\be 
D1 \lra F1\,, \qquad D5 \lra NS5 \,;
\ee 
all others such as W and KK are unaffected, and the D3 goes into
itself.  The effect of this $\bZ_2$ on units is
\be
{\tilde{\gs}} = {{1}\over{\gs}} \quad , \qquad
{\tilde{\gs}}^{\quarter} {\tilde{\ls}}={\gs}^{\quarter} \ls \,.
\ee
{}From this one can easily check that the tensions of \aeg F1
and D1's transform into each other under the $\bZ_2$ flip.

\smallskip
\noindent{\ul{T-duality}} 
\smallskip

\noindent The operation of T-duality on a circle switches winding and
momentum modes of fundamental strings (F1) and exchanges Type IIA and
IIB.  The effect on units is to invert the radius in string units, and
leave the string coupling in one lower dimension unchanged:
\be
{\frac{{\tilde{R}}}{\tilde{\ls}}} =
{\frac{\ls}{R}} \quad , \qquad
{\frac{{\tilde{\gs}}}{\sqrt{{\tilde{R}}/{\tilde{\ls}}}}}  =
{\frac{\gs}{\sqrt{R/\ls}}} \quad , \qquad {\tilde{\ls}} = \ls
\,.
\ee
T-duality does not leave all branes invariant; it changes the
dimension of a D-brane depending on whether the transformation is
performed on a circle parallel ($\parallel$) or perpendicular
($\perp$) to the worldvolume.  It also changes the character of a KK
or NS5; doing T-duality along the isometry direction (isom) of the KK
gives an NS5.  Summarising, we have:
\be\ba{cc}
{\rm{D}}p \lra {\rm{D}}p-1 (\parallel)\ {\rm{or\ D}}p+1 (\perp) \,,
\qquad\qquad & 
{\rm{KK\ (isom)}} \lra {\rm{NS5}} \,;
\ea\ee
Everything else is unaffected.

Let $z$ be the isometry direction.  Then T-duality acts on NS-NS
fields as follows:
\be\label{ztdual}\ba{rl}
\bs
e^{2{\tilde{\Phi}}} = &\! e^{2\Phi}/G_{zz} \,,\quad
{\tilde{G}}_{zz} = 1/G_{zz} \,,\quad
{\tilde{G}}_{\mu z} = B_{\mu z}/G_{zz} \,,\quad
{\tilde{B}}_{\mu z} = G_{\mu z}/G_{zz} \,,\\
\bs
{\tilde{G}}_{\mu\nu} =&\! G_{\mu\nu} - 
\left(G_{\mu z}G_{\nu z}-B_{\mu z}B_{\nu z}\right)/G_{zz}\,,\\
{\tilde{B}}_{\mu\nu} =&\! B_{\mu\nu} - 
\left(B_{\mu z}G_{\nu z}-G_{\mu z}B_{\nu z}\right)/G_{zz} \,.
\ea\ee
T-duality also acts on R-R fields, and the correct formul\ae\ can be
found in \cite{myerspuff}.  For simple situations involving no NS-NS
B-field and no off-diagonal metric components, we have either
${\tilde{C}}_{\mathit n+1}{\eql}C_{\mathit n}\wedge dz$ ($\perp$) or
${\tilde{C}}_{\mathit n}\wedge dz{\eql}C_{\mathit n+1}$ ($\parallel$),
as appropriate.

Note that if we do T-duality on a supergravity D$p$-brane in a
direction perpendicular to its worldvolume, we are dualising in a
direction which is not an isometry, because the metric and other
fields depend on the coordinates transverse to the brane.  But the
T-duality formul\ae\ for supergravity fields apply only when the
direction along which the T-duality is done is an isometry direction.
If it is not, then we should first ``smear'' the D$p$-brane in that
direction to create an isometry and then do T-duality.  We will
discuss smearing explicitly in subsection (\ref{subssmear}) for the
case of BPS branes.

Note also that in the presence of some branes, string momentum or
winding number may not be conserved, \aeg\ string winding number in a
KK background. However, the conserved quantity transforms as expected
under T-duality, as discussed in \cite{unwinding}.  

\subsection{An example of solution-generating}\label{solgeneg}

In general, finding new solutions of supergravity actions can be quite
difficult because the equations of motion are very nonlinear.  The
search for new solutions is aided by classical no-hair theorems, which
say that once the conserved charges of the system of interest are
determined, the spacetime geometry is unique.  It is important for
applicability of the no-hair theorems that any black hole singularity
be hidden behind an event horizon; the theorems fail in spacetimes
with naked singularities.

There is a solution-generating method available in string theory which
is purely algebraic(!).  We will wrap up this section by giving an
explicit example of how easily new solutions can be made using this
method, by starting with a known solution.

Consider a neutral black hole in $(d-1)$ dimensions, which may be
thought of as a higher-dimensional version of $d{\eql}4$ Schwarzschild:
\be\label{ddimschwarzschild}
d{\hat{S}}^2_{d-1} = -\left(1-K(\rho)\right)dt^2 +
\left(1-K(\rho)\right)^{-1}d\rho^2 + \rho^2d\Omega_{d-3}^2\,,
\ee
where 
\be
K(\rho)\equiv \left({{\rH}\over{\rho}}\right)^{d-4} \,.
\ee
There is no gauge field or dilaton turned on, so this is a solution in
string and Einstein frame.

The mass of this spacetime is obtained using the general procedure of
subsection \ref{mandjandq}. The harmonic gauge condition is satisfied
here and so, via
\be
g_{tt} \sim -1 + {{16\pi G_{d-1}
M_{d-1}}\over{(d-3)\Omega_{d-3}\rho^{d-4}}}  \,,
\ee
we extract
\be
M_{d-1} = {{(d-3)\Omega_{d-3} \rH^{d-4}}\over{16\pi G_{d-1}}} \,.
\ee
Since this black hole is a solution of the $d-1$ dimensional Einstein
equations, taking a direct product of it with the real line $\bR$
satisfies the $d$ dimensional Einstein equations (this can be checked
explicitly).  This procedure is called a ``lift'' and we end up with a
configuration in $d$ dimensions with translational invariance in the
$z$ direction:
\be\ba{rl} 
\bs
dS_d^2 = & dz^2 - \left(1-K(\rho)\right)dt^2 +
\left(1-K(\rho)\right)^{-1}d\rho^2 + \rho^2d\Omega_{d-3}^2 \,,\cr 
= & \left(-dt^2 + dz^2\right) + K(\rho)dt^2 +
\left(1-K(\rho)\right)^{-1}d\rho^2 + \rho^2d\Omega_{d-3}^2 \,.  
\ea\ee
Now let us do a boost on this configuration:
\be
\pmatrix{ dt \cr dz} \rightarrow 
\pmatrix{ \cosh\!\gamma & \sinh\!\gamma \cr 
          \sinh\!\gamma & \cosh\!\gamma } \,
\pmatrix{ dt \cr dz } \,.
\ee
This transformation takes solutions to solutions, as can be checked by
substituting into the equations of motion.  Boosting is a general
procedure that can be used to make new solutions, as in
\cite{senmake}.  The metric is affected as
\be\ba{rl}
dS_d^{2\ \prime} = & \left(-dt^2 + dz^2\right) +
K(\rho)\left(\cosh\!\gamma{}dt+\sinh\!\gamma{}dz\right)^2 \cr
\bs
& +\left(1-K(\rho)\right)^{-1}d\rho^2 + \rho^2d\Omega_{d-3}^2 \cr = &
-dt^2 \left(1-K(\rho)\cosh^2\!\gamma\right) +
dz^2 \left(1+K(\rho)\sinh^2\!\gamma\right) \cr & +
2dtdz\cosh\!\gamma\sinh\!\gamma{K(\rho)} +
\left(1-K(\rho)\right)^{-1}d\rho^2   + \rho^2d\Omega_{d-3}^2 \,.
\ea\ee
The horizon, which is at $G^{\rho\rho}\rightarrow0$, occurs when
$K(\rho)=1$ \aie at $\rho=\rH$, not at $G_{tt}=0$.  Now, suppose the
$z$ dimension is compactified on a circle whose radius is $R$ at
$\infty$, \aie in the asymptotically flat region of the geometry.  At
$\rho=\rH$, by contrast, the radius of the circle at the horizon is
$R\,\sqrt{G_{zz}(r{\eql}r_H)}=\,R\cosh\!\gamma\,>\,R$.  Therefore we
see that adding longitudinal momentum makes the compactified dimension
larger at the horizon.  

Now let us KK down again to make new $(d-1)$-dimensional black hole.
We had in subsection (\ref{simplereduction}) the relations
$$\ba{rl}
\bs
dS^2_d = & d{\hat{S}}_{d-1}^2+
e^{2{\hat{\chi}}}\left(dz+{\hat{A}}_\mu{}dz^\mu\right)^2 
\,,\cr
e^{\Phi} = & e^{{\hat{\Phi}}+\half{\hat{\chi}}} \,,
\ea$$
so, for example,
\be
{\hat{G}}_{tt}=G_{tt}-G_{tz}^2/G_{zz} = -1 + K\cosh^2\!\gamma -
{{(K\cosh\!\gamma\sinh\!\gamma)^2}\over{(1+K\sinh^2\!\gamma)}}  \,.
\ee
{}From this we obtain
\be
d{\hat{S}}_{d-1}^{2\ \prime} = {{-\left(1 - K(\rho)\right)}
\over{\left(1+K(\rho)\sinh^2\!\gamma\right)}}dt^2 
+ {{1}\over{\left(1-K(\rho)\right)}}d\rho^2 + \rho^2d\Omega_{d-3}^2
\,,
\ee
and
\be
{\hat{A}}_t = {{K(\rho)\cosh\!\gamma\sinh\!\gamma}\over
{\left(1+K(\rho)\sinh^2\!\gamma\right)}} \,,
\ee 
and 
\be 
e^{\hat{\Phi}} = e^{-\half{\hat{\chi}}} =
\left(1+K(\rho)\sinh^2\!\gamma\right)^{-\quarter}\,.
\ee
The conserved quantum numbers of this new spacetime are
\be\ba{rl}
\bs
M^\prime &\!= {\displaystyle{
{{\Omega_{d-3}\rH^{d-4}}\over{16\pi G_{d-1}}}
\left[(d-3) + (d-4)\sinh^2\!\gamma\right] }} \,,\cr
Q^\prime &\! = {\displaystyle{
R {{\Omega_{d-3}\rH^{d-4}}\over{16\pi G_{d-1}}}
\left[{{1}\over{2}}\sinh(2\gamma)\right] }} \,.
\ea\ee
To regain the original neutral black hole, we simply take the limit
$\gamma\rightarrow 0$.

Now consider taking the opposite limit $\gamma\rightarrow\infty$.  In
order to keep our expressions from blowing up, we must also take the
horizon radius of the original black hole to zero, $\rH\rightarrow 0$,
in such a fashion that
\be
{{1}\over{2}}\rH^{d-4}e^{2\gamma}\equiv k={\rm{fixed\ }} \ \,,
\quad {\rm{so\ }} K(\rho) = {{k}\over{\rho^{d-4}}} \,.
\ee
Defining light-cone coordinates $dz^\pm\equiv(t\pm z)/\sqrt{2}$, we
find in the higher dimension
\be
dS^2_d = -2dz^+\left[dz^- - {{k}\over{\rho^{d-4}}}dz^+\right] +
\left(d\rho^2+\rho^2d\Omega_{d-3}^2\right)\,.
\ee
This is the gravitational wave W, which has zero ADM mass in $d$
dimensions.  If we wanted to create a (NS-NS) charged black string
configuration instead of a gravitational wave, we would use T-duality
as in (\ref{ztdual}) to convert; we would get the fundamental string
F1.  We could then use other dualities to convert that to a D$p$-brane
or NS5-brane spacetime.

Taking the same limit for the $(d-1)$-dimensional black hole gives the
extremal black hole, which has zero Hawking temperature.  The
connection between these two extremal animals is brought into relief
via the relation
\be 
M_d^2 = 0 = M_{d-1}^2 - {{Q^2}\over{R^2}} \,.
\ee
The $d-1$-dimensional charge is the $z$-component of the
$d$-dimensional momentum.

The wave W is one of the purely gravitational BPS objects in string
theory.  The other is the KK monopole.  Labelling the five
longitudinal directions $y_{1\cdots 5}$, and the four transverse
directions $x_i,i=1,2,3$, and $z$; the metric is
\be\ba{l}\label{kkeqn}
\bs
ds^2 = -dt^2 + dy_{1\cdots 5}^2 + 
   H^{-1}(x)\left(dz+A_idx^i\right)^2 + H(x)dx_{1\cdots 3}^2 \,,\cr 
2\partial_{[i}A_{j]}(x) =  \epsilon_{ijk}\partial_k H(x) \,.
\ea\ee
The $A_i$ can be found via the curl equation, given that $H=1+k/|x|$.
The periodicity of the azimuthal angle must be $4\pi$ to avoid conical
singularities.

If we want to put angular momenta $J_i$ on our charged black holes,
strings, or branes, we must start with a Kerr-type black hole, rather
than a Schwarzschild-type one.  In Boyer-Lindqvist-type coordinates,
with one angular momentum $a$ and $G_d$ temporarily set to 1 for
simplicity, the metric in $d>3$ dimensions is \cite{robangmom}
\be\ba{rl}
\bs
ds_d^2= &\! -{\displaystyle{
{{\left(\rho^2+a^2\cos^2\!\theta-2m\rho^{5-d}\right)} \over
{\left(\rho^2+a^2\cos^2\!\theta\right)}} dt^2 + 2dtd\varphi
{{2m\rho^{5-d}a\sin^2\!\theta}\over{\left(\rho^2+a^2\sin^2\!\theta\right)}}
}} \cr
\bs
&\! + {\displaystyle{ 
 {{\sin^2\!\theta}\over{\left(\rho^2+a^2\cos^2\!\theta\right)}}
\left[ \left(\rho^2+a^2\right)\left(\rho^2+a^2\cos^2\!\theta\right)
+ 2ma^2\sin^2\!\theta\rho^{5-d}\right]d\varphi^2 }}
\cr
&\! + {\displaystyle{
{{\left(\rho^2+a^2\cos^2\!\theta\right)}\over{\rho^2+a^2-2m\rho^{5-d}}}
d\rho^2 + \left(\rho^2+a^2\cos^2\!\theta\right)d\theta^2 
+ \rho^2\cos^2\!\theta  d\Omega_{d-4}^2 }}  \,.
\ea\ee
The horizon is at $G^{\rho\rho}\rightarrow 0$, \aie at
\be
\rho^2 + a^2 - 2m\rho^{5-d} = 0 \,.
\ee
There is a behaviour change at $d{\eql}5$.  For $d{\eql}4$,
$r_\pm{\eql}m\pm\sqrt{m^2-a^2}$ and so there is a maximum angular
momentum $a_{\rm max}{\eql}m$.  For $d{\eql}5$, the horizons are present
if $a^2\!\leq\!m$, and the singularity structure is different.  In
addition, angular momentum is consistent with supersymmetry
\cite{BMPV}, unlike for $d{\eql}4$.  Lastly, for $d\!>\!5$, there is
always a solution with $r_+\!>\!0$, so there is no restriction on the
angular momentum for classical rotating black holes.

The equations and the analysis are more complicated if there are two
or more angular momentum parameters.  The details are contained in
\cite{robangmom}.  Note that these higher-$d$ black holes can be used
as the starting point for generating rotating string and brane
solutions using the boosting procedure, in direct analogy to the
example we gave above.  For example, since we obtain a $d{\eql}10$
string by doing boosts and dualities on a $d{\eql}9$ black hole, we see
that there are up to four independent angular momentum parameters for
a black string.

\sect{$p$-branes, extremal and non-extremal}\label{sectthree}

String theory spacetimes with conserved quantum numbers can be black
holes, but more commonly they are black $p$-branes
\cite{horostrom}.  These objects have translational symmetry
in $p$ spatial directions and, as a consequence, their horizon (for
zero angular momenta) is typically topologically
$\bR^{p}\!\times\!{S^{d-1}}$, where $d$ is the number of space
dimensions transverse to the $p$-brane.

Type IIA string theory in the strong coupling limit is
eleven-dimensional supergravity, which has only two fields in its
bosonic sector, the metric tensor and the three-form gauge potential.
We start our discussion of branes with the BPS M-branes.

\subsection{The BPS M-brane and D-brane solutions}

The BPS M2-brane spacetime has worldvolume symmetry group $SO(1,2)$,
and the transverse symmetry group is $SO(8)$.  Let us define the
coordinates parallel and perpendicular to the brane to be
$(t,x_\parallel)\,,x_\perp$, respectively.  Then, using these
symmetries and a no-hair theorem, the spacetime metric turns out to
depend only on $|x_\perp|\equiv r$, and has the form
\be
ds^2_{11} = H_2^{-2/3} dx_\parallel^2 + H_2^{1/3}dx_\perp^2 
\,,\quad
A_{012} = -H_2^{-1} \,.
\ee 
The fact that the same function appears in the metric and gauge field
is a consequence of supersymmetry.  Note that the metric is
automatically in Einstein frame because there is no string frame in
$d{\eql}11$.  It turns out that supersymmetry alone is not enough to give
the equation that the function $H$ must satisfy; rather, the
supergravity equations of motion must be used.  One finds that $H_2$
must be harmonic as it satisfies a Laplace equation in $x_\perp$.  The
solution is
\be 
H_2 = 1 + {{r_2^6}\over{r^6}} \,,\quad {\rm{where}}\ r_2^6=32\pi^2 N_2
\ell_{11}^6 \,,
\ee
where we remind the reader that $\ell_{11}$ is the eleven-dimensional
Planck length.

The BPS M5-brane has symmetry group $SO(1,5)\times SO(5)$, and the
metric is
\be
ds^2_{11} = H_2^{-1/3} dx_\parallel^2 + H_2^{2/3}dx_\perp^2 \,,
\ee
and the harmonic function is this time 
\be
H_5 = 1 + {{r_2^3}\over{r^3}} \,,\quad
{\rm{where}}\  r_5^3=\pi N_5\ell_{11}^3 \,.
\ee
In this case, the gauge field is magnetically coupled, $F_{\mathit 4}$
is proportional to the volume element on the $S^4$ transverse to the
M5-brane.

For the M2, the origin of coordinates $r=0$ is singular and so there
must be a $\delta$-function source there, to wit the fundamental
M2-brane.  This happens essentially because the M2-brane is
electrically coupled. The magnetically coupled BPS M5-brane is
solitonic and nonsingular, in that the geometry admits a maximal
analytic extension without singularities \cite{ght95}.  However, the
nonextremal version of the M5 has a singularity and does need a
source.  Near-horizon, the M2 spacetime is $AdS_4\times S^7$ and the
M5 is $AdS_7\times S^4$.  Since the M2 and M5 are asymptotically flat,
again we have interpolation between 2 highly supersymmetric vacua as
in the case of the Reissner-Nordstr{\"{o}}m black hole.

Let us now move down to ten dimensions.  The symmetry for BPS
D$p$-branes is $SO(1,p)\times SO(9-p)$.  In the string frame, the
solutions are \cite{horostrom}:
\be\ba{rl}\label{dpsugra}
\bs
dS^2 & = H_p(r)^{-\half}\left(-dt^2+dx_\parallel^2\right) 
+ H_p(r)^{+\half}dx_\perp^2 
\,,\cr
\bs
e^{\Phi} & =  H_p(r)^{\quarter(3-p)} \,,\cr
C_{01\cdots p} & = \gs^{-1}\left[ 1 - H_p(r)^{-1} \right] \,.
\ea\ee
The function $H_p(r)$ is harmonic; it satisfies $\partial_\perp^2
H_p(r)=0$,
\be\label{eqnforHp}
H_p = 1 + {{c_p\gs N_p\ls^{7-p}}\over{r^{7-p}}} \,,\qquad 
c_p \equiv 
(2\sqrt{\pi})^{(5-p)}\Gamma\left[\half(7-p)\right] \,.
\ee
Note that the function $H_p$ would still be harmonic if the constant
piece, namely the 1, were missing.  The asymptotically flat part of
the geometry would be absent for this solution.  

The (double) horizon of the D$p$-brane geometry occurs at $r=0$, and
in every case except the D3-branes the singularity is located there as
well.  Hence, for the D$p$-branes with $p\not=3$, the singularity is
null.  Since the singularity and the horizons coincide for these
cases, we may worry that the singularity is not properly hidden behind
an event horizon, and so perhaps it should be classified as naked.  We
therefore demand that a null or timelike geodesic coming from infinity
should not be able to bang into the singularity in finite affine
parameter.  Interestingly, this condition separates out the D6-brane
from the others as being the only one possessing a naked
singularity!\footnote{We first realized this in a conversation with
Donald Marolf, although the observation may not be original.}

For the D3-brane the dilaton is constant, and the spacetime turns out
to be totally nonsingular: all curvature invariants are finite
everywhere.  This allows a smooth analytic extension inside the
horizon, like the case of the M5-brane \cite{ght95}.  The near-horizon
D3-brane spacetime is $AdS_5\times S^5$.  The Penrose diagram for the
D3 is like that of the M5.

The causal structures of the BPS M-branes and D$p$-branes are
summarised in the Penrose diagrams in Fig.\ref{m2m5dppenr}.  Note that
the isotropic coordinates $x_\perp$ cover only part (shaded) of the
maximally extended spacetime.

\bfig
\bmp{0.33}
\hskip0.05\textwidth
\epsfysize=1.75truein
\epsfbox{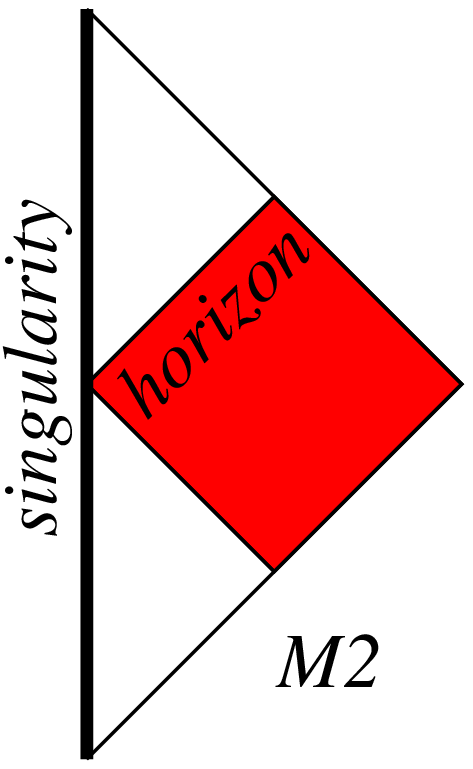}
\emp
\bmp{0.33}
\hskip0.05\textwidth
\epsfysize=1.375truein
\epsfbox{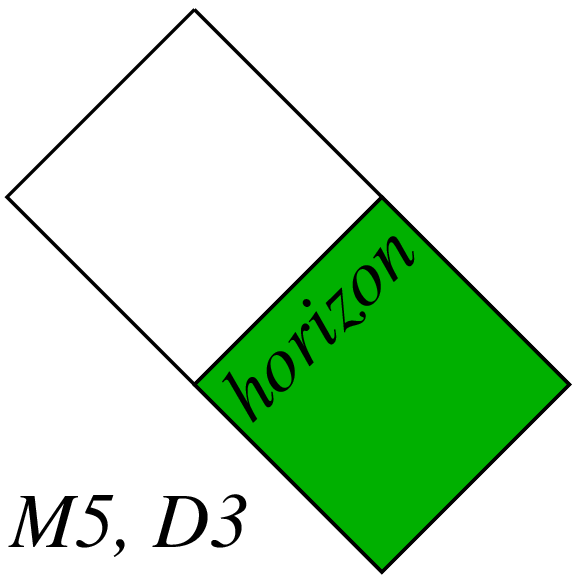}
\emp
\bmp{0.33}
\hskip0.1\textwidth
\epsfysize=0.875truein
\epsfbox{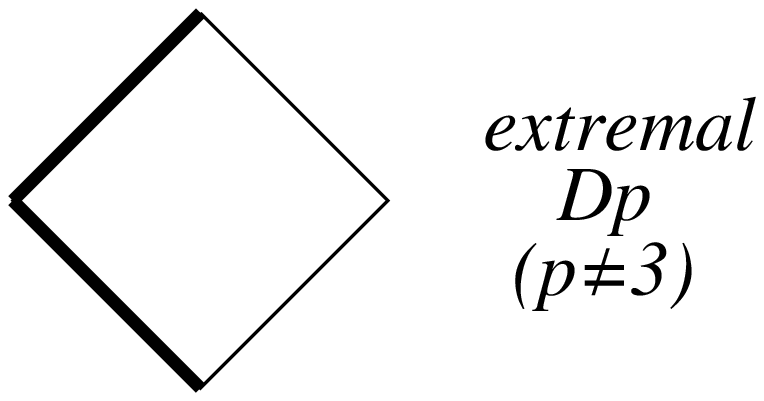}
\emp
\caption{\small{The Penrose diagrams for the extremal M- and 
D$p$-branes.}}
\label{m2m5dppenr}
\efig

The F1 and NS5 spacetimes may be found by using the T- and S-duality
formul\ae\, that we gave in the last subsection.

\subsection{Arraying BPS branes}\label{subssmear}

Consider the BPS D$p$-branes.  They are described by the metric
(\ref{dpsugra}) with a single-centred harmonic function $H_p$.  In
fact, BPS multi-centre solutions are also allowed because  the
equation for $H_p$, $\nabla_\perp^2 H_p=0$, is linear:
\be
H_{\overline{p}} = 1 + c_p\gs N_p\ls^{7-p} 
\sum_{i} {{1}\over{|x_\perp-x_{\perp\ i}|}^{7-p}} \,;
\ee
The physical reason this works is that parallel BPS branes of the same
kind are in static equilibrium: the repulsive gauge forces cancel
against the attractive gravitational and dilatonic forces.

Let us make an infinite array of D$p$-branes along the $x^{p+1}$
direction with periodicity $2\pi{}R$.  Define
\be
r^2\equiv{\hat{r}}^2+(x^{p+1})^2 \,;
\ee
then
\be
H_{\overline{p}} = 1 + c_p\gs N_p\ls^{7-p} 
\sum_{n=-\infty}^{\infty} {{1}\over{
\left[{\hat{r}}^2+\right(x^{p+1}-2\pi{}Rn\left)^2\right]
	}^{\half(7-p)}} \,. 
\ee
Now, if $x_\perp\gg R$, then the summand varies slowly with $n$ and
we can approximate the sum by an integral.  Changing variables to $u$,
\be
x^{p+1} \equiv 2\pi{}Rn - {\hat{r}}u \,,
\ee
we obtain
\be\ba{rl}
H_{\overline{p}} \simeq & {\displaystyle{
 1+ c_p\gs N_p\ls^{7-p} {{1}\over{2\pi{}R}} 
{{1}\over{{\hat{r}}^{7-[p+1]}}} 
{\underbrace{\int du {{1}\over{\sqrt{1+u^2}^{(7-p)}}} }}  }}  \,, \cr
{\ } & \hskip2.0truein \equiv I_p \,.
\ea\ee
The quantity $I_p$ can be easily evaluated, 
\be
I_p = \sqrt{\pi}\Gamma\left[\half(7-\{p+1\})\right]/
		\Gamma\left[\half(7-p)    \right] \,.
\ee
Then using $b_p=(2\sqrt{\pi})^{5-p}\Gamma\left[\half(7-p)\right]$ we
find
\be
H_{\overline{p}} \simeq 1 + \left[{{N_p}\over{(R/\ls)}}\right]
\gs c_{p+1} \left({{\ls}\over{\hat{r}}}\right)^{7-[p+1]} \,.
\ee
We can now take the limit that the arrayed objects make a linear
density of branes.  Then matching the thereby smeared harmonic
function with the $(p+1)$-brane harmonic function $H_{p+1}$ gives
\be
N_{p+1} = {{N_p}\over{\left(R/\ls\right)}} \,.
\ee
We see that the linear density of $p$-branes per unit length in string
units becomes the number of $(p+1)$-branes.

To check the identification we use the T-duality rules (\ref{ztdual}),
with the isometry direction $x^{p+1}=z$, to obtain
\be\ba{rl}
\bs
d{\tilde{S}}^2 &\! = H_{\overline{p}}^{-\half} \left( -dt^2 +
dx_{1\ldots p}^2 + dz^2 \right) + H_{\overline{p}}^{\half} 
\left(d{\hat{r}}^2 + {\hat{r}}^2d\Omega_{[8-(p+1)]}^2\right)  \,; \cr
\bs
e^{\tilde{\Phi}} &\!= 
H_{\overline{p}}^{\quarter(3-p)}/H_{\overline{p}}^{\quarter}
= H_{\overline{p}}^{\quarter[3-(p+1)]} \,,\\
{\tilde{C}}_{01\ldots p+1} &\!= \gs^{-1}\left[ 1 -
H_{\overline{p}}^{-1} \right] \,.
\ea\ee
These agree with our expectations; they are precisely the supergravity
fields appropriate to the D$(p{\pls}1)$-brane.

The procedure of arraying the branes and then taking the limit is
known as ``smearing''; it results in a larger brane.  Unsmearing, on
the other hand, is in general difficult because dependence on the
additional coordinate(s) must be reconstructed.  In the case of a
single type of D-branes we can guess and correctly get known results,
but more generally guessing is not enough.  In some cases with
intersecting branes, unsmeared solutions do not exist, for good
physics reasons \cite{awps99}.

Using dualities and our arraying formul\ae\ we can of course
interconnect all M-branes and D-branes with the NS-branes, W and KK.
In working through this exercise, it is worth remembering that
worldvolume directions are already isometry directions, and so in
reducing along a worldvolume direction of a D$p$-brane we have simply
$N_{p+1}=N_p$.

\subsection{$p$-brane probe actions and kappa symmetry}

We would now like to consider what happens when we probe a D$p$-brane
spacetime, using another D$p$-brane.  We will treat the probe as a
``test'' brane, \aie we will ignore its effect on the background
geometry.  This is a very good approximation provided that $N$, the
number of branes sourcing the spacetime, is large.

The action of a probe brane in a supergravity background has two
pieces,
\be
S_{\rm probe} = S_{\rm DBI} + S_{\rm WZ} \,,
\ee
which are, to lowest order in derivatives,
\be\ba{rl}
\bs
S_{\rm DBI} =&\! - {\displaystyle{ {{1}\over{\gs (2\pi)^p l_s^{p+1}}}
\int}}
d^{p+1}\sigma e^{-\Phi} \sqrt{-\det\bP\left(
G_{\alpha\beta} + \left[2\pi F_{\alpha\beta} + B_{\alpha\beta}\right]
\right) } \,, \cr
S_{\rm WZ} = &\! - {\displaystyle{ {{1}\over{(2\pi)^p l_s^{p+1}}} 
\int}}
\bP\exp\left( 2\pi F_{\mathit 2} + B_{\mathit 2} \right) \wedge 
{\displaystyle{\oplus_n}} C_{\mathit n}
\,.
\ea\ee
where the $\sigma$ are the worldvolume coordinates and $\bP$ denotes
pullback to the worldvolume of bulk fields.  The brane action encodes
both kinetic and potential information, such as which branes can end
on other branes \cite{wittenf1d1,mrd}.  The WZ term, in particular,
encodes the fact that D$p$-branes can carry charge of {\em smaller}
D-branes by having worldvolume field strength $F_{\mathit 2}$ turned
on.

Let us digress a bit on the structure of this action before we do the
actual probe computation.  The action we have written is appropriate
for a brane which is topologically $\bR^{1,p}$, and it also works for
branes wrapped on tori.  If the D-brane is wrapped on a manifold which
is not flat, extra terms arise in the probe action.  An example is the
case of K3, where extra curvature terms appear \cite{BBGetc},
consistent with dualities.

Another interesting piece of physics which this action for a single
probe brane does not capture is the dielectric or ``puffing up''
phenomenon of \cite{myerspuff}.  What happens there is that the
presence of $n$ probe branes allows some non-commutative terms in the
probe branes' action which couple in to {\em higher} R-R form
potentials.  An example is the fact that D0-branes in a constant
4-form field strength background develop a spherical D2-brane aspect.
For details on the modifications to the probe D$p$-brane actions, the
reader is referred to \cite{myerspuff}.  The full action for $n$ probe
branes, which involves a nonabelian $U(n)$ worldvolume gauge field, is
in fact not known explicitly because the derivative expansion and the
expansion in powers of the field strength $F$ can no longer be
unambiguously separated.  See the recent review \cite{aatBIreview}.

The action $S_{\rm probe}$ possesses bulk supersymmetry, but not
world-brane supersymmetry {\em a priori}.  The $U(1)$ gauge field
$F_{\mathit 2}$ lives on the branes, while the metric and $B$-field
are pullbacked to the brane in a supersymmetric way, \aeg
\be
\bP\left(G_{\alpha\beta}\right) = \left( \partial_\alpha X^\mu -
i{\overline{\theta}}\Gamma^\mu \partial_\alpha\theta \right)
\left( \partial_\beta X^\nu -
i{\overline{\theta}}\Gamma^\nu \partial_\beta\theta \right) G_{\mu\nu}
\,.
\ee
After fixing of reparametrisation gauge invariance and on-shell, there
are twice too many fermionic degrees of freedom.  This problem is
familiar already from the Green-Schwarz approach to superstring
quantisation \cite{gswbook}.  The solution lies in an additional
symmetry known as kappa-symmetry, a local fermionic symmetry which
eliminates the unwanted fermionic degrees of freedom via a projection
condition.  In the case of Green-Schwarz quantisation of the
superstring in a flat background, kappa-symmetric actions need a
constant $B_{\mathit 2}$ turned on.  In light-front gauge, the
projection condition which ensues is $\Gamma^+ \theta^{1,2}=0$, and
then via the equations of motion one sees that the erstwhile
worldsheet scalars $\theta$ are in fact worldsheet spinors, and
worldsheet supersymmetry then becomes manifest.  See also the very
recent important work of \cite{berkovits}, in which a manifestly
supersymmetric covariant quantisation of the Green-Schwarz superstring
has been achieved.

A similar procedure works for the D-branes as well.  In this case the
DBI and WZ terms need each other in order to ensure kappa symmetry,
all the while respecting bulk supercovariance.  There is an intricate
consistency \cite{kappa23} between kappa symmetry, the bulk
supergravity constraints\footnote{Here we mean supergravity
constraints in the technical sense; see \aeg \cite{wessbaggerbook}.},
and the bulk supergravity equations of motion.  In a flat target
space, the case of static gauge was worked out in \cite{kappa4}; the
kappa symmetry can be used to eliminate $\theta^2$ and then the other
spinor $\theta^1$ becomes the worldvolume superpartner of the $U(1)$
gauge field and the transverse scalars.  More generally, fixing the
reparametrisation and kappa gauge symmetries to give manifest
worldvolume SUSY is tricky.  There has been some progress in
$AdS\!\times\!S$ spaces, see \aeg \cite{kappaadsxs}.

Now let us get back to using our test D$p$-brane to probe the
supergravity spacetime formed by a large number $N$ of the same type
of brane.  We have for the supergravity background the fields
(\ref{dpsugra}), which we repeat here for ease of reference,
$$\ba{rl} 
\bs
dS^2 &\!= H_p^{-\half}\left(-dt^2
+dx_\parallel^2\right) + H_p^{+\half}dx_\perp^2\,\cr 
\bs
e^{\Phi} &\!= H_p^{\quarter(3-p)} \,,\cr 
C_{01\cdots p} &\!= \gs^{-1} \left[ 1 - H_p^{-1} \right] \,.  
\ea$$
The physics is easiest to interpret in the static gauge, where we fix
the worldvolume reparametrisation invariance by setting
\be
X^{\alpha}=\sigma^{\hat{\alpha}} \,, \quad \alpha = 0,\ldots p \,.
\ee
We also have the $9-p$ transverse scalar fields $X^i$, which for
simplicity we take to be functions of time only,
\be
X^i = X^i(t) \,,\quad i = p+1 \ldots 9 \,.
\ee
We will denote the transverse velocities as $v^i$,
\be
v^i \equiv {{dX^i}\over{dt}} \,.
\ee
Now we can compute the pullback of the metric to the brane.
\be\ba{rl}
\bs
\bP\left(G_{00}\right) &\!=
  (\partial_{0}X^\alpha)(\partial_{0}X^\beta)
  G_{\alpha\beta} 
+ (\partial_{0}X^i)(\partial_{0}X^i) G_{ij} \cr
\bs
&\!\qquad = G_{00} + G_{ij}v^iv^j = -H_p^{-\half} + H_p^{+\half} 
{\vec{v}}^2 \,;\cr
\bP\left(G_{\alpha\beta}\right) &\!=  H_p^{-\half} \,.
\ea\ee
The next ingredient we need is the determinant of the metric.  To
start, notice that 
\be -\det\bP(G_{\alpha\beta}) ({\vec{v}}={\vec{0}})
=H_p^{-\half(p+1)} \,, \ee
so that
\be
\sqrt{-\det\bP(G_{\alpha\beta})}=
H_p^{-\quarter(p+1)}\,\sqrt{1-{\vec{v}}^2H_p} \,.
\ee
Putting this together with the expression for the dilaton and the R-R
field, we obtain
\be
S_{\rm{DBI}}+S_{\rm{WZ}}=  {{1}\over{(2\pi)^{p+1}\gs\ls^{p+1}}}
\int d^{p+1}\sigma \left[
-H_p^{-1}\sqrt{1-{\vec{v}}^2H_p}+H_p^{-1} -1 \right] \,.
\ee
{}From this action we learn that the position-dependent part of the
static potential vanishes, as it must for a supersymmetric system such
as we have here.  The constant piece is of course just the D$p$-brane
tension.  In addition, we can expand out this action in powers of the
transverse velocity.  We see that, to lowest order,
\be
S_{\rm probe} = {{1}\over{(2\pi)^{p+1}\gs\ls^{p+1}}}
\int d^{p+1}\sigma \left[ -1 + \half {\vec{v}}^2
+ {\cal{O}}({\vec{v}}^4) \right] \,,
\ee
and so the metric on moduli space, which is the coefficient of
$v^iv^j$, is flat.  This is in fact a consequence of having sixteen
supercharges preserved by the static system.

\subsection{Nonextremal branes}

In string frame and with a Schwarzschild-type radial coordinate
$\rho$, the metric and dilaton fields of the nonextremal versions of
the D$p$-branes can be written as \cite{duffreview}
\be\ba{rl}
\bs
dS^2 &= -\Delta_+(\rho)\Delta_-(\rho)^{-\half}dt^2 +
\Delta_-(\rho)^{+\half}dx_{\parallel}^2 + \cr
\bs
& \qquad\qquad
\Delta_+(\rho)^{-1}\Delta_-(\rho)^{\half(p-3)/(7-p)-1}d\rho^2 +
\rho^2\Delta_-(\rho)^{\half(p-3)/(7-p)}d\Omega^2 \,,\cr
e^\Phi &=  \Delta_-(\rho)^{\quarter(p-3)} \,,
\ea\ee
where 
\be
\Delta_\pm(\rho)\equiv 1 - \left({{r_\pm}\over{\rho}}\right)^{7-p}\,.
\ee
and the Hodge dual field strength for the R-R potential is directly
proportional  to the volume-form on the $(8{\mns}p)$-sphere.

Defining 
\be
r_+^{7-p}=\rH^{7-p}\cosh^2\!\beta \,,\quad
r_-^{7-p}=\rH^{7-p}\sinh^2\!\beta\,,
\ee
and making a change of coordinates to $r^{7-p}=\rho^{7-p}-r_-^{7-p}$,
the metric turns into a form more easily related to the extremal case
we studied in the last subsection,
\be
dS^2 = D_p(r)^{-\half}\left(-K(r)dt^2+dx_\parallel^2\right) +
D_p(r)^{\half}\left(dr^2/K(r)+r^2d\Omega_{8-p}^2\right) \,,
\ee
where 
\be\label{DandKfns}
D_p(r)=1+(\rH/r)^{7-p}\sinh^2\!\beta, \,,\qquad
K(r)=1-(\rH/r)^{7-p} \,.
\ee
The other fields are
\be\ba{rl}
e^\Phi         =&\! D_p(r)^{(3-p)/4} \,,\\
C_{01\ldots p} =&\! (\coth\!\beta)\gs^{-1}
		     \left[1 - D_p(r)^{-1}\right] \,.
\ea\ee
In these expressions, the boost parameter $\beta$ is given by
\be\label{eqnforbeta}
\sinh^2\!\beta = -\half + 
\sqrt{\quarter + \left[c_p{\gs}N(\ls/\rH)^{7-p}\right]^2}
\ee
Notice in particular that in the extremal limit, where
$\rH\!\rightarrow\!0$, $\beta\!\rightarrow\!\infty$.  Alternatively,
the change in the harmonic function due to nonextremality can be
codified in a parameter $\zeta=\tanh\!\beta$:
\be\label{zetaeqn}
D_p(r) = 1 + \zeta c_p{\gs}N(\ls/r)^{7-p} \,,\qquad
\zeta = \sqrt{1 + 
          \left[{{\rH^{7-p}}\over{2c_p{\gs}N\ls^{7-p}}}\right]^2}
        - \left[{{\rH^{7-p}}\over{2c_p{\gs}N\ls^{7-p}}}\right] \,.
\ee
Then we can express the gauge field as
\be
C_{01\ldots p} = \zeta^{-1} \gs^{-1} \left[1-D_p(r)^{-1}\right] \,.
\ee

The ADM mass per unit $p$-volume  and the charge are
\be\ba{rl}
{{M_p}\over{(2\pi)^pV_p}}  = &\!  {\displaystyle{
{{(\rH/\ls)^{7-p}}\over{c_p\gs^2\left(2\pi\right)^p\ls^{p+1}}}
\left[ \cosh^2\!\beta + {{1}\over{(7-p)}} \right] }} \,, \cr
N_p = &\! {\displaystyle{
{{1}\over{c_p\gs}}\left({{\sqrt{r_+r_-}}\over{\ls}}\right)^{7-p}
}} \,.
\ea
\ee
The Hawking temperature and the Bekenstein-Hawking entropy are,
respectively, 
\be\ba{rl}
\bs
\TH =& {\displaystyle{ {{(7-p)}\over{4\pi\rH\cosh\!\beta}} }}  \,,\cr
\SBH =& {\displaystyle{
	{{\Omega_{8-p}}\over{4G_{10-p}}} 
	\rH^{8-p}\cosh\!\beta }} \,.
\ea\ee
The extremal solution has degenerate horizons $r_+{\eql}r_-$, and zero
Bekenstein-Hawking entropy $\SBH$.  The Hawking temperature $\TH$ of
the extremal brane is also zero.

If we were to wrap this brane on a $T^p$, then by the neat consistency
of $\SBH$ in various dimensions we discussed in section 1, the zero
entropy result is also true of the $d{\eql}(10{\mns}p)$ R-R black hole.
The volume of the torus at the horizon $\propto D_p(\rH)^{-\quarter
p}\rightarrow 0$ at extremality.  This fact is related to zero
entropy, via the field equations.

The causal structure of the uncompactified nonextremal D$p$-brane can
be found by noticing that the inner horizon is singular.  The Penrose
diagram in the $(t,\rho)$ plane then looks like that of a
Schwarzschild black hole.

We close the discussion of the nonextremal D$p$-brane solutions with a
remark on supergravity $p$-brane equations of state.  For
near-extremal $p$-branes, the horizons are nearly degenerate.  In this
limit, $\zeta\!\rightarrow\!1$, the function $D_p(r)\rightarrow
H_p(r)$, and the only alteration of the metric due to nonextremality
is the presence of $K(r)$.  The relation between $\rH$ and the energy
density above extremality $\varepsilon$ is
\be\label{vareps}
\rH^{7-p} = \varepsilon G_{10} 8\pi^{\half(p-7)}\Gamma[\half(7-p)] \,.
\ee
The thermodynamic temperature and entropy are related to
$\varepsilon$, which in the near-extremal limit is much smaller than
the BPS D$p$-brane tension, as
\be
\TH \sim \varepsilon^{\half(5-p)/(7-p)} \,,\quad{\rm{and\ }}\quad
\SBH\sim \varepsilon^{\half(9-p)/(7-p)} \,.
\ee
For general $p$ these relations are not familiar from any field
theory.  Disagreement between free field theory and supergravity
entropies for these non-BPS systems is of course to be expected.
There is however one notable exception, the case $p=3$.  In that case,
a free massless gas gives entropy as a function of energy
$S(\varepsilon)\sim V (\varepsilon V)^{3/4}$.  Comparing this to the
supergravity equations here, we see that the scaling agrees, with
$\TH$ playing the role of the temperature $T$.  There is disagreement
in detail \cite{GKP}, which comes from ignoring interactions
\cite{joegary1}.

Other nonextremal branes, such as NS5, can be obtained from the above
D$p$-brane solutions by duality transformations.  We now move to
discussion of a general instability afflicting nonextremal branes and
black holes.

\subsection{The Gregory-Laflamme instability}

An important instability of nonextremal $p$-branes was discovered in
\cite{gregorylaflamme}.  The simplest example of this phenomenon,
which we now review briefly, occurs for neutral objects.  We start
with a neutral $(d-1)$-dimensional black hole.  It can come from a
neutral configuration in $d$ dimensions in (at least) two different
ways.

The first is from a black string, wrapped on compactified circle of
radius $R$; and the second is from an array of $d$-dimensional black
holes, spaced by a distance $2\pi R$.  These are shown in
Fig.\ref{stringvsarray}.  

\bfig
\hskip0.2\textwidth\epsfysize=1.25truein
\epsfbox{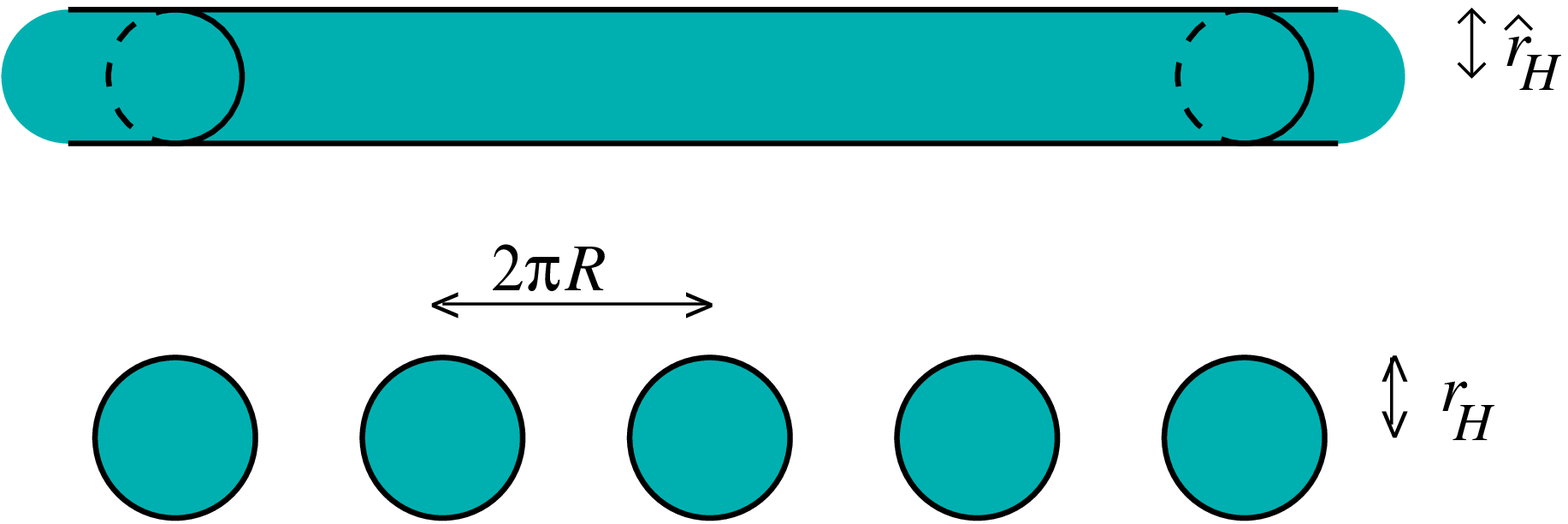}
\caption{\small{A black string versus an array of black holes.}}
\label{stringvsarray}
\efig

The array has to be infinite in order to get a static solution
\cite{robarray}.  It well approximates the metric of the
$(d-1)$-dimensional black hole of interest if the perpendicular
distances from the array are much larger than the spacing $2\pi R$.

The question is then to find out which of the above configurations
actually eventuates.  Let us work in the microcanonical ensemble,
which is appropriate for fixed energy (mass) of the system.  The basic
idea of the Gregory-Laflamme story is that whoever has the biggest
entropy wins.  The physics point is that the array of black holes has
a different entropy than black string, because entropy is proportional
to the area of the horizon, and spheres scale differently than
cylinders.  To see how it goes explicitly, let
\be
M_{\rm{array}} = M_{\rm{string}} \,.
\ee
For the black hole in $d$ dimensions, the properties of which we
showed in detail in subsection {\ref{solgeneg}} on
solution-generating, we have
\be
M \sim {{{\rH}^{d-3}}\over{G_d}} \,,\quad 
S \sim {{{\rH}^{d-2}}\over{G_d}} \,.
\ee
Therefore the mass per unit length of the array scales as
\be
{{M_{\rm{array}}}\over{R}}\sim{{{\rH}^{d-3}}\over{G_d}} \,,\qquad
{{M_{\rm{string}}}\over{R}}\sim{{{{\hat{r}}_H}^{d-4}}\over{G_{d-1}}}
\,,
\ee
and since the masses must be equal we obtain
\be
\rH^{d-3} \sim {\hat{r}}^{d-4}R \,.
\ee
Now we can find which configuration has biggest entropy:
\be
{{S_{\rm{array}}}\over{S_{\rm{string}}}} \sim {{\rH^{d-2}}\over{G_d}}
\, {{G_{d-1}}\over{{\hat{r}}_H^{d-3}}}
\sim \left({{R}\over{{\hat{r}}_H}}\right)^{1/(d-2)} 
\sim \left({{R}\over{\rH}}\right)^{1/(d-3)} \,.
\ee
So the array dominates for small horizon radii, and the black string
dominates for large horizon radii.

Sending $R\rightarrow\infty$, we see that the uncompactified neutral
black string is always unstable.  One can also see that this string is
unstable by doing perturbation theory; there is a tachyonic mode, as
shown in the original paper \cite{gregorylaflamme}.  

Note that the Gregory-Laflamme instability is different from the
Hawking radiation instability.  Let us now consider the possibility
that when a neutral black string falls apart into an array of black
holes, it violates the cosmic censorship hypothesis.  In order for the
cylindrically symmetric horizon of the string to break up into an
array of spherical horizons, the singularity inside the black string
horizon would have to go naked, at least for a while.  In
gravitational collapse, what may well happen instead is that the bits
and pieces will collapse into the configuration preferred by the
maximal entropy condition, obviating the need for temporary nakedness.
However, in situations where the radius $R$ of the compact dimension
varies dynamically in such a way that the string/array transition
boundary is crossed, it is difficult to argue that violation of cosmic
censorship does not occur.

The Gregory-Laflamme result does {\em not} imply instability of the
uncompactified BPS charged $p$-branes; there are several ways to see
this.  The first is that the tachyonic mode found for the neutral
systems disappears in the extremal case; the length scale of the
instability goes to infinity as the nonextremality parameter goes to
zero.  Another way to see it is that the BPS branes are protected by
the Bogomolnyi bound.  Consider what a BPS brane could break up into.
A D$p$-brane, for example, has a conserved charge, with $p$ even for
Type IIA and odd for Type IIB.  Therefore, if for example an
uncompactified D1-brane wanted to break up into an array of D0-branes
it would be out of luck because D0's and D1's do not occur in the same
theory.  If the D1 were wrapped on a circle, there would be a regime
($R\!<\!\ls$)in which we should more properly describe it in the
T-dual theory, i.e. as a D0.  In this case the configuration is still
stable, of course.

In our discussion of supergravity $p$-branes, for simplicity we
avoided those branes of dimension too large for them to be
asymptotically flat.  This was partly because they give rise to
infrared problems, via logarithmic and linear potentials.  We can
however make one remark here about domain walls in the context of the
Gregory-Laflamme instability.  Domain walls separating different vacua
of a theory will be stable even if they are neutral, because it would
cost an infinite amount of energy for them to break up.

In this section we have been concerned with the properties of
$p$-brane geometries as classical spacetimes.  More precisely, we were
interested in semiclassical properties, such as Hawking radiation.
Since the Hawking temperature is proportional to $\hbar$, the
radiation is turned off in the $\hbar{\rightarrow}0$ limit.  Also,
since as $\hbar{\rightarrow}0$ all entropies are strictly infinite,
one can argue that the Gregory-Laflamme instability is also absent in
the classical limit.  On the other hand, in the original paper
exhibiting the tachyonic instability, the analysis was in fact
classical.  But since the dynamics of the instability requires the
singularity to become naked while the horizon rearranges itself, the
classical approximation is hardly a self-consistent analysis.  It
would be very interesting to apply the excision techniques of
\cite{choptuik} in a numerical approach to understanding the
Gregory-Laflamme instability.

We now move away from classical spacetimes by asking where they let us
down.

\sect{When supergravity goes bad, and scaling limits}\label{sectfour}

The supergravity actions such as (\ref{iiaaction}) which we met in
section \ref{secttwo} describe low-energy approximations to string
theory.  As such, they are appropriate for situations where
corrections to the terms in them are small.  In string theory, there
are two expansion parameters which encode corrections to the
lowest-order (supergravity) actions, namely the sigma-model
loop-counting parameter $\alpha^\prime$ and the string loop-counting
parameter $\gs$.  Since $\alpha^\prime\equiv\ls^2$ is a dimensionful
parameter, we need to fold it in with \aeg a measure of spacetime
curvature in order to get a dimensionless measure of the strength of
sigma-model corrections.  The first corrections to the tree level IIA
action shown above occur \cite{mbg} at ${\cal{O}}(\ls^6)$; lower order
corrections are prevented by supersymmetry.  For the string loop
corrections in the supergravity arena, we need the dilaton field,
which typically varies in spacetime.  The measure of how badly string
loop corrections are needed is then $\gs e^\Phi$.

We now discuss how string theory handles the breakdown of classical
spacetime, in a few examples.

\subsection{The black hole correspondence principle}

The basic idea behind the Correspondence Principle is that stringy or
braney degrees of freedom take over when supergravity goes bad.

The first example analysed was that of the $d$-dimensional neutral
black hole, which carries only mass.  As discussed in subsection
\ref{solgeneg} on solution-generating, there is no dilaton so the 
Einstein and string metrics are the same,
\be
dS_d^2 = -\left[1-\left({{\rH}\over{r}}\right)^{d-3}\right]dt^2
+\left[1-\left({{\rH}\over{r}}\right)^{d-3}\right]^{-1}dr^2
+r^2d\Omega_{d-2}^2 \,,
\ee
where
\be
\rH^{d-3} = {{16\pi G_d M}\over{(d-2)\Omega_{d-2}}} 
\sim \gs^2\ls^{d-2} M \,,
\ee
Note that if we fix the mass $M$ and radius $r$ in units of $\ls$,
then the metric becomes flat as $\gs\rightarrow 0$.  (For simplicity
we taken the volume of any internal compact dimensions to be of order
the string scale.  The actual value does not affect the argument.)

The supergravity black hole solution breaks down in the sense of the
correspondence principle \cite{joegary1} when curvature invariants at
the horizon are of order the string scale.  The physical reason why we
concentrate on the horizon, rather than the singularity, is that its
presence is what signals the existence of a black hole.  Using the
horizon also gives rise to sensible answers which fit together in a
coherent fashion under duality maps.  A curvature invariant which is
nonzero for the neutral black hole is
$R^{\mu\nu\lambda\sigma}R_{\mu\nu\lambda\sigma}\!=
\!{{12}\over{\rH^4}}$, so that breakdown of supergravity occurs when
\be\label{corrpt}
\rH\sim\ls \,.
\ee
The thermodynamic temperature and entropy of the black hole scale as
\be
\TH = {{(d-3)}\over{4\pi{}\rH}} \,,\qquad
\SBH = {{\Omega_{d-2}\rH^{d-2}}\over{4G_d}}\,,
\ee
so the Hawking temperature at the correspondence point (\ref{corrpt})
is $\TH\sim1/\ls$.

The simplest string theory object which carries only the conserved
quantum number of mass is the closed fundamental string.  We will
therefore be interested in seeing if we have a fundamental string
description where the black hole description breaks down.  (One reason
why we choose the simplest object, rather than say a spherical
D2-brane, is Occam's razor.  It is also important that the
correspondence point occurs at $\rH\!\sim\!\ls$ which involves no
powers of $\gs$.)  In fact, the idea that black holes might be
fundamental strings dates back to the late '60's.  The idea was put on
a firmer footing by Sen \cite{senbhss} and Susskind \cite{lenny}
before the duality revolution.  The subsequent formulation of the
Correspondence Principle made those ideas more powerful.  One of the
ways it did this was to recognise that black holes and string states
typically do not have identical entropy for all values of parameters;
rather, the transition between black hole and string degrees of
freedom occurs at a transition point, known as the Correspondence
Point.  The existence of a correspondence point for every system
studied is a highly nontrivial fact about string theory and the
degrees of freedom that represent systems in it in different regions
in parameter space.

To progress further, we now need the statistical entropy of closed
string states due to the large degeneracy at high mass.  This is a
standard result in perturbative string theory so we will not review it
here but refer to the texts \cite{joebigbook,gswbook}.  We assume that
the string coupling is weak so that we can use the free spectrum
computation; this assumption will be justified {\em a posteriori}.

Using the relation between the oscillator number $N$ and the mass $m$,
$\ls^2m^2\sim N$, we have for the closed superstring degeneracy of
states at high mass,
\be
d_m \sim e^{m/m_0}\,,\qquad m_0\sim{{1}\over{\ls}}\,.
\ee
With better approximation schemes, one can keep track of power-law
prefactors that depend on the number of large dimensions.  We have
suppressed these because they are not important at large-$m$.

The quantity $m_0$ is the Hagedorn temperature.  At the Hagedorn
temperature, the canonical ensemble is in fact no longer well-defined.
This happens because the partition function diverges,
\be
Z = \int_0^{\infty} dm e^{m/m_0} e^{-m/T} \rightarrow \infty
\quad {\rm{above\ }} T=m_0 \,.
\ee
At the Hagedorn temperature, the excited string becomes very long and
floppy.  The Boltzmann entropy of the string state is the log of the
degeneracy of states,
\be
S_{\rm{string}} =\log(d_m) \sim {{m}\over{\ls}} \,.
\ee

Matching the masses at the correspondence point for general
Schwarzschild radius yields
\be
M \sim {{\rH^{d-3}}\over{\gs^2\ls^{d-2}}} \sim m \,.
\ee
This gives the general entropy ratio
\be
{{\SBH}\over{S_{\rm{string}}}} \sim
{{\rH^{d-2}}\over{\gs^2\ls^{d-2}}}\,
{{\gs^2\ls^{d-3}}\over{\rH^{d-3}}} \quad\sim {{\rH}\over{\ls}} \,.
\ee
We can see four pieces of physics from this formula.  Firstly, the
crossover from the black hole to string state indeed happens at
$\rH\sim\ls$, as suggested earlier.  Secondly, the black hole
dominates for $\rH\gg\ls$ \aie for large mass, while the string
dominates at lower mass.  Thirdly, let us calculate the string
coupling at the correspondence transition point.  Since the entropy at
correspondence is $S\sim m/m_s$, and $\ls{m}\sim\sqrt{N}$, we get
$S\sim\sqrt{N}$.  Also, we have the formula
$S\sim\ls^{d-2}/G_d{\sim}1/\gs^2$.  {}From this we find that
$\gs{\sim}N^{-\quarter}$ at transition.  This is indeed weak coupling
since $N$ is very large.  This justifies our earlier assumption that
we could calculate the string degeneracy by using weak-coupling
results.  Lastly, note that in general $d$, the mass at correspondence
is not the Planck mass $1/\ell_d$.

More work has been done on the physics of the transition between the
black hole and the string state.  The interested reader is referred to
\aeg \cite{joegary2,ramziet} and references
therein.

We have seen that the black hole and string state entropies match in a
scaling analysis at the correspondence point.  The physics
implications of the correspondence principle run even deeper, however.
The conservative direction to run the matching argument tells us that
a string state will collapse to a black hole when it gets heavy
enough.  The radical direction to run the argument is the other way:
the correspondence principle is in fact telling us that the endpoint
of Hawking radiation for a Schwarzschild black hole is a hot string.
The hot string will then subsequently decay by emitting radiation
until we are left with a bath of radiation.  An interesting fact about
this decay of a massive string state in perturbative string theories
is that the spectrum is thermal, when averaged over the degenerate
initial states \cite{amatirusso}.

Overall, we see that the picture of decay of a Schwarzschild black
hole in string theory is in tune with expectations that a truly
unified theory should not allow loss of quantum coherence.  

\subsection{NS-NS charges and correspondence}

The work of Sen \cite{senbhss} on comparing entropy of BPS black holes
and the corresponding string states predated the correspondence
principle, but the results can in fact be considered as additional
evidence for it.  

Black holes with two NS-NS charges in $4\leq d\leq 9$ dimensions can
be constructed using the solution-generating technique
\cite{peetelectric}.  Taking the BPS limit is straightforward, and the
Bekenstein-Hawking entropy is easily obtained.  One hiccough that occurs
is that the entropy of the classical BPS black holes is zero, because
the area of the horizon is zero.  However, as argued by Sen,
\cite{senbhss}, higher order corrections to the equations of motion
will modify this, and make the area of the horizon become of order
string scale rather than zero.  This results in a finite entropy,
which can be compared to the entropy of the stringy state because the
system is BPS and there is a nonrenormalisation theorem for the
degeneracy of states.

The next step is to identify which stringy state the black hole will
turn into at the correspondence point.  Consider the deviation of the
geometry from Minkowski spacetime, as we did for the neutral black
holes.  Corrections to the flat metric go like $\delta
G_{\mu\nu}\!\sim\!G_dM/r^{d-3}$, and as $\gs\!\rightarrow\!0$ this
scales to zero with the Newton constant.  (We have assumed that no
compactified directions scale to zero as a power of $\gs$.)  {}From
this, we can then guess that the black hole will turn into a
perturbative string state at the correspondence point.  In particular,
the BPS black holes correspond to states of the fundamental string
with both momentum and winding charge, wound around a circle.  The
degeneracy of states formula is well known and can be easily compared
to the Bekenstein-Hawking entropy of the black holes.  It is in
scaling agreement with the entropy coming from the statistical
degeneracy of states of the closed string with the same quantum
numbers \cite{senbhss,peetelectric}.

For the case of NS5-brane charge, the physics is more tricky.  The
reason is related to how deviations from the flat metric scale with
$\gs$ for the different branes.  Above, we saw how BPS black holes
carrying string-like charges turned into string states at the
correspondence point, which occurred at weak coupling.  An analogous
phenomenon is not possible for NS5-branes.  We can see this from
combining the scalings (\ref{theas}) in the Bogomolnyi bound $M \geq
a|Z|$ with the generic equation for the deviation from the flat
metric, $\delta G_{\mu\nu} \sim G M / r^n$ for some $n$ appropriate to
the brane.  Since Newton's constant scales as $\gs^2$ at fixed $\ls$,
any brane with an $a$ scaling with two or more negative powers of
$\gs$ will not approach the flat metric as the string coupling is
scaled to zero.  The F1 and D$p$ have $a\sim 1,1/\gs$ respectively,
but the NS5 has $a\sim 1/\gs^2$ and so it is out of luck.  We do not
have space here to discuss the physics of what replaces the
supergravity NS5-brane in regions of parameter space where the
supergravity solution goes bad, but the question has been investigated
in limits different to $\gs\rightarrow 0$; see \aeg
\cite{IMSY,oferreview}.

Quite generally, though, in order to apply the correspondence
principle, we must identify the microscopic degrees of freedom that
will take over from the supergravity description when it breaks down,
\aie where the curvature at the horizon or the dilaton gets too large.
There are two important criteria which these stringy/braney degrees of
freedom must fulfil: they must have same conserved quantum numbers,
and they must be localised near the would-be black hole at the
correspondence point.  The second condition is needed in order to
prevent counting of the wrong states, \aeg we would not count closed
strings far outside the would-be Schwarzschild radius in our original
example of the neutral black hole.

We now move to the case of R-R charged systems.

\subsection{Where BPS D$p$-branes go bad}

Let us begin our discussion of the case with one R-R charge with an
analysis of where the supergravity solutions break down.  The Ricci
scalar is nonzero in the D$p$-brane spacetimes, and we find
\be
R[G] = -\quarter(p^2-4p-17)\left(\partial_r H_p\right)^2
H_p^{-\halff}
\,.
\ee
Let us consider the behaviour of this as $r\rightarrow 0$.  Since the
harmonic function $H_p \sim r^{p-7}$ near $r=0$, we have
\be R[G] \rightarrow {\rm (const)} r^{\halff(7-p)} \left(
r^{p-8}\right)^2 \sim r^{\half(3-p)} \,.
\ee
This blows up for the big $p$-branes, \aie those with $p>3$.  In
addition, we know that the curvature is zero at infinity, and rises as
we come in from infinity.  Therefore the curvature is non-monotonic
for branes with $p<3$.  The information on the curvatures for $p\not
=3$ branes is summarised in Fig.\ref{pbcurv}.

\bfig
\bmp{0.5}
\epsfysize=1.25truein
\epsfbox{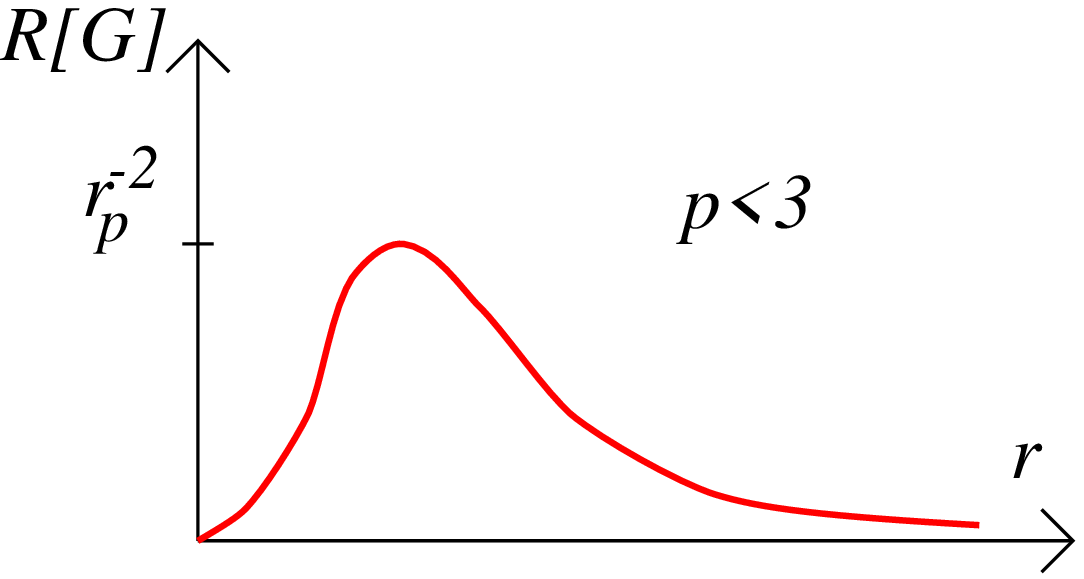}
\emp
\bmp{0.5}
\epsfysize=1.25truein
\epsfbox{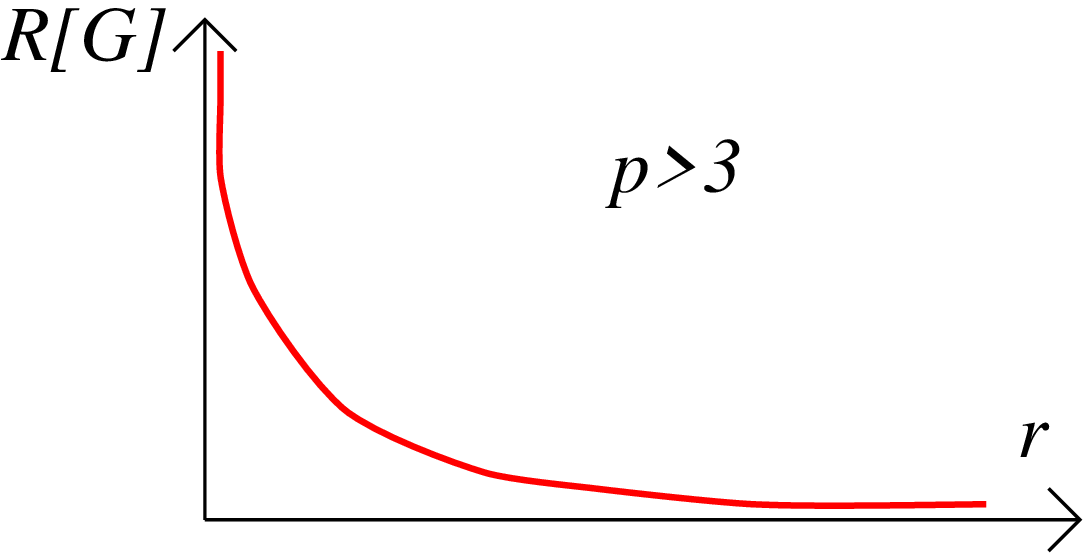}
\emp
\caption{\small{Curvature versus radial coordinate 
for D$p<3$- and D$p>3$-branes.}}
\label{pbcurv}
\efig

The dilaton behaves differently.  We have
\be e^\Phi = 
H_p^{\quarter(3-p)} \rightarrow {\rm (const)} 
r^{\quarter(7-p)(3-p)} \,.
\ee
This blows up at $r=0$ for the small branes, \aie for $p<3$.  
The slope for the dilaton is monotonic, but for $p>3$ there is an
inflection point.  We summarise this information in Fig.\ref{pbdil}.

\bfig
\bmp{0.5}
\epsfysize=1.25truein
\epsfbox{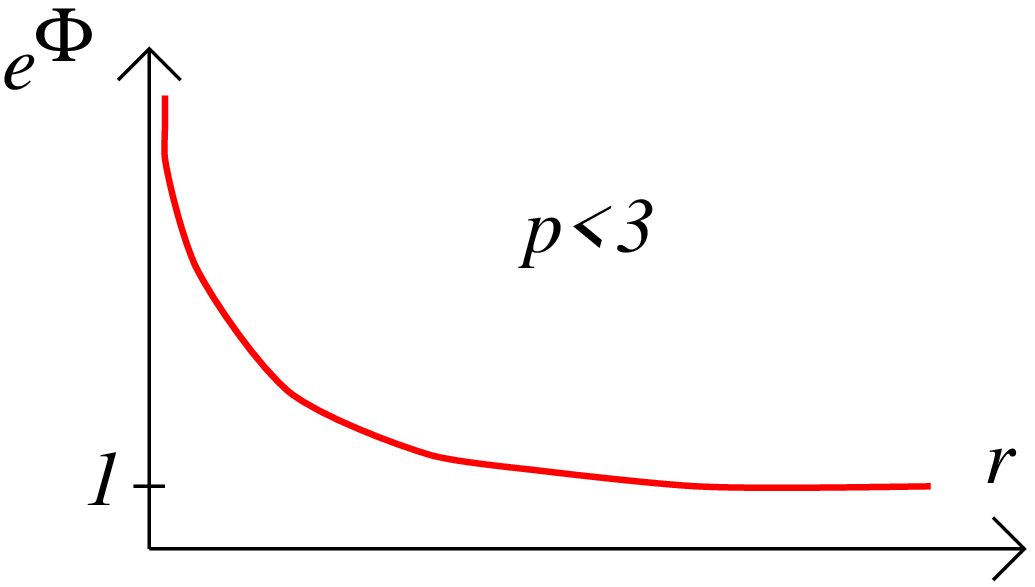}
\emp
\bmp{0.5}
\epsfysize=1.25truein
\epsfbox{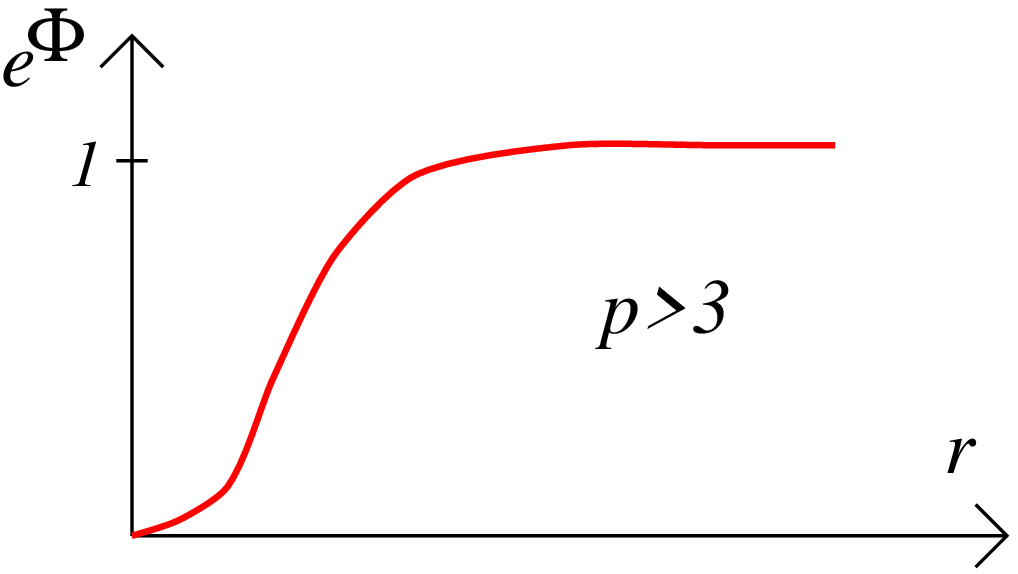}
\emp
\caption{\small{Dilaton versus radial coordinate 
for D$p<3$- and D$p>3$-branes.}}
\label{pbdil}
\efig

Note the interesting fact that, if the asymptotically flat part of the
geometry is removed by losing the constant piece (the 1) in the
harmonic function, then the behaviour of both the curvature and the
derivative of the dilaton becomes monotonic.  This turns out to be a
crucial supergravity fact in the context of the D$p$-brane
gravity/gauge correspondences of \cite{IMSY}.

In our brief discussion of the NS5-brane in the last subsection, we
saw that the D$p$-brane supergravity solutions do approach a flat
metric as $\gs\rightarrow 0$ at fixed $\ls$.  By following the
conserved quantum numbers, we therefore see that the weak-coupling
degrees of freedom are D$p$-branes in their {\em perturbative}
incarnation as hypersurfaces where fundamental strings end.

If we then compactify the D$p$-branes on $T^p$, we find R-R black
holes.  By the structure of Kaluza-Klein reduction formul\ae, we can
see that the resulting supergravity geometries blow up at $r=0$.  R-R
black holes in $d{\eql}4\ldots 10$ with one charge are of course partnered
with wrapped perturbative D-branes \cite{joegary1} .  

In this system, the energy above extremality $\Delta E$ can be carried
by either open or closed fundamental strings, as long as they are
close to the D-branes.  Open and closed strings have different
equations of state.  Again, we assume weak string coupling; this
assumption can be justified a posteriori.  For the open strings,
assuming a free massless gas yields
\be
\Delta E_{o} \sim N_p^2 V_p T^{p+1} \,,\qquad
S_{o} \sim N_p^2 V_p T^{p}\,,
\ee
while for the closed strings the equation of state is
\be
S_{c} \sim \ls\Delta E\,.
\ee
It is found that open strings dominate for near-extremal black holes,
while closed strings dominate for far-from-extremal black holes as
happened in our neutral black hole example.  In addition, the
correspondence points of the single-charge NS-NS and single-charge R-R
black holes are related by duality and they match up.  This is a
general phenomenon; also, in a highly nontrivial fashion it meshes
nicely with the Gregory-Laflamme transition
\cite{joegary1}.

In terms of advances in precise computations of black hole entropy,
the most important examples of the application of the correspondence
principle are systems with two or more R-R charges.  This is the case
both for the BPS and the near-BPS black holes.  The crucial physics
observation is that for these systems, the scaling works in such a way
that there is no correspondence point, and so exact comparisons can be
made to weak-coupling stringy/braney calculations for black holes of
any horizon radius.  We will discuss the spectacular success of these
microscopic calculations in later sections.

\subsection{Limits in parameter space, and singularities}

In figuring out what degrees of freedom replace a fundamental string
or D-brane supergravity geometry when it goes bad, we discussed the
limit $\gs\rightarrow 0$ of the system.  More generally, the idea of
taking limits of parameters, in the context of the correspondence
principle and otherwise, has yielded very powerful results.  These
results have taught us very interesting facts about gravity and about
gauge theories, including non-commutative gauge theories.

A limit of D$p$-brane systems which has been used to great effect is
the decoupling limit, in which interactions between the open strings
ending on the branes and the closed strings in the bulk are turned
off.  The resulting gravity/gauge correspondences are the domain of
other Lecturers at this School, but we cannot resist a few remarks
here.  The main physics behind the limit is to take string tension to
infinity, while holding some physically interesting parameters fixed.
It can be confusing to scale dimensionful quantities to zero, so we
work with dimensionless quantities here.  In units of a typical energy
$E$ of the system, taking the string tension to infinity is then
expressed as $\ls E\rightarrow 0$.  In order to retain a finite
$d{\eql}p{\pls}1$-dimensional gauge coupling on the branes, we hold
fixed $\gYM^2E^{p-3}{\equiv}(2\pi)^{p-2}\gs(E\ls)^{p-3}$.  Also held
fixed is the energy of open strings stretched between different
D-branes separated by distance $r$, i.e. $U/E{\equiv}r\ls^{-2}/E$.  In
the decoupling limit, by definition, the bulk theory and the brane
theory are each a unitary theory on their own.  Maldacena \cite{juan}
argued (initially for certain systems) that the two theories are
actually dual to one another.  This idea has been extended to many
other systems, in \aeg \cite{IMSY}, goes by the name of the
gravity/gauge correspondence, and is very powerful.

Assuming that the gravity/gauge correspondence (conjecture) is true
for all values of loop-counting parameters, then it provides an
explicit realization of information return.  It does this because any
process of black hole formation and evaporation in the supergravity
theory has a dual representation in the unitary quantum field theory.
It is, however, extremely difficult to see how information return
works in practice \cite{bglBDHM}, because the duality between the
gravitational theory and the gauge theory is a strong-weak duality.
The issue of how a semiclassical spacetime picture emerges from the
strongly coupled gauge theory, with (approximate) locality and
causality built in, is one of the most interesting and important
challenges of this field of study.

In the decoupling limit, the supergravity D$p$-brane geometry loses
its asymptotically flat part, as can be seen by plugging the above
scalings into the equation (\ref{eqnforHp}) for the harmonic function.
(Also, the worldvolume coordinates on the brane are the same as the
$x_\parallel$, which are the supergravity worldvolume coordinates in
the asymptotically flat region of the original D$p$-brane geometry.)
So, let us consider the near-horizon geometry of the $0\le
p<5$-branes.  Abbreviate
\be
r_p^{7-p} \equiv c_p \gs N_p \ls^{7-p} \,,
\ee
and look at the geometry for $r\ll r_p$, i.e. let us ignore the 1 in
the harmonic function for the D$p$-brane geometry.  Changing to a
coordinate
\be
z = {{2}\over{(5-p)}} {{r^{\half(p-5)}}\over{r_p^{\half(p-5)}}}
\ee
for the BPS systems yields the structure
\be
dS_{10}^2 \rightarrow {\rm (const)} z^{(3-p)/(7-p)}
\left[
{{-dt^2+dx_\parallel^2+dz^2}\over{z^2}} + {{(5-p)^2}\over{4}}
d\Omega_{8-p}^2 \right] \,,
\ee
which gives a geometry conformal to $AdS_{p+2}\times S^{8-p}$
\cite{adsxs}.  The z-dependent prefactor disappears only for $p=3$.
Since the asymptotically flat part of the geometry is gone in the
decoupling limit, the Penrose diagrams are drastically altered.

In addition, as we saw in our analysis of where D$p$-branes go bad,
the curvature and the dilaton behave monotonically with radius when
the 1 is missing from the harmonic function.  Combining the
supergravity and brane field theory information in the decoupling
limit leads to the construction of the phase diagram \cite{IMSY} for
the D$p$-brane system.  A nice discussion of phase diagrams in more
generality can be found in \cite{emilreview}.  More recent
considerations which include the physics of turning on a $B$-field
(noncommutativity), with emphasis on $d{\eql}1{\pls}1$, may be found
in \cite{igorjuanphase}.  For the non-BPS systems, the only deviation
from the BPS metric in the decoupling limit is the nonextremality
function (the $K$ function in eqn (\ref{DandKfns})) which multiplies
$G_{tt},G_{rr}^{-1}$.  One way to see that the $D$ function is
unmodified from the BPS case is to combine the decoupling limit
scalings with the equation for the energy density above extremality
$\varepsilon$, given in (\ref{vareps}), with the relation for the
boost parameter (\ref{eqnforbeta}).

To finish this section on where supergravity brane geometries go bad,
we now make a few remarks about classical curvature singularities.

In the discussion of the correspondence principle, for cases with
separate horizon and singularity, we used the curvature at the horizon
to determine where the supergravity solution broke down.  The question
of what happens at the singularity is also, of course, a question of
physical interest in string theory.  The general expectation might be
that string theory smoothes out regions of classically infinite
curvature.

However, Horowitz and Myers \cite{garyrob} made the important point
that some singularities are not of the kind that can be smoothed out
because this would give rise to a contradiction.  The prototypical
example is the negative-mass Schwarzschild geometry.  Since $M{<}0$,
the horizon is absent, so the singularity is naked.  If the
singularity were smoothed out by stringy phenomena, the resulting
finite-sized blob would be an allowed object with overall negative
mass.  It would then destabilise the vacuum - via pair production, for
example.  The upshot is that the negative-mass Schwarzschild geometry
is a figment of the classical physicist's imagination.

It is also important to note that the question of whether a geometry
is singular depends on the dimension of the supergravity theory it is
embedded in.  For example, in \cite{higherdres}, it was shown that
some lower-dimensional black holes with singularities could be lifted
to nonsingular solutions in higher dimensions.  For understanding
possible resolution of singularities in terms of basic stringy objects
like D-branes, the best dimension to do the singularity analysis is
$d{\eql}10$, which is the dimension in which D-branes naturally live.
It is generally more confusing to try to do the analysis directly in
lower dimensions.  In addition, one should be sure that any operation
one does in supergravity also makes sense in string theory.

There are spacetimes in string theory with singularities, such as the
fundamental string and the gravitational wave, which appear to be
exact solutions to all orders in $\alpha^\prime$.  In
\cite{arkadysing} it was, however, argued that forgotten source terms
in the action actually do lead to $\alpha^\prime$ corrections, which
smooth out these singularities.  For the string, we can in any case
think of the singularity of the classical geometry as smoothed out by
the source which is the fundamental string itself \cite{dabhar}.  In
addition, for the D$p{\not\!{=}}3$-branes, the phase diagrams of
\cite{IMSY} show that a gauge theory takes over in regions where the
classical geometry has a (null) singularity.  This provides an
understanding of singularity resolution in these systems, which
possess $\cN{\eql}4$ supersymmetry in $d{\eql}4$ language.

A more recently discovered phenomenon known as the enhan\c con
mechanism has provided a stringy resolution of some $\cN{\eql}2$
classical timelike naked singularities \cite{enhancon}.  The essential
physics behind this is that string theory knows what to do when
certain cycles on which D-branes are wrapped become small; previously
irrelevant degrees of freedom become light and enter the dynamics.
Put this way, the enhan\c con phenomenon may in fact be quite general;
work on more applications is in progress.

\sect{Making black holes with branes}\label{sectfive}

Black holes in string theory with macroscopically large entropy can
all be constructed out of various $p$-brane constituents.  We
concentrate in this section mostly on BPS systems where the rules are
simplest.

\subsection{Putting branes together}

Two clumps of parallel BPS $p$-branes are in static equilibrium with
each other.  In addition, BPS $p$-branes and $q$-branes for some
choices of $p,q$ can be in equilibrium with each other under certain
conditions.  One way to find many of the rules is to start with the
fundamental string intersecting a D$p$-brane at a point.

By T- and S-duality, we can infer the following $d{\eql}10$ NS5-, F1-,
and D$p$- brane intersections.  We use the convention that an $A$-type
object intersecting a $B$-type object in $k$ spatial dimensions is
represented by $A{\parallel}B(k)$ or $A{\perp}B(k)$, depending on
whether $A$ and $B$ are parallel or perpendicular to each other.  In
this notation, our fundamental string/D$p$-brane intersection is
denoted F1$\perp$D$p(0)$.  We then get via dualities
\be\ba{l}
\bs
{\rm{D}}m\parallel{\rm{D}}m{\pls}4(m)\,, m=0,1,2 
\quad \rightarrow 
{\rm{D}}p\perp{\rm{D}}q(m)\,,\quad p+q=4+2m \,; \cr
{\rm{F}}1\parallel{\rm{NS}}5\,,\qquad 
{\rm{NS}}5\perp{\rm{NS}}5(3)\,,\qquad 
Dp\perp{\rm{NS}}5(p-1) \,.
\ea\ee
For simplicity we have restricted to $p\leq 6$ $p$-branes whose
geometries are asymptotically flat.  (We have also only listed
pairwise intersections for the same reason; multi-brane intersections
must obey the pairwise rules for each pair.)  In $d{\eql}11$ the rules are
\be\ba{l}
\bs
{\rm{M}}2\perp{\rm{M}}2(0)\,, \quad
{\rm{M}}2\perp{\rm{M}}5(1)\,, \quad
{\rm{M}}5\perp{\rm{M}}5(1)  \ {\rm{or\ }}
{\rm{M}}5\perp{\rm{M}}5(3)\,;\cr
\bs
{\rm{W}}\parallel{\rm{M}}2\,, \quad
{\rm{W}}\parallel{\rm{M}}5\,, \quad
{\rm{M}}2\parallel{\rm{KK}} \ {\rm{or\ }} 
{\rm{M}}2\perp{\rm{KK}}(0)\,, \cr
\bs
{\rm{M}}5\parallel{\rm{KK}} \ {\rm{or\ }} 
{\rm{M}}5\perp{\rm{KK}}(1)\ {\rm{or\ }}
{\rm{M}}5\perp{\rm{KK}}(3)\,;\cr
{\rm{W}}\parallel{\rm{KK}}\,,
{\rm{KK}}\perp{\rm{KK}}(4,2)\,.
\ea\ee
This leads to a set of rules for putting W and KK on $d{\eql}10$ branes.
Recall that for KK, whose spacetime metric was displayed in
eqn.(\ref{kkeqn}), one of the four transverse directions is singled
out as the isometry direction while the metric depends on the other
three coordinates.  Because not all perpendicular directions are
equivalent, the KK intersection rules are rather involved; see for
example \cite{kkint}.

For some brane intersections not displayed above, there is an
additional complication which arises upon careful consideration of
force cancellation, via closed string tree or open string 1-loop
amplitudes.  The prototypical example is the case of a D0-brane and a
D8-brane.  When the D0-brane crosses the D8-brane, a fundamental
string is created; the physics requires this to happen for force
cancellation to be preserved.  A dual situation where this occurs is
in the Hanany-Witten setup where a D3-brane is created when a D5-brane
crosses an NS5-brane.  For a pedagogical discussion of this brane
creation story we refer the reader to \cite{igor9709}.

Another method which emphasises the supergravity aspect of the
intersection rules was explained in \cite{andyopen} and in the
mini-review of \cite{branesurgery}.  We now go over the latter
discussion briefly.

\subsection{Intersection-ology {\`{a}} la supergravity}

The simplest system to study is $d{\eql}11$ supergravity, and studying the
action for the theory gives rise to the intersection rules for
M-branes.  The action for the gauge potential $A_{\mathit 3}$ in the
bosonic sector is
\be
S[A_{\mathit 3}] = {{1}\over{16\pi G_{11}}} \int \left\{ -\left[
d^{11}x
\sqrt{-g}{{|F_{\mathit 4}|^2}\over{2(4!)}}\right] + 
\left[ \# F_{\mathit 4}\wedge F_{\mathit 4}\wedge A_{\mathit 3} 
\right] \right\}  \,.
\ee
The constant $\#$ can be changed by a field redefinition.  The field
strength $F_{\mathit 4}$ is defined as
\be
F_{\mathit 4}=dA_{\mathit 3} \,,
\ee
and so it obeys a Bianchi identity
\be
d F_{\mathit 4} = 0 \,. 
\ee
This implies that the charge
\be
Q_5 = \int_{S^4} F_{\mathit 4} 
\ee
is conserved, where the integral is over a transverse four-sphere.
This is the M5-brane charge.  The Bianchi identity also implies that
the M5-brane cannot end on anything else.  It can, however, have a
funny-shaped worldvolume pointing in different directions.

Then, with a convenient normalisation of \#, the field 
equation for $A_{\mathit 3}$ is
\be\ba{rl}
\bs
d{\,}^* F_{\mathit 4} =&\! - F_{\mathit 4}\wedge F_{\mathit 4} \cr
=&\! - \left(dA_{\mathit 3}\right)\wedge F_{\mathit 4} =
-d\left(A_{\mathit 3}\wedge F_{\mathit 4} \right) \,,
\ea\ee
where we used the Bianchi identity.  This tells us that the conserved
charge, this time the M2-brane charge, is
\be
{\hat{Q}}_2 = \int_{S^7} \left[ {\,}^* F_{\mathit 4} + A_{\mathit
3}\wedge F_{\mathit 4}
\right] \,.
\ee
Consider the M2-brane ending on something.  To picture this, suppress
one of its dimensions and also several of those of the object on which
the M2 ends.  In fact, the surface on which the M2 ends must be the
M5, because nothing else carries $A_{\mathit 3}$.  A diagram is shown
in Fig.\ref{m2intm5}.

\bfig
\hskip0.2\textwidth\epsfysize=1.25truein
\epsfbox{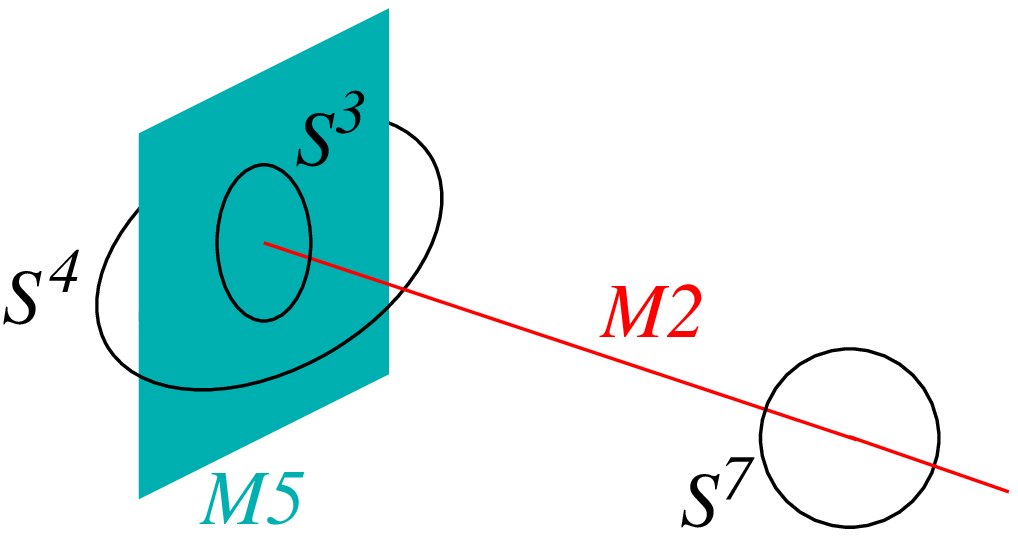}
\caption{\small{An M2-brane intersecting a M5-brane, with 
transverse spheres shown.}}
\label{m2intm5}
\efig

Far from the boundary, only the ${\,}*F_{\mathit 4}$ piece in the
charge ${\hat{Q}}_2$ matters, and so ${\hat{Q}}_2$ is indeed the
membrane charge.  On the other hand, right at (and only at) the place
where the M2 ends on the M5, we can deform $S^7\rightarrow S^4\times
S^3$.  In addition, the components of the field strength $F_{\mathit
4}$ parallel to the M5-brane are approximately zero there, because the
flux threads the $S^4$ in a spherically symmetric way.  As a
consequence, on the M5, one can write the approximate relation
$A_{\mathit 3}\simeq dV_{\mathit 2}$, for some two-form $V$.  Then the
charge factories into
\be\ba{rl}
\bs
{\hat{Q}}_2 &\! \simeq {\underbrace{ \int_{S^3} dV_{\mathit 2} }}
\quad {\underbrace{ \int_{S^4} F_{\mathit 4} }} \cr {\ } &\! {\rm
string\ charge} \qquad Q_5 \,.
\ea\ee
The first factor is the (magnetic) charge of the string which is the
boundary of the M2-brane in the M5-brane worldvolume.  This leads to
the rule M2$\perp$M5(1).

This procedure can be generalised to find other brane intersection
rules in other supergravity theories in various dimensions
\cite{branesurgery}.

\subsection{Making BPS black holes with the harmonic function rule}

BPS black holes in dimensions $d{\eql}4\ldots 9$ may be constructed
from BPS $p$-brane building blocks.  Typically, however, they have
zero horizon area and therefore non-macroscopic entropy.  The
essential reason behind this slightly annoying fact may be distilled
from the supergravity field equations \cite{juanthesis}.  The sizes
and shapes of internal manifolds, as well as the dilaton, turn out to
be controlled by scalar fields, and the horizon area is related to
these scalars.  But in any given dimension $d$, there are only a few
independent charges on a black hole, and mostly these give rise to too
few independent ratios to give all the scalar fields well-behaved vevs
everywhere in spacetime.  For stringy black holes made by
compactifying on tori, the only asymptotically flat BPS black holes
with macroscopic finite-area occur with 3 charges in $d{\eql}5$ and 4
charges in $d{\eql}4$ \cite{finitearea}.  For a survey of
supergravities in various dimensions and the kinds of black objects
that can carry various central charges, relevant to D-brane
comparisons, see \cite{juanferrara,mirjhull}.

A systematic ansatz \cite{hrule} is available for construction of
supergravity solutions corresponding to pairwise intersections of BPS
branes, which is known as the ``harmonic function rule''.  The ansatz
is that the metric factories into a product structure; one simply
``superposes'' the harmonic functions.  This ansatz works for both
parallel and perpendicular intersections, using the construction rules
we reviewed in the last subsection, with the restriction that the
harmonic functions can depend only on the overall transverse
coordinates.  In this way we get only smeared intersecting brane
solutions.  Let us discuss some examples.

We use a convention where $-$ indicates that the brane is extended in
a given dimension, $\cdot$ indicates that it is pointlike, and $\sim$
indicates that, although the brane is not extended in that direction
{\em a priori}, its dependence on those coordinates has been smeared
away.  As an example, consider a D5 with a (smeared) D1:
\be\ba{ccccccccccc}\label{d5sd1}
& 0& 1 & 2 & 3 & 4 & 5 & 6 & 7 & 8 & 9\cr
{\rm{D1}}
& - & - & \sim & \sim & \sim & \sim & \cdot & \cdot & \cdot & \cdot \cr
{\rm{D5}} 
& - & - & -    & -    & -    &  -   & \cdot & \cdot & \cdot & \cdot  
\ea\ee
and D2 perpendicular to D2' (both smeared):
\be\ba{ccccccccccc}
& 0& 1 & 2 & 3 & 4 & 5 & 6 & 7 & 8 & 9 \cr
{\rm{D2}}
& - & -    & -    & \sim & \sim & \cdot & \cdot & \cdot & 
     \cdot & \cdot \cr
{\rm{D2'}} 
& - & \sim & \sim & -    & -    & \cdot & \cdot & \cdot & 
     \cdot & \cdot 
\ea\ee

For the D1-D5 system, let us define
$r^2{\eql}x_\perp^2\!\equiv\!\sum_{i=1}^4(x^i)^2$ to be the overall
transverse coordinate in the setup above in eqn.(\ref{d5sd1}).  Then
the string frame metric is, using the harmonic function rule,
\be\ba{rl}
\bs
dS^2_{10} = & H_1(r)^{-\half}H_5(r)^{-\half}\left(-dt^2 + dx_1^2\right)
+ H_1(r)^{+\half}H_5(r)^{-\half} dx_{2\cdots 5}^2 \cr
& + H_1(r)^{+\half}H_5(r)^{+\half} \left(dr^2+r^2d\Omega_3^2\right)\,,
\ea\ee
and dilaton is
\be
e^\Phi =  H_1(r)^{+\half}H_5(r)^{-\half} \,,
\ee
while the gauge fields are as before,
\be
C_{01} =         \gs^{-1}\left[1-H_1(r)^{-1}\right] \,,\qquad
C_{01\ldots 5} = \gs^{-1}\left[1-H_5(r)^{-1}\right] \,.
\ee
The independent harmonic functions both go like $r^{-2}$ in the
interior, which is natural for a D5-brane and also for a D1-brane
smeared over four coordinates:
\be
H_5(r) = 1 + {{\#'}\over{r^2}} \,,\qquad
H_1(r) = 1 + {{\#}\over{r^2}} \,.
\ee
Notice that if we wrap $x^2\cdots x^5$ on $T^4$, in order to make a
$d{\eql}6$ black hole with two charges, the volume of the $T^4$ is
finite at the event horizon $r=0$:
\be
{{{\rm{Vol}}(T^4)}\over{(2\pi)^4V_4}}
= \sqrt{G_{22}\cdots G_{55}} =
\left({{H_1}\over{H_5}}\right)^{\quarter 4}
\rightarrow\left({{\#}\over{\#'}}\right) \,.
\ee
However, if we compactify the direction along the string, $x^1$, on a
circle, the radius goes to zero at the event horizon no matter how
large its value $R$ at infinity:
\be {{{\rm{Vol}}(S^1)}\over{(2\pi)R}} =
\sqrt{G_{11}}=(H_1H_5)^{-\half}\rightarrow r/\sqrt{\#\#'}\rightarrow 0
\,.
\ee
In addition, the area of the event horizon in Einstein frame, and
therefore the entropy, is zero.

It is interesting to note that not all known supergravity solutions
for intersecting branes are smeared or delocalised in this way.  The
factorised metric ansatz works in some other situations as well.  \aEg
let $H_1$ depend on $x_{6\cdots 9}\equiv x_\perp$, and on $x_{2\cdots
5}\equiv x_\parallel$, and let $H_5$ depend on $x_\perp$.  The
equations of motion are found to be, see \aeg \cite{intint},
\be
\partial_\perp^2 H_5(x) = 0 \,,\qquad
\left[ \partial_\perp^2 + H_5 \partial_\parallel^2 \right] 
H_1(x_\perp,x_\parallel) = 0 \,.
\ee
Therefore $H_5$ is as before in the smeared case, but $H_1$ has extra
dependence, on the coordinates $x_\parallel$ parallel to the D5-brane
but perpendicular to the D1-brane.  $H_1$ cannot be written in terms
of elementary functions but can be written as a
($x_\parallel$-)Fourier transform of known functions.  This is the
case even with transverse separations between the D1's and D5's.  More
generally, there is an interesting delocalisation phenomenon which
occurs as the transverse separation between a D$p$-brane and a
D$p{\pls}4$-brane to which it is parallel goes to zero.  Delocalisation
is found to occur only for $p\!<\!2$; an explanation of these
phenomena in the context of the AdS/CFT correspondence was found.
Some localised solutions are known analytically near the horizon of
the bigger brane, and for some intersecting brane systems the
factorised ansatz is not sufficient.  For a discussion of the above
issues see \cite{awps99}, and for recent advances in constructing
localised intersecting M5-brane solutions in $d{\eql}11$ see
\cite{faysmith}.

\subsection{The 3-charge $d{\eql}5$ black hole}

We saw in the previous subsection that a black hole with only D1- and
D5-brane charges does not have a finite horizon area.  We can now use
our knowledge from solution-generating to puff up this horizon to a
macroscopic size by using a boost in the longitudinal direction $x_9$.

The ingredients for building this black hole are then the previous
branes with the addition of a gravitational wave W:
\be\ba{ccccccccccc}
& 0& 1 & 2 & 3 & 4 & 5 & 6 & 7 & 8 & 9\cr
{\rm{D1}} 
&-& - & \sim & \sim & \sim & \sim & \cdot & \cdot & \cdot & \cdot \cr
{\rm{D5}} 
&-& - & - & - & - & - & \cdot & \cdot & \cdot & \cdot \cr
{\rm{W}}  
&-& \rightarrow & \sim & \sim & \sim & \sim & \cdot & \cdot & \cdot &
\cdot
\ea\ee
The $\rightarrow$ indicates the direction in which the gravitational
wave W moves (at the speed of light).

The BPS metric for this system is obtained from the simpler metric for
the plain D1-D5 system by boosting and taking the extremal limit.  To
get rid of five dimensions to make a $d{\eql}5$ black hole, we then
compactify the D5-brane on a $T^4$ of volume $(2\pi)^4V$, and then the
D1 and the remaining extended dimension of the D5-brane on a $S^1$ of
radius $R$.  The $d{\eql}5$ Einstein frame metric becomes
\be\ba{rl}
\bs
ds_5^2 = &\! 
-\left(H_1(r) H_5(r) \left(1+K(r)\right)\right)^{-2/3} dt^2 
\cr &\!+ \left(H_1(r) H_5(r) \left(1+K(r)\right)\right)^{1/3} 
\left[ dr^2 + r^2 d\Omega_3^2 \right] \,,
\ea\ee
where the harmonic functions are
\be
H_1(r) = 1 + {{r_1^2}\over{r^2}} \,,\qquad
H_5(r) = 1 + {{r_5^2}\over{r^2}} \,,\qquad
K(r) = {{r_m^2}\over{r^2}} \,,
\ee
and using arraying for $H_1$ and $K$ we find
\be\label{gravrads}
r_1^2 = {\frac{\gs N_1\ls^6}{V}} \,, \qquad 
r_5^2 = \gs N_5 \ls^2 \, \qquad
r_m^2 = {\frac{\gs^2 N_m \ls^8}{R^2 V}} \,.
\ee
This supergravity solution has limits to its validity.  If the stringy
$\alpha^\prime$ corrections to geometry are to be small, we need the
curvature invariants small.  Supposing that we keep the volumes $V,R$
fixed in string units, this forces the radius parameters to be large
in string units, $r_{1,5,m}\!\gg\!\ls$.  We can also control string
loop corrections if $\gs\!\ll\!1$.  These two conditions are
compatible if we have large numbers of branes and large momentum
number for the gravitational wave W.  Note from the relations
(\ref{gravrads}) that for all $N$'s of the same order hierarchically,
$N_1\!\sim\!  N_5\!\sim\! N_m$, while for $V/\ls^4\!\sim\! 1$,
$R/\ls\!\geq\!1$ and $\gs$ small, $r_{1,5} \gg r_m$.  On the other
hand, if we want $r_{1,5,m}$ of the same order, $N_m$ must be
hierarchically large: $N_m\!\gg\!N_{1,5}$.

The next properties of this spacetime to compute are the thermodynamic
quantities.  The BPS black hole is extremal and it has $\TH=0$.  For
the Bekenstein-Hawking entropy,
\be\ba{rl}
\bs
\SBH &\!= {\displaystyle{
 {\frac{A}{4 G_5}} \quad = {{1}\over{4G_5}}
\pi^2r^3\left[H_1(r) H_5(r) \left(1+K(r)\right) \right]^{3/6} }}
\ \ {\rm{at\ }} r=0 \,\cr
\bs
=&\! {\displaystyle{
 {{\pi^2}\over{4\left[\eighth\pi/8\gs^2\ls^8/(VR)\right]}}
  \left(r_1 r_5 r_m\right)^{\half} 
= {{2\pi{}VR}\over{\gs^2\ls^8}} 
\left(
{\frac{\gs{}N_1\ls^6}{V}}\,\gs{}N_5\ls^2\, 
{\frac{\gs^2{}N_m\ls^8}{R^2V}} \right)^{\half} }} \cr
= &\!  2\pi\sqrt{N_1 N_5 N_m} \,.
\ea\ee
This entropy is macroscopically large.  Notice that it is also
independent of $R$ and of $V$.  This is to be contrasted with the ADM
mass
\be
M = {{N_m}\over{R}} + {{N_1R}\over{\gs\ls^2}}
+ {{N_5RV}\over{\gs\ls^6}}\,,
\ee
which depends on $R,V$ explicitly.

For the entropy of the black hole just constructed out of D1 D5 and W,
we had $\SBH = 2\pi\sqrt{N_1N_5N_m}$.  More generally, for a more
general black hole solution of the maximal supergravity arising from
compactifying Type II on $T^5$, it is
\be
\SBH = 2\pi\sqrt{{\Delta}\over{48}} \,,
\ee
where the quantity $\Delta$ in the surd is the cubic invariant of the
$E_{6,6}$ duality group,
\be
\Delta = 2 \sum_{i=1}^{4} \lambda_i^3 \,,
\ee
and $\lambda_i$ are the eigenvalues of the central charge matrix $Z$.

A few years ago the claim was made, via classical topological
arguments in Euclidean spacetime signature, that all extremal black
holes have zero entropy.  This result is not trustworthy in the
context of string theory.  For starters, as we mentioned in our
discussion of the Third Law, there is no physical reason why
zero-temperature black holes should have zero entropy.  In any case,
the faulty nature of the classical reasoning in the string theory
context was pointed out in \cite{garyreviews}.  In the Euclidean
geometry, for any periodicity in Euclidean time $\beta$ at infinity,
the presence of the extremal horizon results in a redshift which
forces that periodicity to be substringy very close to the horizon.
Since light strings wound around this tiny circle can condense, a
Hagedorn transition can occur and invalidate the classical
approximation there.  In fact, other Hagedorn-type transitions can
come into play when spatial circles get small near a horizon, as they
do \aeg for $p$-branes compactified on tori \cite{koganetal}.

\subsection{The 4-charge $d{\eql}4$ black hole}

The extremal Reissner-Nordstr{\"{o}}m black hole can be embedded in string
theory using D-branes.  Recall that in the extremal spacetime metric
(\ref{rniso}) we had $H^2(r)$'s appearing in the metric.  This is to
be contrasted with the $H^{\half}$'s to be found in a generic
$p$-brane metric.  {}From this we can guess (correctly) that, in order
to embed the extremal RN black hole in string theory, we will need 4
independent brane constituents.  Restrictions must be obeyed, however,
in order for that black hole to be RN.  To make more general $d{\eql}4$
black holes with four independent charges, we simply lift these
restrictions and allow the charges to be anything - so long as they
are large enough to permit a supergravity description.

For making the $d{\eql}4$ black hole, one set of ingredients would be
\be\ba{ccccccccccc}
& 0& 1 & 2 & 3 & 4 & 5 & 6 & 7 & 8 & 9\cr
{\rm{D2}} 
&-& - & - & \sim & \sim & \sim & \sim & \cdot & \cdot & \cdot \cr
{\rm{D6}} 
&-& - & - & - & - & - & - & \cdot & \cdot & \cdot \cr
{\rm{NS5}}
&-& - & - & - & - & - & \sim & \cdot & \cdot & \cdot \cr
{\rm{W}}  
&-& \rightarrow & \sim & \sim & \sim & \sim & \sim & \cdot & \cdot &
\cdot
\ea\ee
By U-duality, we could consider instead 4 mutually orthogonal
D3-branes, or indeed many other more complicated arrangements
\cite{diamond2}.

In ten dimensions we can construct the BPS solution by using the
harmonic function rule.  So far we have not exhibited the metric for
the NS5-branes but that can be easily obtained using the D5 metric and
using the fact that the Einstein metric (\ref{einsteinstring}) is
invariant under S-duality.  We then have
\be\ba{rl}
\bs
dS^2_{10} = &
H_2(r)^{-\half}H_6(r)^{-\half}
\left[-dt^2+dx_1^2+K(r)(dt+dx_1)^2\right]\cr
\bs
&+ H_5(r)H_2(r)^{-\half}H_6(r)^{-\half}(dx_2^2)  \cr \bs
{} & {\displaystyle{
+H_2(r)^{+\half}H_6(r)^{-\half}H_5(r)(dx_{3\cdots 6}^2) }}\cr
& + {\displaystyle{
H_5(r)H_2(r)^{+\half}H_6(r)^{+\half}(dr^2+r^2d\Omega_2^2) }}\,,
\ea\ee
and
\be e^{\Phi} = H_5^{+\half}H_2^{+\quarter}H_6^{-\quarter(3)}\,.  \ee
After arraying till we are blue in the face, and finding Newton's
constant using
\be
G_4 = {{G_{10}}\over{(2\pi)^6(VR_aR_b)}} =
{{\gs^2\ls^8}\over{8VR_aR_b}} \,,
\ee
we get for the gravitational radii
\be
r_2={{\gs N_2\ls^5}\over{2V}} \,,\ 
r_6={{\gs N_6\ls}\over{2}} \,,\ 
r_5={{N_5\ls^2}\over{2R_b}} \,,\ 
r_m={{\gs^2N_m\ls^8}\over{2VR_a^2R_b}} \,.
\ee
We now use our Kaluza-Klein reduction formul\ae\, to reduce to the
$d{\eql}5$ black string, 
\be\ba{rl}
\bs
dS^2_{5} &\!= 
H_2(r)^{-\half}H_6(r)^{-\half}\left[-dt^2+dx_1^2+K(r)(dt+dx_1)^2\right] \cr
&\!+H_5(r)H_2(r)^{+\half}H_6(r)^{+\half}(dr^2+r^2d\Omega_2^2) \,.
\ea\ee
In this process,  the dilaton gets some factors:
\be
e^{2\Phi_5} = e^{2\Phi_{10}}
{{1}\over{\sqrt{G_{44}\cdots G_{88}}}} =
H_5^{+\half}H_2^{-\quarter}H_6^{-\quarter} \,.
\ee
Using our KK formula ${\hat{G}}_{00}=G_{00}-G_{01}^2/G_{11}$, we
find upon reducing on the last direction
\be\ba{rl}
\bs
dS_4^2 = &\! 
- H_2(r)^{-\half} H_6(r)^{-\half}\left(1+K(r)\right)^{-1} dt^2
\cr 
& + H_2(r)^{+\half} H_6(r)^{+\half}H_5(r) \,(dr^2+r^2d\Omega_2^2)  \,.
\ea\ee
The dilaton gets changed again:
\be
e^{2\Phi_4} =
{{H_5^{+\half}H_2^{-\quarter}H_6^{-\quarter}}\over
{\sqrt{(1+K(r))H_2(r)^{-\half}H_6(r)^{-\half}}}} = 
{{H_5^{\half}}\over{1+K(r)}} \,.
\ee
Finally the Einstein metric in $d{\eql}4$ is
\be\ba{rl}
\bs
ds^2 &\!= -dt^2 \left[{\sqrt{(1+K(r))H_2(r)H_6(r)H_5(r)}}\right]^{-1} 
\cr &\! + (dr^2+r^2d\Omega^2_2)
\left[{\sqrt{(1+K(r))H_2(r)H_6(r)H_5(r)}}\right] \,.
\ea\ee
The Bekenstein-Hawking entropy  is then easily read off to be
\be
\SBH = 2\pi\sqrt{N_2N_6N_5N_m} \,.
\ee
More generally, in the surd is the quantity $\diamondsuit/256$, where
$\diamondsuit$ is the quartic invariant of $E_{7,7}$ \cite{diamond1},
\be
\diamondsuit = \sum_{i=1}^{4}\left|\lambda_i\right|^2 -
2\sum_{i<j}^{4} \left|\lambda_i\right|^2\left|\lambda_j\right|^2
+ 4\left( {\overline{\lambda_1}}{\overline{\lambda_2}}
{\overline{\lambda_3}}{\overline{\lambda_4}} + 
\lambda_1\lambda_2\lambda_3\lambda_4\right) \,,
\ee
where $\lambda_i$ are the (complex) eigenvalues of $Z$; see \aeg
\cite{mirjhull,diamond2}.  More recent further progress on
entropy-counting for these black holes may be found in
\cite{diamond3}.

The connection to the $d{\eql}4$ Reissner-Nordstr{\"{o}}m black hole
is obtained by setting all four gravitational radii to be identical:
$r_2{\eql}r_6{\eql}r_5{\eql}r_m$.

Although we have not discussed nonextremality explicitly here, it can
be achieved by adding extra energy to the system of branes.  Generic
nonextremal branes cannot be in static equilibrium with each other, as
they typically want to fall towards each other, and they do not
satisfy the simple harmonic function superposition rule.  The least
confusing way to construct nonextremal multi-charge solutions is to
start with the appropriate higher-$d$ neutral Schwarzschild or Kerr
type solution, and to use multiple boostings and duality
transformations to generate the required charges.

\sect{BPS systems and entropy agreement}\label{sectsix}

In this section we review the D-brane computation of the entropy of
BPS systems with macroscopic entropy, with its many facets.  For BPS
systems there is a theorem protecting the degeneracy of states, and so
the entropy computed in different pictures will agree.

\subsection{The Strominger-Vafa entropy matching: $d{\eql}5$}

Since we have already built the black holes with the relevant D1 and
D5 charges, and worked out their macroscopic Bekenstein-Hawking
entropy, we turn to the microscopic computations of the entropy from
the string theory point of view.  We will discuss the D-brane method
of \cite{stromvafa} (earlier ideas for a microscopic accounting for
$\SBH$ of BPS black holes \cite{mirj1} with macroscopic entropy
included \cite{mirj2}).  A more detailed review of some aspects of the
D1-D5 system can be found in the recent lecture notes of
\cite{spenta}.

Our setup of branes was
\be\ba{ccccccccccc}
& 0& 1 & 2 & 3 & 4 & 5 & 6 & 7 & 8 & 9\cr
{\rm{D1}} 
&-& - & \sim & \sim & \sim & \sim & \cdot & \cdot & \cdot & \cdot \cr
{\rm{D5}} 
&-& - & - & - & - & - & \cdot & \cdot & \cdot & \cdot \cr
{\rm{W}}  
&-& \rightarrow & \sim & \sim & \sim & \sim & \cdot & \cdot & \cdot &
\cdot
\ea\ee
This system preserves 4 real supercharges, or $\cN{\eql}1$ in
$d{\eql}5$.  This can be seen from the constituent brane SUSY
conditions; each constituent breaks half of SUSYs.  It is necessary
for SUSY to orient the branes in a relatively supersymmetric way; if
this is not done, \aeg if an orientation is reversed, the D-brane
system corresponds to a black hole that is extremal but has no SUSY.

By using D1- and D5-brane ingredients we have two kinds of quantum
number so far, $N_1$ and $N_5$.  The degrees of freedom carrying the
remaining momentum number, and the angular momentum, are as yet
unidentified.  Now, the smeared D1-branes plus D5-branes have a
symmetry group
$SO(1,1)\!\times\!SO(4)_\parallel\!\times\!SO(4)_\perp$.  This
symmetry forbids the (rigid) branes from carrying linear or angular
momentum, and so we need something else.  The obvious modes in the
system to try are the massless 1-1, 5-5 and 1-5 strings, which come in
both bosonic and fermionic varieties.  The momentum $N_m/R$ is indeed
carried by the bosonic and fermionic strings, in units of $1/R$.  The
angular momentum is carried only by the fermionic strings,
$\half\hbar$ each.  Both the linear and the angular momenta can be
built up to macroscopic levels.

The next step is to identify the degeneracy of states of this system.
The simplification made by \cite{stromvafa} is to choose the
four-volume to be small by comparison to the radius of the circle,
\be
V^{\quarter}\ll R \,,
\ee
so that the theory on the D-branes is a $d{\eql}1{\pls}1$ theory.  This
theory has $(4,4)$ SUSY.  

Because the D1-branes are instantons in the D5-brane theory, the
low-energy theory of interest is in fact a $\sigma$-model on the
moduli space of instantons ${\cal{M}}=S^{N_1N_5}(T^4)$.  The central
charge of this $d{\eql}1{\pls}1$ theory is $c{\eql}n_{\rm
bose}{\pls}{\half}n_{\rm fermi}{\eql}6N_1N_5$.  Roughly, this central
charge $c$ can be thought of as coming from having $N_1N_5$ 1-5
strings that can move in the 4 directions of the torus.
Alternatively, $c$ can be thought of as roughly coming from having
$N_1$ instantons in the $U(N_5)$ gauge theory, and $N_5$ orientations
to point them in.

The other ingredient needed to compute the degeneracy of states, apart
from the central charge, is the energy.  Now, since the system is
supersymmetric, we have to put the right-movers in their groundstates.
The left-movers, however, can be highly excited.  Since the excited
states are BPS in $1{\pls}1$ dimensions, their energy and momentum must
be related by $E{\eql}N_m/R$.

The partition function of this system is the partition function for
$n_b{\eql}4N_1N_5$ bosons and an equal number of fermions
\be
Z = \left[ \prod_{N_m=1}^\infty {\frac{1+w^{N_m}}{1-w^{N_m}}}
\right]^{4N_1N_5} \equiv \sum \Omega(N_m) w^{N_m} \,,
\ee
where $\Omega(N_m)$ is the degeneracy of states at $d{\eql}1{\pls}1$
energy $E{\eql}N_m/R$.   At large-$N_m$ we can use the Cardy formula
\be
\Omega(N_m)\!\sim\!\exp\sqrt{{\pi\,c\,E\,(2\pi{}R)}\over{3}}
= \exp\left(2\pi\sqrt{{{c}\over{6}}\,ER}\right) \,.
\ee
This formula assumes that the lowest eigenvalue of the energy operator
is zero, as it is in our system.  (Otherwise we would need to subtract
$24\Delta_0$ from $c$ to get the effective central charge, where
$\Delta_0$ is the ground state energy).

Therefore the microscopic D-brane statistical entropy is
\be
S_{\rm{micro}} = \log\left(\Omega(N_m)\right) = 2\pi\sqrt{N_1N_5N_m}
\,.
\ee
This agrees exactly with the black hole result.  

Subleading contributions to both the semiclassical Bekenstein-Hawking
black hole entropy and to the stringy D-brane degeneracy of states
have been calculated, both on the black hole side and on the D-brane
side, and they have been found to match.  See for example the
beautiful work of \cite{beautifulcorrections}.  At this point we
mention that there is another method using M-theory available for
counting the entropy of these black holes, as discussed in
\cite{mmethod}, which we do not have space to cover here.

\subsection{Rotation}

In $d{\eql}5$ there are two independent angular momentum parameters,
because the rotation group transverse to the D1's and D5's splits up
as $SO(4)_\perp\simeq SU(2)\otimes SU(2)$.  The metrics for general
rotating black holes are algebraically rather messy and we will not
write them here.  We will simply quote the result for the BPS entropy
\cite{BMPV}:
\be
\SBH = 2\pi\sqrt{N_1N_5N_m-J^2}\,.
\ee
The BPS black holes have a nonextremal generalisation, in which the
two angular momenta are independent.  However, in the extremal limit
something interesting happens: the two angular momenta are forced to
be equal and opposite, $J_\phi{\eql}-J_\psi\!\equiv\!J$.  There is also
a bound on the angular momentum,
\be
|J_{\rm{max}}|=\sqrt{N_1N_5N_m}\,.
\ee
Beyond $J_{\rm max}$, closed timelike curves develop, and the entropy
walks off into the complex plane.  Another notable feature of this
black hole is that the funny cross-terms in the R-R sector of the
supergravity Lagrangian like (\ref{iiaaction}) are turned on; this
black hole is {\em not} a solution of $d{\eql}5$ Einstein-Maxwell
theory.  The charges are, however, unmodified by the funny cross-terms
which fall off too quickly at infinity to contribute.

Let us now move to the D-brane field theory.  It is hyperK{\"{a}}hler
due to $(4,4)$ supersymmetry, so let us break up the $\cN{\eql}4$ into
left- and right-moving $\cN{\eql}2$ superconformal algebras, each of
which has a $U(1)$ subgroup.  The corresponding charges $F_{L,R}$ can
be identified \cite{BMPV} as:
\be\label{jpm}
J_{\phi,\psi} = {{1}\over{2}}\left(F_L \pm F_R\right) \,.
\ee
Recall that the BPS system is in the R-moving groundstate, and at
left-moving energy $N_m/R$.  These facts give rise to a bound on
$F_{L,R}$ w.r.t. $L_0, {\overline{L_0}}$.  The essential physics
behind this is simply that in order to build angular momenta we have
to spend oscillators, and for a fixed energy our funds are limited.
For the details let us bosonise the $U(1)$ currents:
\be
J_L = \sqrt{\hat{c}} \partial \xi\,,
\ee
where ${\hat{c}}$ is the complex dimension of ${\cal{M}}$, i.e.
${\hat{c}}=2N_1N_5$. Then a state with charge $F_L$ is represented by
an operator
\be
{\cal{O}} = \exp\left({{iF_L\xi}\over{\sqrt{\hat{c}}}}\right)\Xi\,,
\ee
where $\Xi$ an operator from the rest of the CFT.  The construction is
entirely similar for $F_R$.  Now, the operator ${\cal{O}}$ has
positive dimension overall, so we get a bound on the $U(1)$ charges
\be
L_0 \geq {{F_L^2}\over{2{\hat{c}}}} \quad {\rm{and}}\quad
{\overline{L_0}} \geq {{F_R^2}\over{2{\hat{c}}}} \,.
\ee
Since the R-movers are in their groundstate, $F_R$ is small and fixed.
However, $F_L$ can be macroscopically large.  In the supergravity
description we are only sensitive to macroscopic quantities, and so we
will be unable to `see' $F_R$, only $F_L$.  Since the angular momenta
are the sum and difference of $F_R$ and $F_L$ respectively, as in
(\ref{jpm}), we find agreement with the black hole result in the BPS
limit: $J_\phi=-J_\psi$.

The next item on the agenda is to compute the effect of the angular
momenta on the D-brane degeneracy of states.  We had that the total
eigenvalue of $L_0$ is $N_m$.  However, we spent some of this,
$F_L^2/(2{\hat{c}})$, on angular momenta.  So the available $L_0$ for
making degeneracy of states is
\be
L_0(\Xi) = N_m - {{F_L^2}\over{4N_1N_5}} \equiv
{\tilde{N}}\equiv{\tilde{E}}R\,.
\ee
Notice that for small $F_L$ we have an excitation energy gap
\be
\omega_{\rm{gap}} \sim {{1}\over{N_1N_5R}}\,.
\ee
Then the microscopic entropy is 
\be\ba{rl}
\bs
S_{\rm{micro}} = &\! {\displaystyle{
\log(d_n) = 
2\pi\sqrt{{\tilde{N}}{{\hat{c}}\over{2}}}
= 2\pi\sqrt{\left(N_m-{{F_L^2}\over{4N_1N_5}}\right)N_1N_5} }} \cr 
& = 2\pi\sqrt{N_1N_5N_m - J^2} \,.
\ea\ee
This again agrees explicitly with the black hole calculation.

\subsection{Fractionation}

An important subtlety arises in the use of the exponential
approximation to the degeneracy of states formula.  The approximation
is valid only for energies $E$ such that $\Omega(E)$ is large; this
turns out to be true only for $N_m\!\gg\!N_{1,5}$.  We may ask what goes
wrong if all $N_i$ are of the same order.

The simplest way to see the approximation break down physically is to
picture \cite{lennyjuan} the left-movers as a $d{\eql}1{\pls}1$ gas of
1-5 strings, with order $N_1N_5$ massless species of average energy
$N_m/R$.  Let us introduce a temperature $T_L$ for the left-movers.
Note that doing this does not screw up supersymmetry of the system,
because the BPS condition is a condition on right-movers.  The BPS
system simply has zero right-moving temperature.  It is legitimate to
have different temperatures for left and right movers because there is
a net momentum.

Assuming extensivity gives
\be
E \sim (R) (N_1 N_5) T_L^2  = {{N_m}\over{R}}  \,,
\ee
and entropy
\be
S \sim (R) (N_1 N_5) T_L \,.
\ee
Eliminating $T_L$ between these relations gives
\be
S \sim (N_1 N_5 N_m)^{\half} \,,
\ee
as required.  However, substituting back to find $T_L$ we
find
\be
{{1}\over{T_L}} \sim R \left({{N_1N_5}\over{N_m}}\right)^{\half}
\gg R  \quad {\rm{since\ }} N_{1,5}\sim{}N_m   \,,
\ee
\aie  the inverse temperature is longer than the wavelength of a
typical quantum in a box of size $R$.  So our gas is too cold for
thermodynamics to be applicable, and we cannot trust our equations.

In fact, if the three $N_i$ are of the same order, then the strings
``fractionate'' \cite{lennyjuan}.  A recent analysis \cite{juanhirosi}
of the physics of this system in a CFT approach has yielded a rigorous
explanation of fractionation.  As an example of the basic idea,
consider $N_5=1$; what happens is that the D1's join up to make a long
string, as shown in Fig.\ref{shortvslong}.

\bfig
\hskip0.2\textwidth\epsfysize=0.5truein
\epsfbox{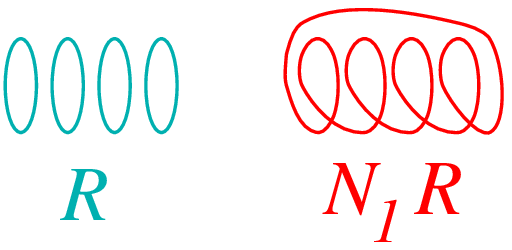}
\caption{\small{$N$ short strings versus  a single string $N$ times
as long.}}
\label{shortvslong}
\efig
Then the energy gap, instead of being $1/R$, is $1/(RN_1N_5)$, which
is much smaller.  As a consequence, there are now plenty of low-energy
states.  Notice that this is the same gap as we saw above in our study
of rotation.  With the smaller gap, we have just one species instead
of $N_1N_5$ species, and the energy is
\be
E \sim (R N_1 N_5) (1) T_L^2  = {{N_m}\over{R}}  \,.
\ee
The entropy is
\be
S \sim (R N_1 N_5) (1) T_L \,.
\ee
Therefore, the temperature is as before,
\be
{{1}\over{T_L}} \sim {{(RN_1N_5)}\over{(N_1N_5N_m)^{\half}}} 
\ll RN_1N_5 \,;
\ee
but this time it is plenty hot enough for the equation of state to be
valid because the box size is bigger by a factor of $N_1N_5$.

The entropy counting now proceeds in a similar manner as before, but
the central charge and the radius are modified as
$c=6N_1N_5\rightarrow c=6$ and $R\rightarrow RN_1N_5$.  The result is
identical.

\subsection{$d{\eql}4$ entropy counting}

A canonical set of ingredients for building the $d{\eql}4$ system is
what we had previously in building the black hole:
\be\ba{ccccccccccc}
& 0& 1 & 2 & 3 & 4 & 5 & 6 & 7 & 8 & 9\cr
{\rm{D2}} 
&-& - & - & \sim & \sim & \sim & \sim & \cdot & \cdot & \cdot \cr
{\rm{D6}} 
&-& - & - & - & - & - & - & \cdot & \cdot & \cdot \cr
{\rm{NS5}}
&-& - & - & - & - & - & \sim & \cdot & \cdot & \cdot \cr
{\rm{W}}  
&-& \rightarrow & \sim & \sim & \sim & \sim & \sim & \cdot & \cdot &
\cdot
\ea\ee
The new feature of this system compared to the previous one is that
D2-branes can end on NS5-branes.  It costs zero energy to break up a
D2-brane as shown in Fig.\ref{ns5splitd2}.

\bfig
\hskip0.2\textwidth\epsfysize=1.125truein
\epsfbox{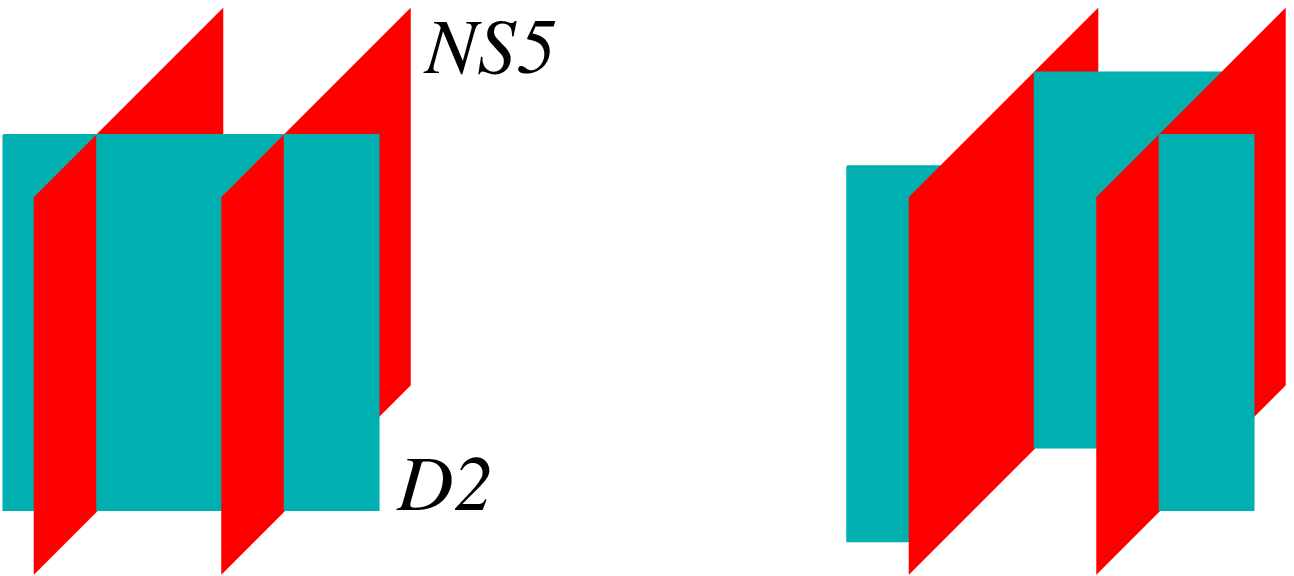}
\caption{\small{How D2's can split in the presence of NS5's.}}
\label{ns5splitd2}
\efig

These extra massless degrees of freedom in the system lead to an extra
label on the 2-6 strings, giving rise to an extra factor of $N_{\rm
NS5}$ in the degeneracy.  The entropy counting proceeds just as
before, and yields \cite{cliffetal}
\be
S_{\rm micro} = 2\pi\sqrt{N_2N_6N_{\rm NS5}N_m} \,,
\ee
which again agrees exactly with the Bekenstein-Hawking black hole
entropy.  A major difference between this and the $d{\eql}5$ case is that
rotation is incompatible with supersymmetry; in addition, there can be
only one angular momentum $J$.

\sect{Non-BPS systems, and Hawking radiation}\label{sectseven}

Among black holes and black branes, BPS systems are the systems under
the greatest theoretical control because supersymmetry implies the
presence of nonrenormalisation theorems for quantities including
entropy.  Their non-BPS counterparts are also very much worthy of
study and we now turn to a discussion of their properties.

\subsection{Nonextremality}

The nonextremal black hole metric for the D1 D5 W system comes
from the $d{\eql}10$  string frame metric
\be\ba{rl}\label{tendd1d5w}
\bs
dS_{10}^2 &\! = D_1(r)^{-\half} D_5(r)^{-\half} \left[ -dt^2 + dz^2
+ K(r)\left(\cosh\!\alpha_m dt + \sinh\!\alpha_m dz\right)^2\right] 
\cr &\! + D_1(r)^{+\half} D_5(r)^{-\half} dx_\parallel^2 
+ D_1(r)^{+\half} D_5(r)^{+\half} 
\left[ {\displaystyle{ {{dr^2}\over{\left(1-K(r)\right)}} }} 
+ r^2 d\Omega_3^2 \right] 
\ea\ee
where
\be
K(r) = {{\rH^2}\over{r^2}} \,,\qquad
f_{1,5}(r)= 1+ K(r)\sinh^2\!\alpha_{1,5}\,,
\ee
and $\alpha$'s are the boost parameters used to make this solution.
The conserved charges are given by
\be
N_1 = {{   V\rH^2}\over{\gs\ls^6  }}{{\sinh(2\alpha_1)}\over{2}}
\,,\quad
N_5 = {{    \rH^2}\over{\gs\ls^2  }}{{\sinh(2\alpha_5)}\over{2}}
\,,\quad
N_m = {{R^2V\rH^2}\over{\gs^2\ls^8}}{{\sinh(2\alpha_m)}\over{2}}
\,,;
\ee
and the mass and thermodynamic quantities are
\be\ba{rl}
\bs
M_{ADM} = & {\displaystyle{
{{RV\rH^2}\over{\gs^2\ls^8}}\left[
{{\cosh(2\alpha_1)}\over{2}}+{{\cosh(2\alpha_5)}\over{2}} + 
{{\cosh(2\alpha_m)}\over{2}} \right]  }} \,;\cr
\SBH = & {\displaystyle{
{{2\pi{}RV\rH^3}\over{\gs^2\ls^8}}\left[
\cosh\!\alpha_1\cosh\!\alpha_5\cosh\!\alpha_m\right]   }} \,;\cr
\TH = & {\displaystyle{
{{\ls}\over{2\pi\rH\left[
\cosh\!\alpha_1\cosh\!\alpha_5\cosh\!\alpha_m\right]}} }} \,.
\ea\ee
In the limit
\be\label{dilgas5d}
\rH^2\sinh^2\!\alpha_{1,5}\equiv r_{1,5}^2 \gg
r_m^2\equiv \rH^2\sinh^2\!\alpha_{m}\quad\gg\ls^2 \,,
\ee
the expression for the ADM mass simplifies; the energy above
extremality becomes
\be
\Delta E \equiv M - \left({{N_5RV}\over{\gs\ls^6}}
+{{N_1R}\over{\gs\ls^2}}\right)
\simeq {{RV\rH^2}\over{\gs^2\ls^8}}{{\cosh(2\alpha_m)}\over{2}}\,.
\ee
We also had that
\be
{{N_m}\over{R}} =
{{RV\rH^2}\over{\gs^2\ls^8}}{{\sinh(2\alpha_m)}\over{2}} \,;
\ee
now define
\be
N_{L,R} = {{R^2V\rH^2}\over{4\gs^2\ls^8}}e^{\pm 2\alpha_m} \,.
\ee
{}From this we can see that the system has effectively split into
independent gases of left- and right-movers:
\be
\Delta E = {{1}\over{2\ls}}\left(N_R+N_L\right) \,,\qquad
N_m = N_L - N_R \,.
\ee
This regime is dubbed the ``dilute gas regime'' because the energies
and momenta are additive.  This regime is, in fact, exactly what is
selected by taking the decoupling limit we discussed in section
(\ref{sectfour}).

Let us proceed to compute the Bekenstein-Hawking entropy.  In the
dilute gas limit (\ref{dilgas5d}), the only boost parameter which is
still effectively non-infinite is $\alpha_m$.  The entropy is then
proportional to
\be
\cosh\!\alpha_m = 
   {{1}\over{2}}\left(e^{\alpha_m}+e^{-\alpha_m}\right)
 = {{1}\over{2}} \left(\sqrt{{4\gs^2\ls^8N_L}\over{R^2V\rH^2}}
 + \sqrt{{4\gs^2\ls^8N_R}\over{R^2V\rH^2}}\right) \,.
\ee
The dilute gas entropy becomes \cite{snonext}
\be\ba{rl}
\bs\bs
\SBH = & {\displaystyle{
{{2\pi{}RV\rH^3}\over{\gs^2\ls^8}}
\left({{\gs N_1\ls^6}\over{V\rH^2}}\right)^{\half}
\left({{\gs N_5\ls^2}\over{\rH^2}}\right)^{\half}
\left({{\gs^2\ls^8}\over{R^2V\rH^2}}\right)^{\half}
\left[\sqrt{N_L}+\sqrt{N_R}\right] }}  \cr
= & 2\pi\left(\sqrt{N_1N_5N_L}+\sqrt{N_1N_5N_R}\right) \,.
\ea\ee
Thus the entropy is additive.  By very similar calculations as before
we can see that the D-brane entropy counting gives exactly the same
result as the Bekenstein-Hawking entropy in the dilute gas regime.
For the nonextremal system the agreement persists even with the
introduction of rotation \cite{blmpsv}.  In the case of the $d{\eql}4$
four-charge black holes, similar results ensue \cite{hlmne4d}.

We may ask at this stage why the entropy of these near-extremal
supergravity and perturbative D-brane systems agree, as there is no
theorem protecting the degeneracy of non-BPS states.  What is going on
physically is that conformal symmetry possessed by the
$d{\eql}1{\pls}1$ theory is sufficiently restrictive, even when it is
broken by finite temperature, for the black hole entropy to be
reproduced by the field theory.  

\subsection{The BTZ black hole and the connection to D1-D5}

In three spacetime dimensions, the rule
\be
{\rm{``}} g_{tt} = -1 + \left({{\rH}\over{r}}\right)^{d-3} {\rm{''}}
\ee
for spacetimes without cosmological constant no longer applies because
of logarithmic divergence problems.  If, however, there is a negative
cosmological constant, then there are well-behaved black holes, the
BTZ black holes \cite{btz}.  They are solutions of the action
\be
S = {{1}\over{16\pi G_3}}\int d^3x
\sqrt{-g}\left(R_g+{{2}\over{\ell^2}}\right) \,,
\ee
\aie the cosmological constant is $\Lambda=-1/\ell^2$.  The
metric is
\be\ba{rl}
\bs
ds_{\rm{BTZ}}^2 = &\! {\displaystyle{
-{{(w^2-w_+^2)(w^2-w_-^2)}\over{\ell^2w^2}}dt^2
+ {{\ell^2w^2}\over{(w^2-w_+^2)(w^2-w_-^2)}}dw^2 }} \cr
& {\displaystyle{
+ w^2\left(d\varphi+{{w_+w_-}\over{\ell w^2}}dt\right)^2 }} \,.
\ea\ee
The coordinate $\varphi$ is periodic, with period $2\pi$.

The event horizons are at $w=w_\pm$, and the mass and angular momenta
are given by
\be
M = {{(w_+^2-w_-^2)}\over{8\ell^2G_3}} \,,\qquad
J = {{(w_+w_-)}\over{4\ell{}G_3}} \,.
\ee
The thermodynamic entropy and temperature are
\be
\SBH = {{2\pi{}w_+}\over{4G_3}} \,,\qquad
\TH = {{(w_+^2-w_-^2)}\over{2\pi{}w_+\ell^2}} \,.
\ee

Consider the object with the following specific negative value of the
mass parameter:
\be
J=0\,,\qquad M= - {{1}\over{8\ell^2G_3}}\,.
\ee
This animal is not a black hole, but the metric becomes
\be
ds^2 = -{{(r^2+1)}\over{\ell^2}}dt^2 +
{{\ell^2}\over{(r^2+1)}}dr^2 + r^2d\varphi^2\,.
\ee
This is $AdS_3$ in global coordinates.  In fact, due to the properties
of $d{\eql}3$ gravity, the BTZ spacetime is everywhere locally $AdS_3$.
There is, however a global obstruction: $\varphi$ is compact.

We are mentioning the BTZ spacetime because in many earlier papers on
D-branes and entropy counting, a so-called ``effective string'' model
kept popping up in descriptions of the physics.  In fact, this
effective string story amounted to having a BTZ black hole lurking in
the geometry in each case.  This is intimately related to the
$AdS_3/CFT_2$ and $AdS_2/CFT_1$ correspondences; see the review
\cite{magoo} for more details.

Now, let us work on the connection \cite{hyun} between the BTZ black
holes we have just studied, and the spacetime metric for the D1-D5-W
system.  The nonextremal 3-charge $d{\eql}5$ black hole descends from
the $d{\eql}10$ metric (\ref{tendd1d5w}) we displayed in the last
subsection.  Let us wrap the four dimensions $x_\parallel$ of the D5
not parallel to the D1 on a $T^4$.  Reducing on $x_\parallel$, we get
a $d{\eql}6$ black string
\be\ba{rl}
\bs
dS^2_6 = &\! D_1(r)^{-\half}D_5(r)^{-\half}\left[
-dt^2+dz^2+
K(r)\left(\cosh\!\alpha_mdt+\sinh\!\alpha_mdz\right)^2\right]
\cr & +{\displaystyle{
 D_1(r)^{+\half}D_5(r)^{+\half}\left[
{{dr^2}\over{\left(1-K(r)\right)}}+r^2d\Omega_3^2\right] 
}}\,.
\ea\ee
Now let us define the near-horizon limit.  We will take
\be
r^2 \ll r^2_{1,5}\equiv\rH^2\sinh^2\!\alpha_{1,5} \,,
\ee
but we will not demand a similar condition on $r_m$.  (This is the
dilute gas condition all over again.)  In this limit, the volume of
the internal $T^4$ goes to a constant at the horizon,
\be
{\rm{Vol}}(T^4) \rightarrow V_4 
\left({{r_1^2}\over{r_5^2}}\right) \,,
\ee
and so does the dilaton:
\be
e^\Phi \rightarrow \left({{r_1}\over{r_5}}\right) \,.
\ee
These two scalars are examples of ``fixed scalars''.  They are not
minimally coupled.

Since the dilaton is constant near-horizon, the near-horizon string
and Einstein metrics differ only by a constant (which we now
suppress).  The angular piece of the metric also dramatically
simplifies:
\be
G_{\Omega\Omega} = 
r^2\sqrt{1+{{r_1^2}\over{r^2}}}\sqrt{1+{{r_5^2}\over{r^2}}}
 \longrightarrow\  r_1 r_5 \equiv \lambda^2 \,;
\ee
we get a 3-sphere of constant radius $\lambda$.  For the other piece
of the metric
\be
ds_{t,z,r}^2 \rightarrow
{{r^2}\over{\lambda^2}}\left[-dt^2+dz^2+
K(r)\left(\cosh\!\alpha_mdt+\sinh\!\alpha_mdz\right)^2\right]
+ {{\lambda^2dr^2}\over{r^2\left(1-K(r)\right)}} \,.
\ee
Defining
\be
w_+^2 \equiv \rH^2 \cosh^2\!\alpha_m \,, \quad
w_-^2 \equiv \rH^2 \sinh^2\!\alpha_m \,, \quad
\ee
we get
\be\ba{rl}
ds_{t,r,z}^2 = &\! {\displaystyle{
{{1}\over{\lambda^2}}
\left[-dt^2(r^2-w_+^2)+dz^2(r^2+w_-^2)+2dtdzw_+w_-\right] }} \cr
& {\displaystyle{
+ {{\lambda^2dr^2}\over{r^2\left(1-(w_+^2-w_-^2)/r^2\right)}} }} \,.
\ea\ee
Changing coordinates to
\be
w^2\equiv r^2+w_-^2 \,,
\ee
and doing some algebra the $d{\eql}6$ metric can be rearranged to
\be\ba{rl}
\bs
ds^2 = & {\displaystyle{
-dt^2{{(w^2-w_+^2)(w^2-w_-^2)}\over{\lambda^2w^2}} +
 {{w^2\lambda^2dw^2}\over{(w^2-w_+^2)(w^2-w_-^2)}} }} \cr
& + {\displaystyle{
{{w^2}\over{\lambda^2}}\left(dz+{{w_+w_-}\over{w^2}}dt\right)^2
+ \lambda^2 d\Omega_3^2 }} \,.
\ea\ee
This is recognisable as the direct product of $S^3$ and a BTZ black
hole, if we simply rescale coordinates as
\be
z\rightarrow {{z}\over{R}} \equiv\varphi   \,,\qquad
w\rightarrow {{wR}\over{\lambda}} \,\qquad
t\rightarrow {{t\lambda}\over{R}} \,.
\ee
{}From this it appears that only remnant of the D1,D5 data goes into
the cosmological constant $\lambda=\ell$ for the BTZ black hole; this
is a consequence of having taken the near-horizon limit.  In fact,
there is an overall constant $r_1/r_5$ differentiating the Einstein
metric from the string metric, which we suppressed.  In addition, we
are required to compactify $z$, the direction along the D1-brane, in
order to make the identification precise.

Now note that only the momentum charge controls extremality, because
\be
r_\pm^2 \equiv \rH^2 \ba{c}\cosh^2\cr\sinh^2\ea\alpha_m  \,,
\ee
and so we get the relations
\be\ba{rr}
{\rm{wrapped\ extremal\  black\  string\ }} 
\longrightarrow& {\rm{extremal\  BTZ}}\times{\rm{S}}^3 \,, \cr
{\rm{wrapped\ nonextremal\  black\  string\ }} 
\longrightarrow& {\rm{nonextremal\  BTZ}}\times{\rm{S}}^3 \,.
\ea\ee

For $d{\eql}4$ black holes, the structure is BTZ$\times S^2$, which can
be seen by considering a $d{\eql}5$ black string.  A BTZ spacetime also
appears even for rotating black holes, but it is only a local
identification; there is a global obstruction.  In addition, one has
to go to a rotating coordinate system to see the BTZ structure
\cite{finn}.

There are entropy-counting methods available which use only the
properties of three-dimensional gravity, see \aeg \cite{btzentropy}
for a discussion of some of the physics issues.  We do not have space
to discuss these methods here.

\subsection{A universal result for black hole absorption}

We would now like to review the calculation of \cite{DGM} of the
absorption cross-section for a spherically symmetric black hole.  We
will then go on to study the analogous process in the D-brane picture.

The semiclassical black hole calculations involve several steps.  One
begins with a wave equation for the ingoing mode of interest, which
can be complicated due to mixing of modes.  This wave equation is not
always separable.  Typically it is necessary to use approximations to
find the behaviour of wavefunctions in different regions of the
geometry.  The last step is to match the approximate solutions to get
the absorption probability, and thereby also the absorption
cross-section.  For emission we use detailed balance.

In performing the calculations it is found that the absorption
probability is not unity because the curved geometry outside the
horizon backscatters part of the incoming wave.  Also, the dominant
mode at low-energy turns out to be the $s$-wave.  The result of
\cite{DGM} is that the low-energy $s$-wave cross-section for
absorption of minimally coupled scalars by a $d$ dimensional
spherically symmetric black hole is universal, the area of the event
horizon.  Let us review this calculation, to illustrate how very
different it is from the D-brane computation.

The $d$-dimensional spherically symmetric black hole metric takes the
form in Einstein frame
\be
ds^2 = -f(\rho)dt^2 + g(\rho)\left[d\rho^2 +
\rho^2d\Omega_{d-2}^2\right] \,.
\ee
If the metric is not already in this form, a coordinate transformation
can always be found to bring it so.  Then $\sqrt{-g} =
\sqrt{f(\rho)g(\rho)^{d-1}}\rho^{d-2}$.  For minimally coupled
scalars, the wave equation is
\be
\nabla^\mu\nabla_\mu \Psi = {{1}\over{\sqrt{-g}}}\partial_\mu
\left(\sqrt{-g}g^{\mu\nu}\partial_\nu\right) \Psi =0 \,.
\ee
For the $s$-wave,  let $\Psi=\Psi_\omega(\rho) e^{-i\omega t}$,
and so
\be
\partial_t\left(g^{tt}\partial_t\right)\Psi_\omega 
- {{1}\over{\sqrt{f(\rho)g(\rho)^{d-1}}\rho^{d-2}}}\partial_\rho
\left(\sqrt{f(\rho)g(\rho)^{d-1}}\rho^{d-2}\, g(\rho)^{-1}
\partial_\rho\right)\Psi_\omega = 0  \,.
\ee
Take the frequency of the wave $\omega$ to be much smaller than any
energy scale set by the black hole.  This is the definition of
``low-energy''.  Now, defining
\be\label{wotsit}
\partial_\sigma \equiv 
\sqrt{f(\rho)g(\rho)^{d-3}}\rho^{d-2}\partial_\rho \,,
\ee
leads to the wave equation
\be
\left( \partial_\sigma^2 + 
\left[\rho^2(\sigma)g(\rho(\sigma))^{d-2} \,\omega^2\, \right]\right)
\Psi_\omega(\sigma)=0\,.
\ee
Let the horizon be at $\rho=\rH$; then the entropy is in these
conventions
\be
\SBH = {{\Omega_{d-2}\rH^{d-2}g(\rH)^{\half(d-2)}}\over{4G_d}}
\equiv {{\Omega_{d-2}}\over{4G_d}} \RH^{d-2} \,,
\ee
Consider now the function in front of $\omega^2$ in the previous
equation.  Near the horizon, (in the ``near zone'') the wave equation
is
\be
\left[ \partial_\sigma^2 + \omega^2 \RH^{2(d-2)}\right]
\Psi^{\rm{near}}_\omega(\sigma) = 0\,.
\ee
The solution must be purely ingoing at the horizon and so
\be
\Psi^{\rm{near}}_\omega(\sigma) = e^{-i\omega\RH^{d-2}\sigma}\,.
\ee
We need to know how far out in $\rho$ this solution is good.  It works
when the above approximation we used in the wave equation is good, and
that will be for $\rho$'s such that the area of the sphere is still of
order the horizon area.  By studying (\ref{wotsit}) very carefully, we
can see that this is in fact far enough out that the small-$\sigma$
approximation is roughly valid.  This turns out to be enough to
guarantee that there is a region of overlapping validity of this
near-zone wavefunction with the far-zone wavefunction which we will
get to shortly.

So at the edge of its region of validity the near-zone wavefunction is
\be
\Psi^{\rm{near}}_\omega(\rho) \Bigr|_{\rm edge} \sim
1-i\omega\RH^{d-2}{{\rho^{3-d}}\over{(3-d)}} \,.
\ee
The next item on the agenda is the ``far-zone'' wavefunction.  Far
away, $\rho$ is the smarter variable to use: 
\be
\left[\rho^{d-2}\partial_\rho\left(\rho^{d-2}\partial_\rho\right) +
\omega^2\rho^{2(d-2)}\right]\Psi^{\rm{far}}_\omega =0\,; 
\ee
changing variables to eliminate the linear derivative
\be
\Psi^{\rm{far}}_\omega\equiv\rho^{-\half(d-2)}\chi_\omega \,,
\ee
and defining
\be
z\equiv \omega\rho \,,
\ee
gives
\be
\left[\partial_z^2 + 
1 - {{(d-2)(d-4)}\over{4z^2}}\right]\chi_\omega=0\,.
\ee
Solutions to this equation are Bessel functions for $\chi_\omega(z)$,
so that
\be
\Psi^{\rm{far}}_\omega(z) = z^{\half(3-d)}
\left[AJ_{\half(d-3)}(z)+BJ_{-\half(d-3)}(z)\right] \,.
\ee
In order to find the behaviour of this wavefunction on the edge of its
region of validity, use the small-$z$ series expansions
\be
J_\nu(z)\rightarrow
\left({{z}\over{2}}\right)^\nu{{1}\over{\Gamma(\nu+1)}} \,,
\ee
to get
\be
\Psi^{\rm{far}}(\rho) \Bigr|_{\rm edge} \sim
{{2^{\half(3-d)}}\over{\Gamma[\half(d-1)]}}A +
{{2^{\half(d-3)}}\over{\Gamma[\half(5-d)](\omega\rho)^{d-3}}} B \,,
\ee
Matching to the near-zone wavefunction on its edge yields
\be
A = \Gamma[\half(d-1)]2^{\half(d-3)} \qquad B = i
{{\Gamma[\half(5-d)]2^{\half(3-d)}(\omega\RH)^{d-2}}\over{(3-d)}}\,.
\ee
Far away, we use the $z\rightarrow\infty$ expansion of the Bessel
functions\footnote{If $d$ is odd, the Bessel functions $J_{\pm\nu}$
are not independent; the result is unaffected but the details are
slightly different.} and the behaviour is oscillatory,
\be
J_\nu(z)\rightarrow \sqrt{{2}\over{\pi z}}\left[\cos\left(z -
{{\pi\nu}\over{2}} - {{\pi}\over{4}} \right)\right]\,, 
\ee
as we would expect for a wave.  Then
\be\ba{rl}
\bs
\Psi_\omega^{\rm{far}}(\omega\rho) \rightarrow &\!
{\displaystyle{
{\sqrt{{2}\over{\pi (\omega\rho)^{d-2}}}} \left(
e^{+i(\omega\rho-\quarter\pi)}\left[ e^{-i\quarter(d-3)} \half A +
e^{+i\quarter(d-3)} \half B \right]
\right.  }} \cr &\!+ {\displaystyle{
\left.  e^{-i(\omega\rho-\quarter\pi)}\left[ e^{+i\quarter(d-3)} \half A 
+ e^{-i\quarter(d-3)}
\half B \right] \right) }} \,.
\ea\ee
Now, the absorption probability is
\be\ba{rl}
\bs
\Gamma = & 1 - \left| {\rm{Reflection\ coefficient\ }} \right|^2 
\cr
= & {\displaystyle{ 1 - 
\left| {{A + Be^{+i\half(d-3)}}\over{A + Be^{-i\half(d-3)}}} 
\right|^2 }}
\,.
\ea\ee
Lastly, the fluxes need normalising because ingoing plane waves are
used rather than ingoing spherical waves,
\be
e^{ikz} \equiv N {{e^{-i\omega \rho}}\over{\rho^{\half(d-2)}}} 
\left(Y_{0\cdots 0} = {{1}\over{\sqrt{\Omega_{d-2}}}}\right) \,.
\ee
Putting it all together yields
\be
\sigma_{\rm{abs}} = \Gamma|N|^2 = 
{{2\sqrt{\pi}^{d-1}\RH^{d-2}}\over{\Gamma[\half(d-1)]}}
\equiv A_{\rm{H}} \,.
\ee
This result for low-energy minimally coupled scalar $s$-waves is
completely universal for spherically symmetric black holes.  To our
knowledge, it is an interesting open problem to find whether this
result carries over to black holes with angular momentum.

\subsection{Emission from D-branes}

The BPS D1-D5 system with momentum has no right-movers at all; this
was necessary for it to be supersymmetric.  Adding a little
nonextremality gives a few right-movers, $N_R{\ll}N_L$.  Using the gas
picture which we used in our entropy discussion and in explaining
fractionation, we get \cite{juanthesis}
\be\label{tltr}
T_{L,R} = {{1}\over{\pi R}} {{\sqrt{N_{L,R}}}\over{\sqrt{N_1N_5}}}\,.
\ee
These temperatures are related to the Hawking temperature $\TH$ as
\be
T_L^{-1} + T_R^{-1} = 2 T_H^{-1}\,.
\ee
Since in the dilute gas approximation the right-movers are far less
numerous than the left-movers, the temperatures (\ref{tltr}) satisfy
$T_L{\gg}T_R$.  Therefore, to a good approximation, $T_H\simeq T_R$.

Consider low-energy left- and right-moving quanta, with frequencies
$\omega$ $n$ times the gap frequency
$\omega_{\rm{gap}}{\sim}1/(N_1N_5R)$.  Using the relation (\ref{tltr})
for the temperature, we can see that the frequencies satisfy
$\omega{\ll}T_L$.  If we consider nontrivial scattering, the dominant
process at low energy will be the collision of two open strings
joining up to make a closed string which then moves off into the bulk.
\be
\hskip-1.0truein
\left(\omega_L={{+n}\over{RN_1N_5}}\right) +
\left(\omega_R={{-n}\over{RN_1N_5}}\right)
\quad\longrightarrow\quad 
\left(\omega_c={{2n}\over{RN_1N_5}}\right) \,.
\ee
This emission from the brane is the D-brane analogue of Hawking
radiation.

For all but very near-BPS cases, $N_R$ is macroscopically small but
still microscopically large, so we use the canonical ensemble.  (When
$N_R\rightarrow 0$, the thermality approximation will break down, and
on the black hole side we will be in trouble with the third law.)

The rate for the emission process is \cite{dasmathur}
\be\ba{rcccc}
d\Gamma \sim 
& {\underbrace{ {{d^4k}\over{\omega_c}} }}
& {\underbrace{ {{\ls^5}\over{RV \omega_L\omega_R}}  }}
& {\underbrace{ \delta\left(\omega_c - (\omega_L+\omega_R)\right) }}
& \left| {\cal{A}} \right|^2 \,, \cr
{\ }
& \perp\ {\rm{phase\ space\ }} 
& {\rm{normalizations\ }} 
& {\rm{momentum\ conservation\ }} 
& {\rm{coupling\ }} \,.
\ea\ee 
For simplicity, let us consider emission of a quantum corresponding to
a minimally coupled bulk scalar, such as an internal component of
$G_{\mu\nu}$.  The calculation of the amplitude was first done in
\cite{dasmathur} and we now review it for purposes of illustration
here.  The computation proceeds by considering the D-string worldsheet
theory in a supergravity background which is to first approximation
taken to be Minkowski space with no gauge fields and constant dilaton.
The piece of the brane action we need is
\be
S_{\rm DBI} = -{{1}\over{(2\pi)\gs\ls^2}} \int d^2\sigma e^{-\Phi}
\sqrt{-\det(\bP(G_{\alpha\beta}))} + \ldots \,,
\ee
Let us pick static gauge, and expand the spacetime string metric as
\be
G_{\mu\nu}=\eta_{\mu\nu} + 2\kappa_{10} h_{\mu\nu}(X) \,.
\ee
The $\kappa_{10}$ in this relation is the same one appearing in the
Einstein-Hilbert action
\be
S_{\rm bulk} = {{1}\over{2\kappa_{10}^2}} \int d^{10}x \sqrt{-G}
e^{-2\Phi} R[G] + \ldots \,.
\ee
Then the kinetic term for $h$ is canonically normalised
\be
{\cal{L}}_{\rm bulk} \sim \half \left(\partial
h_{ij}\right)\left(\partial h_{ij}\right) \,,
\ee
while the brane action yields
\be
{\cal{L}} \sim \left(\delta_{ij} + 2\kappa_{10} h_{ij}\right) 
\partial_\alpha X^i\partial^\alpha X^j \,.
\ee
To get this expression, we soaked up a factor of the string tension in
the $X^i$'s to get canonically normalised kinetic energies for them.
This rescaling will not affect our answer because, to lowest order,
the Lagrangian is only quadratic and is therefore independent of the
tension.  For the interaction Lagrangian we then have
\be
{\cal{L}}_{\rm int} \sim \kappa h_{ij} \partial_{\hat{\alpha}} X^i
\partial^{\hat{\alpha}} X^j \,.
\ee
At this point we use the relation that $\kappa_{10}\sim \gs\ls^4$.
Assuming for simplicity that the outgoing graviton momentum is
perpendicular to the D-string, this gives rise to the amplitude
\be
{\cal{A}}\sim\gs\omega^2\ls^2 \,.
\ee  
We use this amplitude as our basic starting point for computing the
emission probability.  Averaging over initial states and and summing
over final gives rise to the occupation factors
\be
\rho_{L,R}(\omega) = {{1}\over{e^{\omega/(2T_{L,R})}-1}} \,;
\ee
in our case in the dilute gas approximation we have
\be
\rho_L(\omega)\simeq{{2T_L}\over{\omega}} \,,\quad
\rho_R(\omega)\simeq{{1}\over{e^{\omega/\TH}-1}} \,.
\ee
Then the emission rate goes as
\be
d\Gamma \propto {{d^4k\ls^7}\over{\omega^3RV}} (N_1N_5R)
\gs^2\omega^4 {{2T_L}\over{\omega}}{{1}\over{e^{\omega/\TH}-1}} \,.
\ee
Computing the exact coefficient gives the precise relation
\be
d\Gamma= A_{\rm{H}} {{1}\over{e^{\omega/\TH}-1}}
{{d^4k\ls^4}\over{(2\pi)^4}} \,.
\ee
This tells us that emission is thermal at the Hawking temperature.
Physically, thermality is a consequence of our having averaged over
initial states.  Using detailed balance to convert emission to
absorption, we find the absorption cross-section to be
\be
\sigma = A_{\rm{H}} \,,
\ee
the area of the event horizon.  This agrees precisely with the result
obtained for the black hole from semiclassical gravity, which we saw
in the last subsection.

The agreement is in fact a many-parameter affair, in that the actual
result for the horizon area depends on many different conserved
quantum numbers.  The agreement depends heavily on the presence of
greybody factors, previously thought to be a nuisance but now seen to
contain interesting physics.  The reader following the normalisation
factors precisely will also have noticed that it was crucial that we
used the length of the circle given by fractionation physics; we would
have been off by powers of $N_{1,5}$ if we had failed to do so.

This is just one example of a more general class of D-brane
calculations which agrees precisely with black hole emission and
absorption rates.  Results obtained in the dilute gas regime turn out
to agree between the supergravity and perturbative D-brane pictures,
whereas for other regimes the agreement is typically less precise.  In
some cases, an appeal to the correspondence principle was necessary in
order to track down missing degrees of freedom giving rise to
contributions to emission/absorption processes.

A great deal of work has been done on comparing decay rates for black
holes and D-brane/strings via computation of scattering amplitudes.
We cannot give a representative or complete list of references here,
but we suggest \cite{dasmathur,maldastromabs,igorabs}.  This general
body of work contributed to identification of operators in the gauge
theory corresponding to bulk supergravity modes in the AdS/CFT
correspondence.

\vskip1.0truein
\section*{Acknowledgements}

This work was supported in part by NSF grant PHY94-07194.  

During construction of this material, I benefited from useful
discussions with many people, including Shanta de Alwis, Mike Duff,
Steve Giddings, David Gross, Steve Gubser, Aki Hashimoto, Simeon
Hellerman, Gary Horowitz, Sunny Itzhaki, Clifford Johnson, Jason
Kumar, Finn Larsen, Don Marolf, Lubos Motl, Joe Polchinski, Simon
Ross, and Eva Silverstein.  I thank L{\'{a}}rus Thorlacius especially
for a critical reading of an earlier version of the manuscript.

I would also like to thank Kayll Lake and his colleagues at Queen's
University for their excellent research tool GRTensorII.  

\vfill

\boxit{0.8\textwidth}{\it Send error corrections, reasonable reference
requests, and excellent suggestions for improvement to\\
\hbox{\hspace{7em}}{\tt peet@physics.utoronto.ca} \\ but please be
sure to put ``TASI-99'' in the subject line to ensure that your
message gets the proper attention.  If you are going to send many
corrections, PLEASE send them all at once.  Thanks.}

\vfill

\addcontentsline{toc}{section}{\protect\numberline{8}{References}}

\end{document}